\documentclass[11pt,a4paper]{article}
\usepackage[a4paper, width = 150mm, top = 30mm, bottom = 30mm]{geometry}
\usepackage{jheppub}
\usepackage{mathtools}
\usepackage{mathrsfs}
\usepackage{float}
\usepackage{pgfplots}
\usetikzlibrary{hobby}
\usetikzlibrary{decorations.markings}
\usepackage{makecell}
\usepackage{circuitikz}
\usepackage{subcaption}
\usepackage{ifthen}
\usepackage{wasysym}
\usepackage[bbgreekl]{mathbbol}
\usepackage{feynmp-auto}
\usepackage{braket}

\pdfoutput=1

\newcommand{\verteq}{\rotatebox{90}{$\,=$}}
\newcommand{\equalto}[2]{\underset{\overset{\verteq}{#2}}{#1}}

\newcommand{\dJ}[1]{\frac{1}{i}\frac{\delta}{\delta \mathfrak{J}(\zeta,#1)}}
\newcommand{\dJh}[1]{-i\frac{\delta}{\delta \mathfrak{J}(#1)}}

\newcommand{\phiavJ}[1]{\braket{\phi(#1)}_{\Jbksmall}}

\newcommand{\A}{\scalebox{.9}{$\scriptscriptstyle A$}}
\newcommand{\B}{\scalebox{.9}{$\scriptscriptstyle B$}}
\newcommand{\C}{\scalebox{.9}{$\scriptscriptstyle C$}}
\newcommand{\D}{\scalebox{.9}{$\scriptscriptstyle D$}}

\newcommand{\bbG}{\mathbb{G}}
\newcommand{\bbGA}{\mathbb{G}_{\text{adv}}}
\newcommand{\bbGR}{\mathbb{G}_{\text{ret}}}
\newcommand{\bbGK}{\mathbb{G}_{\text{K}}}
\newcommand{\WR}{W_{\text{ret}}}
\newcommand{\WA}{W_{\text{adv}}}
\newcommand{\Gin}{G^{\text{in}}}
\newcommand{\Gout}{G^{\text{out}}}
\newcommand{\Kin}{K^{\text{in}}}

\newcommand{\Pplus}{\mathbb{P}_{+}}
\newcommand{\Pmin}{\mathbb{P}_{-}}
\newcommand{\G}{\Gamma}
\newcommand{\Gdag}{\Gamma^{\dagger}}
\newcommand{\Psibar}{\overline{\Psi}}
\newcommand{\psibar}{\overline{\psi}}
\newcommand{\psiP}{\psi_{\bar{\Psmall}}}
\newcommand{\psiF}{\psi_{\bar{\F}}}
\newcommand{\psibP}{\overline{\psi}_{\bar{\Psmall}}}
\newcommand{\psibF}{\overline{\psi}_{\bar{\F}}}

\newcommand{\bbS}{\mathbb{S}}
\newcommand{\bbSA}{\mathbb{S}_{\text{adv}}}
\newcommand{\bbSR}{\mathbb{S}_{\text{ret}}}
\newcommand{\bbSDD}{\mathbb{S}_{{}_\text{DD}}}
\newcommand{\bbSK}{\mathbb{S}_{\text{K}}}
\newcommand{\bbW}{\mathfrak{W}}
\newcommand{\frakW}{\mathbb{W}}
\newcommand{\Wf}{\mathbb{W}_{\mathbb{g}}}
\newcommand{\WfR}{\mathbb{W}_{\mathbb{g}\text{ret}}}
\newcommand{\WfA}{\mathbb{W}_{\mathbb{g}\text{adv}}}
\newcommand{\Sbo}{\overline{S}_0}
\newcommand{\Sin}{S^{\text{in}}}
\newcommand{\Sout}{S^{\text{out}}}
\newcommand{\Sbin}{\overline{S}^{\text{in}}}
\newcommand{\Sbout}{\overline{S}^{\text{out}}}

\newcommand{\bulkalpha}{\scalebox{1.2}{$\scriptscriptstyle \bbalpha$}}
\newcommand{\bulkbeta}{\scalebox{1.2}{$\scriptscriptstyle \bbbeta$}}
\newcommand{\bulkgamma}{\scalebox{1.2}{$\scriptscriptstyle \bbgamma$}}
\newcommand{\bulkdelta}{\scalebox{1.2}{$\scriptscriptstyle \bbdelta$}}

\newcommand{\Psmall}{\scalebox{.9}{$\scriptscriptstyle \rm P$}}
\newcommand{\Fsmall}{\scalebox{.9}{$\scriptscriptstyle \rm F$}}
\newcommand{\Jbksmall}{\scalebox{.9}{$\scriptscriptstyle \mathfrak{J}$}}

\newcommand{\Pb}{\bar{\Psmall}}
\newcommand{\Fb}{\bar{\Fsmall}}

\newcommand{\sR}{{}_{\text{R}}}
\newcommand{\sL}{{}_{\text{L}}}

\newcommand{\z}{\zeta}
\newcommand{\ka}{\kappa}

\newcommand{\eF}{e_{\bar{\F}}}
\newcommand{\eP}{e_{\bar{\Psmall}}}

\newcommand{\Jblk}{\mathfrak{J}}
\newcommand{\PhiPI}{\Phi_{\rm 1PI}}
\newcommand{\GPI}{\Gamma_{\rm 1PI}}

\newcommand{\red}[1]{\textcolor{red}{#1}}

\newcommand{\Cnst}{\kappa}
\newcommand{\CnstZero}{\kappa^0}

\newcommand{\nbe}{n}
\newcommand{\nfd}{n^{{\scalebox{.9}{$\scriptscriptstyle \rm FD$}}}}

\newcommand*\diff{\mathop{}\!\mathrm{d}}

\title{Loops Outside a Black Hole}
\author[a]{R. Loganayagam,}
\author[a]{Godwin Martin,}
\author[a,b]{Shivam K. Sharma}

\affiliation[a]{International Centre for Theoretical Sciences (ICTS-TIFR)\\ 
Tata Institute of Fundamental Research, Shivakote, Hesaraghatta Hobli, Bengaluru 560089, India.}

\affiliation[b]{The Institute of Mathematical Sciences, \\
IV Cross Road, C.I.T. Campus, Taramani, Chennai 600113, India.}

\emailAdd{nayagam@icts.res.in}
\emailAdd{godwin.martin@icts.res.in}
\emailAdd{shivam.sharma@icts.res.in}

\abstract{
We present a general conjecture for evaluating multiple discontinuity integrals arising from bulk loop diagrams in the gravitational Schwinger-Keldysh geometry. This generalises earlier tree-level results in \href{https://arxiv.org/abs/2403.10654}{arXiv:2403.10654} to arbitrary bulk loops with no tadpoles (for scalar non-derivative interactions). The conjectured result takes the form of loop integrals performed in a real-time finite-temperature field theory living on the exterior of the black hole. We check our conjecture against all one-loop and many two and three-loop contributions to two, three, and four-point functions. Our diagrammatic rules for the exterior field theory are consistent with microscopic unitarity and thermality at arbitrary loop level. We also remark on a novel approach to real-time finite-temperature holography based on bulk Schwinger-Dyson equations, with the vertices integrated over the black hole exterior.}


\begin{document}

\maketitle

\section{Introduction}

The gravitational Schwinger-Keldysh (grSK) geometry \cite{Skenderis:2008dg, Skenderis:2008dh, Glorioso:2018mmw}, has, in the last few years, allowed the systematic computation of real-time correlators in AdS black holes. This computation of the Schwinger-Keldysh correlators of the boundary theory has, in turn, enabled the study of open quantum field theories using holographic methods \cite{Jana:2020vyx, Ghosh:2020lel}.

The essential idea is to consider a quantum field theory (the system) coupling to a holographic, finite temperature conformal field theory (the bath). We are then interested in the effective dynamics of the system after the bath is integrated out. This is the \emph{open} question that the grSK geometry answers. It does so by turning the computation of the influence functional \cite{Feynman:1963fq, Breuer:2007juk, Banerjee2018} of the open quantum field theory into a holographic computation of the bulk on-shell action.

Since typical holographic baths are strongly coupled, the grSK geometry allows us to study the effects of strongly coupled baths on open quantum field theories. In \cite{Chakrabarty:2019aeu,Jana:2020vyx}, the authors show that the resulting influence functionals are sensible. In particular, these influence functionals naturally account for a full tower of non-linear fluctuation-dissipation relations (FDRs) arising due to the thermality of the bath. As far as we know, this is the first example of a derivation of non-linear FDRs in open field theory, from an \emph{ab-initio} integration-out of a bath field theory. This demonstrates the power of the grSK technique.

In the last few years, several works have studied interactions on the grSK geometry, following \cite{Jana:2020vyx}. In particular, for bulk theories with no derivative interactions, it has recently been found that the grSK computation can be greatly simplified to yield a physically transparent bulk picture. In \cite{Loganayagam:2024mnj, Martin:2024mdm, Sharma:2025hbk}, the authors show that the bulk picture is, in fact, of a unitary field theory placed in a thermal background in the exterior of the black hole. They christened this the \emph{Black Hole exterior EFT}. Moreover, this exterior EFT comes with a set of  Feynman rules that directly give the grSK result,
sidestepping the intermediate step involving multiple monodromy integrals over the complex radial contour. The exterior EFT results for boundary correlators manifestly satisfy strong physical constraints like unitarity, thermality, locality, and causality.

The exterior EFT, so far, has only been tested at the level of tree Witten diagrams in the grSK geometry. In other words, it has been shown that tree-level Witten diagrams in the grSK geometry give the same answers as tree-level Witten diagrams in the black hole exterior with  EFT Feynman rules. In the present note, we intend to generalise this statement to include bulk loops.\\

\noindent \textbf{Why loops? Boundary perspective}
\medskip

Our motivations to include bulk loops are manifold. There are many interesting physical questions whose answers come dominantly from the loop diagrams in the bulk. We will describe some of these now. We begin with a foundational problem in the theory of hydrodynamics: that of understanding fluctuations, i.e., setting up the theory of fluctuating hydrodynamics. The low-energy, low-momentum behaviour of the boundary CFT allows us to study hydrodynamics as has been seen in the Fluid/Gravity literature \cite{Policastro:2001yc, Kovtun:2004de, Janik:2005zt, Nakamura:2006ih, Janik:2006ft, Chesler:2007sv, Bhattacharyya:2007vjd,Baier:2007ix, Son:2007vk,Rangamani:2009xk,Hubeny:2011hd}. The study of the fluctuations in the boundary hydrodynamics requires the inclusion of bulk loops. This is because fluctuations in hydrodynamics are Avogadro-suppressed and come from finite-$N$ effects. In the AdS/CFT correspondence, finite-$N$ corrections arise from bulk loops.

A concrete question in fluctuating hydrodynamics that requires bulk loops is the phenomenon of long-time tails in current-current correlators \cite{Caron-Huot:2009kyg}.
In the bulk, this question requires the computation of bulk loop Witten diagrams \cite{Kovtun:2003vj}. It is not difficult to see why. Working in the large-$N$ limit for the boundary CFT, infinite $N$ contributions give rise to the classical hydrodynamics. It is only after the finite $N$ corrections are included that we obtain a fluctuating hydrodynamic description with long-time tails. 
In \cite{Caron-Huot:2009kyg}, the authors showed by a one-loop bulk computation that this expectation is indeed borne out in fluid-gravity correspondence: the bulk loops lead to long-time tails.

The second question of interest is the origin of Coleman-Mermin-Wagner-Hohenberg (CMWH) theorem in holography. This is the statement that at finite temperature, a (2+1)-dimensional system does not exhibit a spontaneous symmetry breaking of a continuous symmetry. Breaking of continuous symmetries always results in Goldstone modes, and the Goldstone fluctuations wipe out any ordered state. If we consider this (2+1)-dimensional theory to be one that has a holographic dual, there is then the question of how this mechanism works in the bulk. The finite $N$ fluctuations of the boundary theory, as we have discussed, get mapped to the loop corrections in the bulk. Thus, one would expect that in the bulk, the one-loop corrections restore the symmetry. This is indeed the case as shown by the authors of \cite{Anninos:2010sq}. Further,  dominant non-trivial contributions to many 
transport coefficients come from bulk loop effects \cite{Faulkner:2013bna,10.1093/oso/9780198505921.001.0001}.

Finally, our motivations also come from the open QFT picture we have discussed earlier in this introduction. All the earlier studies \cite{Jana:2020vyx, Ghosh:2020lel, Loganayagam:2020iol, Loganayagam:2020eue, Loganayagam:2022teq, Loganayagam:2022zmq, He:2022jnc, Loganayagam:2024mnj, Martin:2024mdm, Sharma:2025hbk} that addressed this question have considered \emph{ideal baths}, i.e., baths with an infinite number of degrees of freedom. In other words, they have all worked at large-$N$. All realistic baths, though, are at finite $N$. For example, if one considers a heavy quark (the system) streaming through a quark-gluon plasma (the bath), its dynamics would be described by an open theory with the bath having $N=3$. Thus, it is necessary to study finite-$N$ corrections, and bulk loops allow us to do exactly this.\\

\noindent\textbf{Why loops? Bulk perspective}
\medskip

So far, we have discussed why the study of bulk loops in grSK is essential from the point of view of the boundary theory and its probe system. But, there certainly are strong physical motivations to study bulk loops purely from the bulk observer's perspective. From the perspective of the bulk observer, the field theory on the grSK (and equivalently the exterior field theory) models the physics of a black hole \cite{Jana:2020vyx, Loganayagam:2024mnj}. 

The grSK theory, as well as the exterior field theory, naturally account for the dissipative physics of falling into the black hole. This is as expected. But they do much more. They account for the Hawking fluctuations of the black hole as well \cite{Jana:2020vyx, Loganayagam:2024mnj}. As explained in \cite{Loganayagam:2024mnj}, the physics of Hawking radiation is clearest in the exterior field theory picture. The exterior field theory has built in not just the quasinormal mode accounting for dissipation, but also the outgoing mode, which accounts for Hawking radiation.

Loop graphs in this exterior field theory (or grSK) then open up many questions concerning quantum field theories in a Hawking radiating background \cite{Banados:2022nhj,MunozSandoval:2023hix}. For example, we ask: what is the Hawking thermal mass inherited by particles in the black hole spacetime?  How does the Hawking radiation Debye screen fields in the exterior? What is the effect of Hawking radiation on renormalisation running? How do we think about beta functions in the presence of Hawking radiation? How should we understand phase transitions caused by BH heating? We note that answering any of these questions requires the study of loops. With this broad set of motivations, we initiate this study.\\

\noindent\textbf{Our setup}
\medskip

In what follows, we will only be interested in scalar theories in the bulk. Even in this simple case of scalar theories, there are many interesting questions that can be studied in the bulk. Scalar theories of the kind we study here do not exhibit any non-Markovian behaviour; they are completely Markovian \cite{Ghosh:2020lel} and show no memory effects. One interesting question here is how the fluctuations affect even the Markovian (short-lived) physics. This is akin to asking how the thermal fluctuations of water affect the dissolution of salt in water. Note that the dissolution of salt in water is a Markovian process, i.e., it is short-lived.

Purely from the perspective of the bulk physics, one of our goals is to initiate a systematic study of \emph{in-in} loops in a black hole spacetime. We view the grSK geometry as a laboratory to do this. In particular, we would like to present an explicit diagrammatics in the exterior of black holes which would allow the computation of boundary Schwinger-Keldysh correlators. Importantly, we would like this diagrammatics to make manifest the microscopic unitarity as well as the thermality of the black hole. In other words, the diagrammatics should make manifest the Veltman largest time equation and Cutkosky cutting rules \cite{Veltman:1994wz}, as well as satisfying the Kubo-Martin-Schwinger conditions \cite{Kubo:1957mj, Martin:1959jp}.

In this work, we develop the exterior quantum field theory using a Schwinger–Dyson equation approach. We do not ruminate on the details of this approach to quantum field theory in the main text; instead, we relegate these to the appendices, keeping the main text entirely self-contained. The main point to note is that two \emph{a priori} distinct quantum field theories---the grSK QFT, and the exterior EFT---provide the same results for boundary correlators even at loop-orders (see Sec.\ref{sec:Equivalence}). We check this for many loop diagrams, all the way up to three loops.


Furthermore, we show that these quantum fields are consistent with microscopic unitarity and thermality, via the SK collapse and KMS conditions, at arbitrary loop orders. This is another important result that we would like the reader to note. As is well known, the Cutkosky rules, as well as the KMS conditions, find their home in the Schwinger-Keldysh formalism. Until recently, showing unitarity of AdS loop amplitudes has been a complicated endeavour requiring explicit radial integrals \cite{Meltzer:2019nbs, Meltzer:2020qbr}. The power of our formalism is seen here by the fact that we can show unitarity and thermality at an arbitrarily high number of loops by simple graph-theoretic arguments. These arguments are also well-known in the study of open systems \cite{Baidya:2017eho}.\\

\noindent \textbf{What has been done before with bulk loops}
\medskip

The question of bulk loops is, of course, not novel \cite{Aharony:2016dwx,Banados:2022nhj, MunozSandoval:2023hix}. Expectedly, loops are much more studied in the case of pure AdS (see, e.g., \cite{Banados:2022nhj, MunozSandoval:2023hix} for some explicit computations).  AdS loops can be computed using several techniques, including spectral representation \cite{Carmi:2018qzm, Carmi:2019ocp, Giombi:2017hpr}, the split representation \cite{Costantino:2020vdu, Herderschee:2021jbi}, by unitarity methods \cite{Ponomarev:2019ofr, Meltzer:2020qbr}, or by working in Euclidean AdS \cite{Bertan:2018khc, Bertan:2018afl}.  One of the main limitation of all these earlier works is that they have been restricted to vacuum correlators, and usually study a small number of loops (typically one or two).

In this work, we will be interested in the more challenging question of loops at finite temperature, i.e., loop contributions in the presence of a black hole. In this context, we face the issue that there is no clear prescription or diagrammatics that works in the presence of a horizon and a singularity. One needs a clear way to impose boundary conditions at the horizon and fix the domain of integration for the vertices, etc.\footnote{See  \cite{Caron-Huot:2009kyg} for a real-time computation in the bulk which cleverly evades this issue.} 

One possible way out is to use analytic continuation from the Euclidean cigar (see e.g., \cite{Anninos:2010sq,Faulkner:2013bna} for  computations of this type), but this is too onerous in practice for multiple loops. Multi-loop contributions in the bulk can especially have quite an intricate analytic structure,\footnote{We remind the reader that bulk loops correspond, in the dual CFT, to sub leading corrections in the large $N$ expansion. Such sub-leading contributions have a much more complicated analytic structure than the leading large $N$ answers, and are thus not meromorphic. We thank the anonymous referee for emphasising this fact.} making it practically impossible to keep track of all the branch cuts in the external momenta. In fact, even in weakly coupled field theories like $\phi^4$ or QED, the only practical way to derive, say, the kinetic theory is to work directly with real-time perturbation theory.  A central aim of our work is  to provide a systematic real-time bulk framework valid to arbitrary loop order. \\

\noindent 
\textbf{How we differ}
\medskip

Our focus in this work will also be to systematise the computation of bulk loops, and thus provide a framework with which one can compute any loop diagram in the bulk theory. 

The primary reason we would like to set up such a unifying formalism is to understand higher-point functions in fluctuating hydrodynamics. The bulk computation of two-point functions is aided greatly by the KMS condition and the SK collapse condition. This is because a single (thermal) K\"all\'en--Lehmann spectral function \cite{Weinberg:1995mt, Chaudhuri:2018ymp} determines all two-point correlators. Thus, the computation of two-point functions can be done in many different ways, including by analytic continuation from the Euclidean correlators. 

This statement is no longer true for higher-point correlators. At three-point for example, there are two independent spectral functions to compute, and not all the structures are entirely fixed by the KMS and SK collapse conditions \cite{Chakrabarty:2019qcp, Chaudhuri:2018ymp}. Furthermore, the analytic continuation from Euclidean correlators is, at best, impractical. Thus, we would like a way to compute these correlators directly from the bulk in a real-time formalism using Witten diagrammatics. 

Further we go further than all the works cited above. Our formalism can compute an arbitrary Schwinger-Keldysh correlator, not just retarded correlators.

\subsection*{Outline}
We begin by reviewing the basic idea of the Schwinger-Keldysh formalism in quantum field theory, and providing a brief review of the gravitational Schwinger-Keldysh prescription in section \ref{sec:grSKeFTReview}. In the latter half of the same section, we will also review the exterior field theory that arises out of the grSK prescription. We then proceed to set up the diagrammatics for the quantum field theory on the grSK spacetime in section \ref{sec:grSKQFT}. Here, we introduce the Schwinger-Dyson equation and the associated diagrammatics. In section \ref{sec:ExtQFT}, we then adapt this discussion to the exterior quantum field theory. 

In section \ref{sec:Equivalence}, we show the equivalence of these two distinct quantum field theories. Finally, in section \ref{sec:MicUniTherm}, we show that the quantum field theories we have defined satisfy microscopic unitarity and thermality at all loop orders. The details of the various monodromy integrals used in section \ref{sec:Equivalence} can be found in Appendix \ref{app:MonInt}. The reader interested in the derivation of the various Schwinger-Dyson equations in this work from a path integral viewpoint may consult Appendix \ref{app:SDEsfromPI}.

\section{Review of grSK and exterior field theory}\label{sec:grSKeFTReview}

In this section, we begin with a brief review of the Schwinger-Keldysh formalism in quantum field theory. Here, we introduce the notational conventions that will be used throughout the rest of the discussion. We then turn to its gravitational counterpart, often referred to as the \emph{gravitational Schwinger-Keldysh} (grSK) geometry. At the end of this section, we will also briefly review how the grSK geometry gives rise to an exterior field theory that lives solely outside the black hole, and highlight its key features.

The Schwinger–Keldysh \cite{Schwinger:1960qe, Feynman:1963fq, Keldysh:1964ud} formalism replaces the usual path integral contour with a closed contour in the complex time plane, commonly known as the \emph{Schwinger–Keldysh contour}. Unlike standard approaches that compute transition amplitudes between initial and final states, this formalism is designed to compute expectation values and correlation functions. It does so without referring to a final state, which is why it is often called the \emph{`in-in'} formalism.\footnote{For further details on the formalism, we refer to the reviews \cite{Chou:1984es, Landsman:1986uw, Haehl:2025zfn} and the textbooks \cite{Rammer:2007zz, Calzetta:2008iqa, kamenev_2011, Bellac:2011kqa, Gelis:2019yfm}.} The central idea is to evolve the system forward and then backwards in time along a closed contour $\mathcal{C}$ in the complex time plane (See Fig.~\ref{fig:SKcontourtemp}).

When working with a thermal initial state, this contour must also include a Euclidean (imaginary-time) segment of length $\beta$, where $\beta$ is the inverse temperature. This extension ensures that the formalism captures thermal correlations correctly. Thus, the Schwinger–Keldysh (SK) contour at finite temperature can be seen in Figure~\ref{fig:SKcontourtemp} below:
\begin{figure}[H]
\centering
\begin{tikzpicture}[scale=1.6]
    \definecolor{darkred}{rgb}{0.6,0.1,0.1}
    \definecolor{darkblue}{rgb}{0,0,0.8}
    
    \draw[gray,thick,->] (-0.5,0.5) -- (6,0.5) node[right] {\small Re $t$};;

    \draw[gray, thick,->] (0,-2) -- (0,2) node[above] {\small Im $t$};;
    
    \draw[line width=1.2pt, darkred] (0,1) -- (5,1) arc (90:-90:0.5) -- (0,0)--(0,-1.8) ;
    
    \fill[darkred] (0,1) circle (2pt);
    \fill[darkred] (0,-1.8) circle (2pt);
    
    \draw[thick, darkred,->] (2.5,1) -- (2.7,1);
    \draw[thick, darkred,->] (2.5,0) -- (2.3,0);
    \draw[thick, darkred,->] (5,0.8) arc (90:50:0.5);
    
    \node[left, darkred] at (0,1) {\small $t_0+i \epsilon$};
    \node[left, darkred] at (0,0) {\small $t_0-i \epsilon$};
    \node[left, darkred] at (0,-1.8) {\small $t_0+i (\epsilon-\beta)$};
    \node[above,darkred] at (2,1) {\Large $\mathcal{C}$};
    \node[thick] at (5.6,2.1) {\small $t$};

    \fill[darkblue] (3,0) circle (2pt);
    \fill[darkblue] (4,1) circle (2pt);
    
    \node[above, darkblue] at (4,1.05) {\small $\mathcal{O}_R$};
    \node[below, darkblue] at (3,-0.05) {\small $\mathcal{O}_L$};

    \draw (5.5,2.2) -- (5.5,2) -- (5.7,2);
\end{tikzpicture}
\caption{The SK contour at finite temperature $T=\frac{1}{\beta}$ where starting and end points of the contour are identified. The direction of the arrow represents the direction of contour time $t_{\mathcal{C}}$, involving both forward (R) and backwards (L) time-evolving parts. }
\label{fig:SKcontourtemp}
\end{figure}
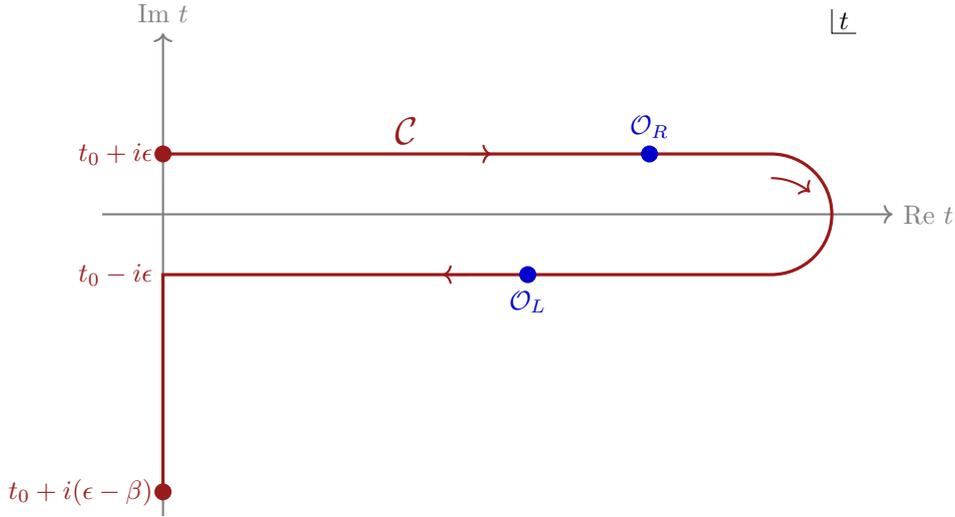

With an eye towards AdS/CFT, let us consider a holographic conformal field theory in $d$ spacetime dimensions (CFT$_{d}$), with action $S_{\text{CFT}}$. Let $\mathcal{O}$ be a bosonic operator in this theory, and let $J$ be a source coupled linearly to it. Both $\mathcal{O}$ and $J$ are defined on the SK contour. The SK generating functional $\mathcal{Z}_{\text{SK}}$ is then given by the path integral:
\begin{equation}
    \mathcal{Z}_{\text{SK}} \left[J\right]=  \int \mathcal{D} \mathcal{O} \   \exp \bigg( i S_{\text{CFT}} +i \oint \diff t_{_{\mathcal{C}}} \diff^{d-1} \textbf{x} \ J(x) \mathcal{O}(x)\bigg)  \ ,
\end{equation}
where $t_{_{\mathcal{C}}}$ denotes the contour time. This is distinct from the usual in-out path integral since both the terms in the exponent are integrated over the full SK contour in the complex plane rather than the in-out contour (which is just the first forward half of the SK contour).


It is convenient to split the fields and sources according to the \emph{forward} (right) and \emph{backwards} (left) time-evolving segments of the contour. In this so-called \textit{right-left} (RL) basis, we define:
\begin{equation}
\begin{split}
    \mathcal{O}(t+i \epsilon) &\equiv \ \mathcal{O}_{\sR}(t)\ ,  \qquad \qquad \mathcal{O}(t-i \epsilon) \equiv \ \mathcal{O}_{\sL}(t)\ , \quad  \forall \quad t \in \mathbb{R} \ ,\\
    J(t+i \epsilon) &\equiv \ J_{\sR}(t)\ ,  \qquad \qquad J(t-i \epsilon) \equiv \ J_{\sL}(t)\ , \quad  \forall \quad t \in \mathbb{R} \ .
\end{split}    
\end{equation}
This allows us to write the SK generating functional in the right-left basis as:
\begin{equation}
    \mathcal{Z}_{\text{SK}}[J_{\sR},J_{\sL}]= \Bigg \langle \exp \Bigg\{i \int \diff^d x \  \Big[ J_{\sR}(x) \mathcal{O}_{\sR}(x)-J_{\sL}(x) \mathcal{O}_{\sL}(x) \Big]  \ \Bigg\} \Bigg \rangle_{\rm CFT}    \ . 
\end{equation}
This representation of the generating functional makes the real-time structure of the contour manifest and serves as the natural starting point for computing real-time (or Schwinger-Keldysh) correlation functions.

We can now perform functional differentiation with respect to the external sources in the SK generating functional $\mathcal{Z}_{\rm SK}$. Explicitly, one can differentiate  $\mathcal{Z}_{\rm SK}$ as,
\begin{equation}
    \frac{1}{\mathcal{Z}_{\rm SK}}\prod_{i=1}^{n}\left(\frac{-i \ \delta   }{\delta J_{\sR}(x_i)} \right)\prod_{i=n+1}^{m} \left( \frac{i \ \delta   }{\delta J_{\sL}(x_i)}\right) \mathcal{Z}_{\rm SK} \Bigg|_{ J_{\sR}= J_{\sL} =0 }   \ , 
\end{equation}
to obtain $m$-point SK correlators of the following type,
\begin{equation}
      \text{Tr } \left[ \hat{\rho}_{\rm initial} \ \mathbb{T}  \Big( \mathcal{O}_{1} \mathcal{O}_{2} \ldots \mathcal{O}_{n} \Big) \ \mathbb{T}^\ast  \Big( \mathcal{O}_{n+1} \mathcal{O}_{n+2} \ldots \mathcal{O}_{m} \Big) \right] \ ,
\end{equation}
where we have denoted $\mathcal{O}_{i} \equiv \mathcal{O}(x_i)$, and $\mathbb{T}, \ \mathbb{T}^\ast$ denote time-ordering and anti-time-ordering, respectively. These orderings reflect the path taken along the SK contour. The symbol $\hat{\rho}_{\rm initial}$ denotes the initial density matrix, which we have taken to be thermal with the inverse temperature $\beta$ in this discussion.

Before turning to the gravitational realisation of the SK formalism, it is important to highlight its relevance in the study of open quantum systems. In particular, the Schwinger-Keldysh generating functional $\mathcal{Z}_{\rm SK}[J]$ naturally doubles as the \emph{influence functional} \cite{Feynman:1963fq}, a central object in the framework of open quantum field theory. The generating functional $\mathcal{Z}_{\rm SK}[J]$ is the influence functional of the probe system $J$. The influence functional encapsulates the effect of integrating out the environmental degrees of freedom on the dynamics of a system. It encodes both dissipation and noise, and often leads to a stochastic effective theory describing the system.

This framework becomes especially powerful in the context of holography, where the environment is effectively modelled by a black hole. In such scenarios, the bulk geometry provides a non-perturbative, gravitational description of the thermal bath, and the path integral over an appropriately constructed SK-type spacetime computes the influence functional for the boundary field theory. This raises a natural question:
\begin{center}
\emph{What is the gravitational dual of the Schwinger-Keldysh formalism in holography?}
\end{center}

Early attempts to address this question involved glueing together the Euclidean and Lorentzian regions of black hole spacetime in a way that mimics the structure of the SK contour (see, e.g., \cite{Skenderis:2008dh, Skenderis:2008dg}). More recently, the so-called \emph{grSK} geometry \cite{Glorioso:2018mmw} has emerged as the appropriate dual spacetime that gives the SK contour at the boundary. This is the prescription we will now review and use throughout this note.

\subsection{Classical Field Theory on grSK}\label{sec:ClassicalFieldTheorygrSK}

Consider a Conformal Field Theory (CFT) at finite temperature in the SK formalism. According to the AdS/CFT correspondence, thermal states of the boundary CFT are dual to black hole (or black brane) geometries in asymptotically AdS spacetimes \cite{Witten:1998zw}. The SK contour, which encodes real-time evolution, introduces a doubling of the field degrees of freedom in the CFT. This doubling is mirrored in the bulk by a corresponding duplication of the AdS black hole geometry, as discussed in \cite{Skenderis:2008dg, Skenderis:2008dh}.

To capture the full real-time dynamics holographically, one must construct a gravitational Schwinger-Keldysh (grSK) geometry --- a bulk spacetime composed of two copies of the AdS black hole geometry, glued together in a way that reproduces the SK contour at the asymptotic boundary. This construction was first systematically proposed in \cite{Glorioso:2018mmw}, and we briefly outline its key features below.\footnote{For a detailed discussion of the grSK geometry and its applications, see \cite{Jana:2020vyx}.}

Let us begin by considering a black brane solution in an asymptotically AdS$_{d+1}$ spacetime, expressed in ingoing Eddington–Finkelstein (EF) coordinates:
\begin{equation}
    \diff s^2 = -r^2 f(r) \diff v^2 + 2 \diff v \diff r + r^2 \diff \mathbf{x}^2  \ , \qquad \qquad f(r)=1-\left( \frac{r_{\rm h}}{r}\right)^d   \ ,
    \label{eq:Metric}
\end{equation}
where $r_{\rm h}$ is the horizon radius. 

The construction of the grSK geometry proceeds by complexifying the radial coordinate and selecting an appropriate codimension-1 slice within this complexified manifold. This slice, referred to as the \emph{grSK contour}, connects the two Lorentzian geometries smoothly reflecting the SK timefold structure at the boundary. The essential features of this contour are illustrated schematically in the figure below.
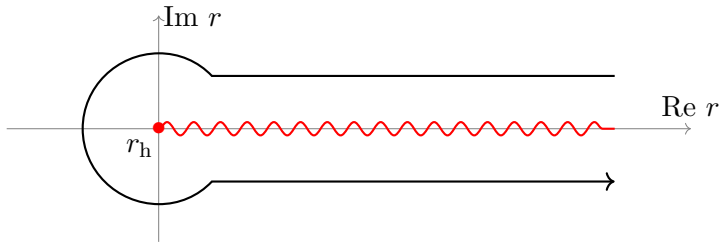
\begin{figure}[H]
\centering
\begin{tikzpicture}
\draw[help lines,->] (-2,0) -- (7,0) coordinate (xaxis);
\draw[help lines,->] (0,-1.5)--(0,1.5) coordinate (yaxis);
\node at (0.45,1.5) {$\text{Im}\ r$};
\node at (7,0.3) {$\text{Re}\ r$};
\draw [thick,->]  (6,0.7) -- (0.7,0.7)
    to [curve through={(0,1)(-1,0) (0,-1)}]
    (0.7,-0.7)--(6,-0.7);
\draw[decoration={snake},decorate, thick, red] (0,0)--(6,0);
\node[red] at (0,0) {$\bullet$};
\node at (-0.25,-0.26) {$r_{\rm h}$};
\end{tikzpicture}
\caption{\small{The gravitational Schwinger-Keldysh contour on the complex $r$ plane.}}
\label{fig:grSKsaddle}
\end{figure}
It is convenient to this radial contour by a complexified coordinate $\z$ defined through
\begin{equation}
\frac{\diff \z}{\diff r } = \frac{2}{i \beta \, r^{2} f(r)} \ ,   
\label{eq:DefZeta}
\end{equation}
where $\beta$ is the inverse Hawking temperature of the black brane, given by
\begin{equation}
    \beta = \frac{4 \pi}{d r_{\rm h}} \ .
\end{equation}
The coordinate $\z$  has a branch cut along the exterior region and is normalised to have unit monodromy around the horizon branch point at $r=r_{\rm h}$. This branch cut is depicted as the red wavy line in Fig.~\ref{fig:grSKsaddle}.

With this new coordinate, we define the grSK geometry as one constructed by taking the black brane exterior and replacing the radial interval extending from the horizon to infinity by the doubled contour enveloping the branch cut in $\z$, as indicated in Fig.~\ref{fig:grSKsaddle}. Thus, the metric for the grSK geometry is given by:
\begin{equation}
    \diff s^2 = -r^2 f(r) \diff v^2 + i \beta r^2 f(r) \diff v \diff \z + r^2 \diff \mathbf{x}^2 \ ,
    \label{eq:MetricgrSK}
\end{equation}
where $r$ should be thought of as a function of $\zeta$ specified by Eq.\eqref{eq:DefZeta}.

Having established the geometric background, a natural next question arises: how do quantum fields behave in this setting? While the dynamics of free fields on such geometries have been studied extensively across various matter sectors \cite{Jana:2020vyx, Ghosh:2020lel, Loganayagam:2020eue, He:2022jnc, Bu:2025zad}, interacting fields have only recently received attention. Recent work \cite{Jana:2020vyx, Loganayagam:2022zmq, Rangamani:2023mok, Loganayagam:2024mnj, Martin:2024mdm} has shown that for theories with non-derivative interactions,\footnote{For derivative interactions, there are extra horizon localised contributions \cite{Chakrabarty:2019aeu, Loganayagam:2022teq}.} the dynamics defined on the grSK geometry induces an effective field theory supported outside the black hole horizon.\footnote{See \cite{Loganayagam:2020iol} for the Schwinger–Keldysh gravity dual of a charged black hole, and \cite{Sharma:2025hbk} for the inclusion of interactions and the associated exterior field theory. } This so-called \emph{exterior field theory} captures the influence of the black hole background on observable degrees of freedom and provides a powerful framework for understanding dissipation, fluctuation, and other open-system phenomena from a holographic perspective. We will now explain the setup of the classical field theory on the grSK spacetime. Once this is done, we will show that this field theory naturally reduces to an exterior field theory, which we will explain in Sec~(\ref{sec:EFTwithFeynmanRules}).

Consider, for concreteness, a massless scalar theory with a cubic interaction on the grSK geometry, described by the action
\begin{equation}
    S =  - \oint_{Y} \left[ \frac{1}{2}g^{\A\B} \partial_{\A} \phi \partial_{\B} \phi +  \frac{\lambda_{3 \rm B}}{3!}\phi^3\right] \  , \qquad \qquad \oint_{Y} \equiv \oint \diff \z \diff^d y \sqrt{-g} \  .
    \label{eq:ActionPhi3grSK}
\end{equation}
Here, $Y$ represents a point in the grSK geometry which is equally well described by $(\z, y)$. Furthermore, $g_{\A\B}$ is the metric on the grSK geometry, given in Eq.~\eqref{eq:MetricgrSK}, and $\lambda_{3 \rm B}$ is the bulk coupling constant. Varying this action, we find the Euler-Lagrange equation
\begin{equation}
    \nabla^2 \phi = \frac{\lambda_{3\text{B}}}{2!} \phi^2 \ ,
    \label{eq:EOMClassical}
\end{equation}
where $\nabla^2$ is the negative of the wave operator, defined as
\begin{equation}
    \nabla^2 \equiv \nabla_{\A}\nabla^{\A} \equiv \frac{1}{\sqrt{-g}} \partial_{\B} \Big(\sqrt{-g} g^{\A \B} \partial_{\A} \Big) \ .
    \label{eq:WaveOperatorCurvedSpacetime}
\end{equation}

As usual, we will solve this equation using perturbation theory in the bulk coupling constant $\lambda_{3 \rm B}$. In other words, we will write the solution in a perturbative expansion of the form
\begin{equation}
\phi = \phi_{(0)} + \lambda_{3\text{B}} \phi_{(1)} + \lambda^2_{3\rm B} \phi_{(2)} + \ldots \ .
\label{eq:PhiPerturbativeExpansion}
\end{equation}
We impose GKPW boundary conditions on $\phi$. Note that there are two boundaries now compared to the single-copy geometry:
\begin{equation}\label{eq:BCs}
    \lim_{\zeta \to 0} \phi(Y) = J_{\sL}   \  ,  \qquad \text{and} \quad \lim_{\zeta \to 1} \phi(Y) = J_{\sR}   \ , 
\end{equation}
where $\zeta=0$ and $\zeta=1$ correspond to the left and the right boundaries, respectively.

We will find it convenient to take the leading-order contribution $\phi_{(0)}$ to satisfy these boundary conditions, i.e.,
\begin{equation}
    \lim_{\zeta \to 0} \phi_{(0)}(Y) = J_{\sL}   \  ,  \qquad \text{and} \quad \lim_{\zeta \to 1} \phi_{(0)}(Y) = J_{\sR}   \ .
\end{equation}
These boundary conditions fix the leading-order solution completely. We will now write down this solution more explicitly. To this end, we introduce the two independent solutions in the ingoing-outgoing basis. We denote the ingoing boundary-to-bulk Green function by $\Gin(r,k)$.\footnote{Note that here we are working in the Fourier domain of the boundary coordinates. The symbol $k$ denotes the boundary momentum, which is the Fourier conjugate of $x$.} This is the solution to the free Klein-Gordon equation, $\nabla^2 \phi = 0$, with the boundary condition
\begin{equation}
    \lim_{r \to r_{\rm c}} \Gin(r,k) = 1 \ , 
\end{equation}
and regularity at the horizon $r_{\rm h}$. Here, $r_{\rm c}$ is the radial cutoff boundary of the spacetime. Note that the regularity at the horizon in ingoing Eddington-Finkelstein coordinates is equivalent to ingoing boundary conditions. Next, we have an outgoing boundary-to-bulk Green function, which is the time-reversal of the ingoing boundary-to-bulk Green function. Thus, it is regular at the past horizon and satisfies 
\begin{equation}
    \lim_{r \to r_{\rm c}} \Gout(r,k) = 1 \ . 
\end{equation}
An explicit expression for the outgoing Green function can be given in terms of the ingoing Green function by using the time-reversal involution of the grSK geometry:
\begin{equation}
    \Gout(r,k) = e^{-\beta k^0 \zeta(r)} \Gin (r,-k)\ .
\end{equation}

In terms of the ingoing and the outgoing boundary-to-bulk Green functions, we can write the solution explicitly in the Son-Teaney form \cite{Son:2009vu, Jana:2020vyx}
\begin{equation}
    \phi_{(0)} (\zeta,k) = -G^{\rm in}(\zeta,k) J_{\Fb}(k) + e^{\beta k^0} G^{\rm out}(\zeta,k) J_{\Pb}(k) \ ,
    \label{eq:LeadingOrderSolutionPF}
\end{equation}
where we have used the past-future basis for the boundary sources, defined as
\begin{equation}
    \begin{split}
        J_{\Fb}(k) &\equiv - \Big[(1+n_{k}) J_{\sR}(k) - n_k J_{\sL}(k)\Big] \ ,\\
        J_{\Pb}(k) &\equiv -n_{k} \Big[J_{\sR}(k) - J_{\sL}(k)\Big] \ ,
    \end{split}
    \label{eq:DefBasisPF}
\end{equation}
where $n_{k}$ is the Bose-Einstein factor, given by
\begin{equation}
    n_k \equiv \frac{1}{e^{\beta k^0}-1}\ .
\end{equation}
Note that the past-future basis arises naturally when the grSK solution is written in the ingoing-outgoing basis of Green functions. This basis is well-known in thermal field theory for making thermal real-time computations simpler \cite{Gelis:1997zv, Chaudhuri:2018ymp, Gelis:2019yfm}. We will see a reappearance of these source when we discuss the exterior field theory soon.

Given the leading-solution that already satisfies the full GKPW boundary conditions, we are left with solving the subleading corrections.
The higher-order corrections to the solution are all normalisable at the boundaries, i.e., 
\begin{equation}
    \lim_{\zeta \to 0,1} \phi_{(i)}(Y) = 0  \quad \forall \quad i>0  \ .
\end{equation}
Since all the higher-order solutions solve the sourced Klein-Gordon equation with the same boundary condition, it is convenient to solve them all at once using the Green function technique. We introduce a \textit{grSK contour-ordered} bulk-to-bulk Green function $\bbG$ that satisfies the sourced Klein-Gordon equation with a point source 
\begin{equation}
    \nabla_Y^2 \bbG (Y|Y_0) = -\frac{\delta^{d+1}(Y-Y_0)}{\sqrt{-g}} \ ,
    \label{eq:PDEFormalBlkBlk}
\end{equation}
and satisfying the bi-normalisable boundary conditions
\begin{equation}
    \lim_{\zeta \to 0} \bbG(Y|Y_0) = 0  \  ,  \qquad \text{and} \quad   \lim_{\zeta \to 1} \bbG(Y|Y_0) = 0 \ ,
    \label{eq:BCBlkBlk}
\end{equation}  
where $Y \equiv (\zeta, y)$ and $Y_0 \equiv (\zeta_0, y_0)$. We will always find it convenient to go to the Fourier domain in the boundary coordinates, in which the above partial differential equation reduces to an ordinary differential equation. Explicitly, we have
\begin{equation}
    \begin{split}
        &-\left[\left(\frac{\diff}{\diff \zeta} +\frac{\beta k^0}{2}\right) r^{d-1}\left(\frac{\diff}{\diff \zeta} +\frac{\beta k^0}{2}\right)+\left(\frac{i\beta}{2}\right)^2 r^{d-1} \left((k^0)^2 - \mathbf{k}^2 f(r)\right)\right]\bbG(\zeta|\zeta_0, k)\\
        &\hspace{11cm}= \frac{i \beta}{2} \delta (\zeta-\zeta_0)\ .
    \end{split}
\end{equation}
Note that the bulk-to-bulk Green function $\bbG(Y|Y_0)$, when written in the boundary Fourier coordinates, only depends on the radial coordinates and the transferred boundary momentum $k$.

Note that the first-order correction to the solution $\phi_{(1)}(X)$ satisfies a Klein-Gordon equation whose source is dependent only on the leading-order solution $\phi_{(0)}(X)$. Thus solving this equation gives the first-order correction $\phi_{(1)}(X)$ in terms of the bulk-to-bulk Green function and $\phi_{(0)}$. Similarly, the source of the second-order correction depends only on the first-order correction $\phi_{(1)}(X)$ and the leading solution. But, we have already expressed $\phi_{(1)}(X)$ in terms of $\bbG(X|Y)$ and $\phi_{(0)}(X)$. Thus, the second-order source depends only on the bulk-to-bulk Green function $\bbG(X|Y)$ and the leading solution $\phi_{(0)}$. One can easily extend this logic to an arbitrary high order in perturbation theory. Thus, the whole perturbative expansion is arranged purely in terms of two ingredients: $\bbG(X|Y)$ and $\phi_{(0)}(X)$. We can now set up a diagrammatics with these two ingredients to compute the solution, as well as any function of it.

Recall that the goal of the grSK prescription is to compute the Schwinger-Keldysh (SK) generating functional of the boundary CFT. This can now be achieved by the GKPW prescription: the SK generating functional of the boundary theory is the bulk on-shell action found by imposing GKPW boundary conditions on the grSK geometry. We have already imposed these boundary conditions on our solution, and thus, all we are left to do is to substitute the solution into the action in Eq.~\eqref{eq:ActionPhi3grSK}.

We will now lay down the diagrammatics to compute this on-shell action.
We need only draw Witten diagrams on the full grSK geometry. The Feynman rules to compute the on-shell action are:
\begin{enumerate}
    \item Multiply every Witten diagram by $-i$.
    \item There is one boundary-to-bulk propagator and one bulk-to-bulk propagator, as expected. And a single vertex, which should be integrated over the full grSK spacetime. The propagators and vertices are given by the rules:
    \begin{equation}
    \tikzset{every picture/.style={line width=0.75pt}} 
    \begin{tikzpicture}[x=0.75pt,y=0.75pt,yscale=-1,xscale=1]
    \centering
    
    \draw    (425.29,104) -- (454.29,104) ;
    \draw    (454.29,104) -- (470.81,81.93) ;
    \draw    (454.29,104) -- (472.57,124.9) ;
    \draw    (62.33,103) -- (106.33,102.92) ;
    \draw    (229.33,103) -- (285.33,103.92) ;
    
    \draw (100,97.4) node [anchor=north west][inner sep=0.75pt]  [font=\scriptsize]  {$\otimes $};
    \draw (58,104.4) node [anchor=north west][inner sep=0.75pt]  [font=\footnotesize]  {$Y$};
    \draw (223,104.4) node [anchor=north west][inner sep=0.75pt]  [font=\footnotesize]  {$Y$};
    \draw (285.37,104.96) node [anchor=north west][inner sep=0.75pt]  [font=\footnotesize]  {$Y_0$};
    \draw (115,93.4) node [anchor=north west][inner sep=0.75pt]    {$=\phi _{(}{}_{0}{}_{)}(Y) \ ,\ $};
    \draw (295,93.4) node [anchor=north west][inner sep=0.75pt]    {$= -i \bbG(Y|Y_0)\ ,\ $};
    \draw (482,90.4) node [anchor=north west][inner sep=0.75pt]    {$=-i\lambda_{3 \rm B} \ .$};
    \end{tikzpicture} 
    \label{eq:FeynRulesPosition}
\end{equation}
Here, the symbol $\otimes$ denotes a point on the boundary.
\item Draw all tree diagrams.
\item Each diagram must be weighed by an appropriate symmetry factor.
\end{enumerate}

Note that the same rules can also be interpreted in both position and Fourier domains for the boundary theory. To avoid confusion, we reproduce the rules explicitly in the boundary Fourier space as well:
\begin{equation}
\tikzset{every picture/.style={line width=0.75pt}} 
\begin{tikzpicture}[x=0.75pt,y=0.75pt,yscale=-0.85,xscale=0.85]
\centering

\draw    (118.29,204.67) -- (212.96,204.67) ;
\draw    (335.5,202.7) -- (430.17,202.7) ;
\draw    (569.1,206.65) -- (608.03,206.65) ;
\draw    (608.03,206.65) -- (627.7,179) ;
\draw    (608.03,206.65) -- (626.88,231) ;

\draw (206.65,195.75) node [anchor=north west][inner sep=0.75pt]    {$\otimes $};
\draw (219.74,195.41) node [anchor=north west][inner sep=0.75pt]    {$=\phi _{( 0)}( \zeta ,k) ,$};
\draw (105.74,202.34) node [anchor=north west][inner sep=0.75pt]    {$\zeta $};
\draw (147.19,187.2) node [anchor=north west][inner sep=0.75pt]    {$\leftarrow k$};
\draw (438.6,195.41) node [anchor=north west][inner sep=0.75pt]    {$=-i\bbG(\zeta _{2} |\zeta _{1} ,k) ,$};
\draw (322.4,200.02) node [anchor=north west][inner sep=0.75pt]    {$\zeta _{1}$};
\draw (364.4,185.22) node [anchor=north west][inner sep=0.75pt]    {$k\rightarrow $};
\draw (424.86,200.68) node [anchor=north west][inner sep=0.75pt]    {$\zeta _{2}$};
\draw (621.8,195.41) node [anchor=north west][inner sep=0.75pt]    {$=-i \lambda _{3\rm B} ,$};
\end{tikzpicture}
\end{equation}
Our convention for the momentum flow directions here is: (i) momenta always flow from the boundary to the bulk along the boundary-to-bulk propagators, (ii) momenta always flow from the right to the left along the bulk-to-bulk propagators, (iii) at a vertex, all the momenta flow into the vertex. In what follows, we will not explicitly give the momentum space rules as we have done here. As we can see, going between the two domains is straightforward at the level of diagrams.

We conclude our discussion of the grSK classical field theory with a word about the diagrammatic notation. We use $\otimes$ in our boundary-to-bulk propagator lines to denote the boundary value of the field $\phi$. This foreshadows our discussion of the exterior quantum field theory, where there will be two such boundary values.
 
\subsection{Exterior Field Theory with Feynman rules}\label{sec:EFTwithFeynmanRules}

The $\phi^3$ theory defined on the full grSK spacetime can be reduced to a theory defined on a single copy of the exterior spacetime, as was shown for scalars in \cite{Rangamani:2023mok, Loganayagam:2024mnj, Sharma:2025hbk} and spinors in \cite{Martin:2024mdm}. In other words, in perturbation theory, we can define an exterior field theory whose vertices are only on one copy of the exterior spacetime. This was shown for theories with no derivative interaction, and also only at tree-level. In \cite{Loganayagam:2024mnj}, the authors showed this for scalar theories with no derivative interactions, which was generalised to Yukawa theories including spinor fields in \cite{Martin:2024mdm}.

To summarise, it has been shown that for scalar and spinor field theories without derivative interactions, the Schwinger-Keldysh correlators of the boundary theory can be computed entirely from Witten diagrams drawn in the exterior region of a black brane spacetime. This result is in line with physical intuition: for an observer restricted to the exterior of the black hole, the relevant physics should resemble that of a thermal field theory at the Hawking temperature of the black brane. Remarkably, the grSK construction precisely realizes this expectation, as we now discuss in more detail.

The effective field theory defined in the exterior exhibits several key features:
\begin{enumerate}

\item 
\textbf{Exterior localisation:} All dynamical degrees of freedom and interactions relevant for computing observables are confined to the region outside the black hole horizon.
\item 
\textbf{Microscopic unitarity and causality:} The theory respects the fundamental principles, ensuring that evolution is unitary and causally consistent within the exterior region.

\item 
\textbf{Thermal structure of interactions:} The interaction vertices are appropriately dressed with statistical distribution functions, such as Bose-Einstein or Fermi-Dirac factors, consistent with the expectations for a thermal field theory.

\item 
\textbf{Dissipation and fluctuation:} Most crucially, the exterior field theory captures both dissipative dynamics---arising from the loss of information to infalling modes---as well as fluctuations, which are sourced by outgoing Hawking radiation. 

\item 
\textbf{Unitary Vertices:} The vertices of this exterior field theory are, in fact, unitary. By this, we mean that the vertices here are the same as those obtained from a unitary theory that is factorised on the two legs of the SK contour. More precisely, there are no extra Feynman-Vernon \cite{Feynman:1963fq} type vertices that couple the two legs.\footnote{Generically, as pointed out by Feynman-Vernon, one expects such extra vertices to appear once we integrate out an environment.} 

\end{enumerate}

Having established the structure of the exterior theory, we now turn to specifying the Feynman rules for such an exterior field theory with cubic interactions.
To introduce the Feynman rules, we require one new ingredient: the retarded bulk-to-bulk Green function. We define the \emph{retarded} bulk-to-bulk Green function $\bbGR(\z|\z_0, k)$ to solve the delta-sourced Klein-Gordon equation
\begin{equation}
    \begin{split}
        &-\left[\left(\frac{\diff}{\diff \zeta} +\frac{\beta k^0}{2}\right) r^{d-1}\left(\frac{\diff}{\diff \zeta} +\frac{\beta k^0}{2}\right)+\left(\frac{i\beta}{2}\right)^2 r^{d-1} \left((k^0)^2 - \mathbf{k}^2 f(r)\right)\right]\bbGR(\zeta|\zeta_0, k)\\
        &\hspace{11cm}= \frac{i \beta}{2} \delta (\zeta-\zeta_0)\ .
    \end{split}
\end{equation}
Here $\zeta=\zeta(r)$. The boundary conditions are that the Green function is retarded in $v$, i.e., analytic in the upper-half plane of $k^0$. Equivalently, this Green function satisfies ingoing boundary conditions at the future horizon. It is also normalisable at the boundary, i.e., 
\begin{equation}
    \lim_{r \to r_{\rm c}} \bbGR(\zeta|\zeta_0, k) = 0 \ .
\end{equation}
This Green function has no direct relation to the grSK contour-ordered Green function defined in the earlier section. In particular, the retarded Green function is not obtained by restricting the end-points of the aforementioned binormalisable propagators to a single copy; such a restriction would result in the Feynman propagator in the exterior, and not the retarded.

We now have all the ingredients to present the Feynman rules of the exterior field theory. These Feynman rules compute the Schwinger-Keldysh generating functional of the boundary CFT:
\begin{enumerate}
    \item Multiply every diagram by $-i$.
    \item The boundary-to-bulk propagators are given by
    \begin{equation}
    \tikzset{every picture/.style={line width=0.75pt}} 
    \begin{tikzpicture}[x=0.75pt,y=0.75pt,yscale=-1,xscale=1, baseline=(current bounding box.center)]
    
    \draw    (441.6,85.23) -- (441.93,125.2) ;
    \draw [shift={(441.96,128.2)}, rotate = 269.53] [fill={rgb, 255:red, 0; green, 0; blue, 0 }  ][line width=0.08]  [draw opacity=0] (8.93,-4.29) -- (0,0) -- (8.93,4.29) -- cycle    ;
    \draw    (441.96,128.2) -- (442.45,165.23) ;
    \draw [shift={(442.45,165.23)}, rotate = 89.24] [color={rgb, 255:red, 0; green, 0; blue, 0 }  ][fill={rgb, 255:red, 0; green, 0; blue, 0 }  ][line width=0.75]      (0, 0) circle [x radius= 3.35, y radius= 3.35]   ;
    \draw [shift={(441.96,128.2)}, rotate = 269.24] [color={rgb, 255:red, 0; green, 0; blue, 0 }  ][line width=0.75]    (0,5.59) -- (0,-5.59)   ;
    
    \draw (448,73.66) node [anchor=north west][inner sep=0.75pt]  [rotate=-88.58]  {$\otimes $};
    \draw (448,164.4) node [anchor=north west][inner sep=0.75pt]    {$r$};
    \draw (420,99.4) node [anchor=north west][inner sep=0.75pt]    {$k$};
    \end{tikzpicture}
    = \frac{\Gin(\zeta,k)}{(1+n_k)} J_{\Fb}(k)  \ ,
    \hspace{2cm}
    \tikzset{every picture/.style={line width=0.75pt}} 
    \begin{tikzpicture}[x=0.75pt,y=0.75pt,yscale=-1,xscale=1, baseline=(current bounding box.center)]
    \draw    (441.6,85.23) -- (441.96,128.2) ;
    \draw [shift={(441.96,128.2)}, rotate = 269.53] [color={rgb, 255:red, 0; green, 0; blue, 0 }  ][line width=0.75]    (0,5.59) -- (0,-5.59)   ;
    \draw    (442,131.2) -- (442.45,165.23) ;
    \draw [shift={(442.45,165.23)}, rotate = 89.24] [color={rgb, 255:red, 0; green, 0; blue, 0 }  ][fill={rgb, 255:red, 0; green, 0; blue, 0 }  ][line width=0.75]      (0, 0) circle [x radius= 3.35, y radius= 3.35]   ;
    \draw [shift={(441.96,128.2)}, rotate = 89.24] [fill={rgb, 255:red, 0; green, 0; blue, 0 }  ][line width=0.08]  [draw opacity=0] (8.93,-4.29) -- (0,0) -- (8.93,4.29) -- cycle    ;
    
    \draw (447.85,73.66) node [anchor=north west][inner sep=0.75pt]  [rotate=-88.58]  {$\otimes $};
    \draw (448,164.4) node [anchor=north west][inner sep=0.75pt]    {$r$};
    \draw (420,99.4) node [anchor=north west][inner sep=0.75pt]    {$k$};
    \end{tikzpicture}
    = G^{\rm out}(\zeta,k ) J_{\Pb}(k) \ .
    \label{eq:BndryBlkPropEFT}
\end{equation}
Here, the symbol $\otimes$ denotes the AdS boundary. The bullet $\bullet$ denotes a bulk point. The momenta $k$ are conventionally taken to flow from the boundary to the bulk. To set up some nomenclature, we call the arrow $\blacktriangleright$ a semi-diode and the line $|$ a semi-capacitor.
    \item The only bulk-to-bulk propagator is given by
    \begin{equation}
    \tikzset{every picture/.style={line width=0.75pt}}
    \begin{tikzpicture}[x=0.75pt,y=0.75pt,yscale=-1,xscale=1, baseline=(current bounding box.center)]
    \draw    (481,124) -- (441,124) ;
    \draw [shift={(441,124)}, rotate = 360] [color={rgb, 255:red, 0; green, 0; blue, 0 }  ][line width=0.75]    (0,5.59) -- (0,-5.59)   ;
    \draw [shift={(481,124)}, rotate = 180] [color={rgb, 255:red, 0; green, 0; blue, 0 }  ][fill={rgb, 255:red, 0; green, 0; blue, 0 }  ][line width=0.75]      (0, 0) circle [x radius= 3.35, y radius= 3.35]   ;
    \draw    (438,124.02) -- (402.04,124.26) ;
    \draw [shift={(402.04,124.26)}, rotate = 179.61] [color={rgb, 255:red, 0; green, 0; blue, 0 }  ][fill={rgb, 255:red, 0; green, 0; blue, 0 }  ][line width=0.75]      (0, 0) circle [x radius= 3.35, y radius= 3.35]   ;
    \draw [shift={(441,124)}, rotate = 179.61] [fill={rgb, 255:red, 0; green, 0; blue, 0 }  ][line width=0.08]  [draw opacity=0] (8.93,-4.29) -- (0,0) -- (8.93,4.29) -- cycle    ;
    
    \draw (430,133.9) node [anchor=north west][inner sep=0.75pt]    {$k$};
    \draw (378,116.9) node [anchor=north west][inner sep=0.75pt]    {$\z_{1}$};
    \draw (488,117.9) node [anchor=north west][inner sep=0.75pt]    {$\z_{2}$};
    \end{tikzpicture}= -i \bbGR (\zeta_2|\zeta_1,k) \ ,
    \label{eq:BlkBlkPropEFT}
\end{equation}
where momentum $k$ flows from left to right.

    \item Next, the vertex factors are given by
    \begin{table}[H]
    \centering
    \begin{tabular}{|c|c|}
 \hline     & \\
    \tikzset{every picture/.style={line width=0.75pt}} 
    \begin{tikzpicture}[x=0.75pt,y=0.75pt,yscale=-0.8,xscale=0.8, baseline=(current bounding box.center)]
    \draw    (415.31,103) -- (415.31,152.95) ;
    \draw [shift={(415.31,100)}, rotate = 90] [fill={rgb, 255:red, 0; green, 0; blue, 0 }  ][line width=0.08]  [draw opacity=0] (8.93,-4.29) -- (0,0) -- (8.93,4.29) -- cycle    ;
    \draw    (415.31,152.95) -- (372.45,183.27) ;
    \draw [shift={(370,185)}, rotate = 324.73] [fill={rgb, 255:red, 0; green, 0; blue, 0 }  ][line width=0.08]  [draw opacity=0] (8.93,-4.29) -- (0,0) -- (8.93,4.29) -- cycle    ;
    \draw    (464,182) -- (415.31,152.95) ;
    \draw [shift={(415.31,152.95)}, rotate = 210.82] [color={rgb, 255:red, 0; green, 0; blue, 0 }  ][fill={rgb, 255:red, 0; green, 0; blue, 0 }  ][line width=0.75]      (0, 0) circle [x radius= 3.35, y radius= 3.35]   ;
    \draw [shift={(464,182)}, rotate = 30.82] [color={rgb, 255:red, 0; green, 0; blue, 0 }  ][line width=0.75]    (0,5.59) -- (0,-5.59)   ;
    
    \draw (425,93.4) node [anchor=north west][inner sep=0.75pt]    {$k_{1}$};
    \draw (362,152.4) node [anchor=north west][inner sep=0.75pt]    {$k_{2}$};
    \draw (460,153.4) node [anchor=north west][inner sep=0.75pt]    {$k_{3}$};

    \end{tikzpicture}  &  $i \lambda_{3\rm B} $  \\   & \\
    \hline   & \\
    \tikzset{every picture/.style={line width=0.75pt}} 
    \begin{tikzpicture}[x=0.75pt,y=0.75pt,yscale=-0.8,xscale=0.8, baseline=(current bounding box.center)]
    \draw    (435.31,123) -- (435.31,172.95) ;
    \draw [shift={(435.31,120)}, rotate = 90] [fill={rgb, 255:red, 0; green, 0; blue, 0 }  ][line width=0.08]  [draw opacity=0] (8.93,-4.29) -- (0,0) -- (8.93,4.29) -- cycle    ;
    \draw    (435.31,172.95) -- (390,205) ;
    \draw [shift={(390,205)}, rotate = 324.73] [color={rgb, 255:red, 0; green, 0; blue, 0 }  ][line width=0.75]    (0,5.59) -- (0,-5.59)   ;
    \draw    (484,202) -- (435.31,172.95) ;
    \draw [shift={(435.31,172.95)}, rotate = 210.82] [color={rgb, 255:red, 0; green, 0; blue, 0 }  ][fill={rgb, 255:red, 0; green, 0; blue, 0 }  ][line width=0.75]      (0, 0) circle [x radius= 3.35, y radius= 3.35]   ;
    \draw [shift={(484,202)}, rotate = 30.82] [color={rgb, 255:red, 0; green, 0; blue, 0 }  ][line width=0.75]    (0,5.59) -- (0,-5.59)   ;
    
    \draw (445,113.4) node [anchor=north west][inner sep=0.75pt]    {$k_{1}$};
    \draw (382,172.4) node [anchor=north west][inner sep=0.75pt]    {$k_{2}$};
    \draw (480,173.4) node [anchor=north west][inner sep=0.75pt]    {$k_{3}$};
    \end{tikzpicture} & $ i \lambda_{3\rm B}  \frac{n_{-k_2} n_{-k_3}}{n_{k_1}} $\\  & \\
    \hline  
    \end{tabular}
    \caption{Feynman rules for the three-point vertices.}
    \label{tab:Vertices}
\end{table}
    The boundary momenta are conserved at all the bulk vertices, and all the bulk vertices are integrated over the exterior of the black hole:
\begin{equation}
    \int_{\rm Ext} \equiv \int_{r_{\rm h}}^{r_{\rm c}} \diff r \  r^{d-1} \ .
\end{equation}
\item Weigh each diagram by an appropriate symmetry factor.
\end{enumerate}

Note that all the propagators in these rules are causal (ingoing, outgoing, and retarded). Thus, the arrows on the propagators can be viewed as time-flow arrows, making causality manifest. The time-flow referred to here is along the ingoing Eddington-Finkelstein time $v$. Equivalently, we have appropriate analyticity in the frequency that is Fourier conjugate to $v$. Next, notice that the vertices given above are exactly those that arise in thermal field theory \cite{Gelis:1997zv}.

Over here, we have only presented the Feynman rules for the cubic interactions. We can generalise to the case of an $n$-point interaction. For an $n$-point vertex, the temperature dependence is given by
\begin{equation}
    \frac{\prod_{i \in |} n_{-k_i}}{ n_{k_\blacktriangleright}} \ , \quad \text{where} \quad k_{\blacktriangleright} \equiv \sum_{j \in\\ \blacktriangleright} k_j\ .
\end{equation}
Here, $|$ denotes the set of all semi-propagators emanating from that vertex and ending in a semi-capacitor (i.e., $|$). Similarly, $\blacktriangleright$ denotes the set of all semi-propagators emanating from that vertex and ending in a semi-diode (i.e., $\blacktriangleright$). This structure of vertices is well known in thermal field theory \cite{vanEijck:1992mq, vanEijck:1994rw, Gelis:1997zv, Carrington:2006xj, Gelis:2019yfm}.

These Feynman rules are sufficient to compute the boundary Schwinger-Keldysh generating functional $\mathcal{Z}_{\rm SK}$ to any order in bulk coupling constant, at tree-level in the bulk. In the following sections, we will explore how these rules apply when loop corrections in the bulk are introduced.

\section{grSK QFT and exterior QFT}\label{sec:grSKQFT}

We now begin our discussion of quantum field theory on the grSK spacetime. Before diving into the details, we pause for a moment to remind the reader of our goals in this section. Recall that the quantity that we are finally interested in computing is the generating functional of boundary Schwinger-Keldysh correlators. The grSK geometry gives us a way to compute these by computing the bulk on-shell action with the appropriate GKPW boundary conditions. This is the story at the tree-level approximation (or at the level of classical field theory) in the bulk. 


In the present note, our goal is to extend this tree-level bulk analysis to include loop corrections. In particular, we aim to obtain a systematic formalism that can be used for calculating the boundary Schwinger-Keldysh generating functional in a loop expansion in the bulk. We will do this in two ways. First, we will set up the quantum field theory and the loop expansion in the grSK spacetime. Secondly, we will extend the exterior field theory picture to include loop corrections. We will then show that both these extensions result in the same boundary correlators. The equivalence of these two procedures will be one of the main results of this note. Furthermore, we show that these loop extensions manifestly satisfy microscopic unitarity and thermality conditions.

There are several equivalent routes to setting up a quantum field theory, familiar from standard discussions of quantum field theory. One way is to perform canonical quantisation, another is to perform path integral quantisation (See \cite{Weinberg:1995mt}). Yet another way to quantise is to simply follow a diagrammatic approach. In this work, we will employ the diagrammatic approach to quantum field theory presented in many textbooks \cite{Cvitanovic:1983eb, Rammer:2007zz, Banks:2014twn, Rivers:1987hi, Kleiss:2021pih}. In particular, we will closely follow \cite{Rammer:2007zz} and \cite{Cvitanovic:1983eb}. We prefer the diagrammatic approach for a very good reason: we already know the diagrammatics at the tree-level for the grSK QFT as well as the exterior QFT (Sec.~(\ref{sec:grSKeFTReview})).

Before specialising to the grSK spacetime, we now present a quick review of the diagrammatic approach in quantum field theory. The basic objects that we want to compute using diagrammatics are correlators.\footnote{Our discussion will be for the bulk quantum field theory. Note that throughout this note, we work in units such that in the bulk $\hbar=1$. We will explain the boundary limit when we get to the discussion of grSK.} For convenience, instead of computing the correlators one by one, we will instead focus on studying their generating functional, which we will denote by $Z[\mathfrak{J}]$,\footnote{The reader familiar with the path integral will recognise this as the sourced path integral.} where $\mathfrak{J}$ is a source of the field. Suitable differentiations of $Z[\mathfrak{J}]$ with respect to the source results in the correlators of the field that one is interested in. Diagrammatically, we represent the generating functional $Z[\mathfrak{J}]$ by a shaded blob:
\begin{equation}
\tikzset{every picture/.style={line width=0.75pt}} 
\begin{tikzpicture}[baseline={([yshift=-0.5ex]current bounding box.center)},x=0.75pt,y=0.75pt,yscale=-0.5,xscale=0.5]

\draw  [fill={rgb, 255:red, 155; green, 155; blue, 155 }  ,fill opacity=1 ][line width=0.75]  (389.35,140.31) .. controls (389.33,124.12) and (402.45,110.98) .. (418.64,110.96) .. controls (434.83,110.95) and (447.98,124.06) .. (447.99,140.25) .. controls (448.01,156.45) and (434.9,169.59) .. (418.7,169.61) .. controls (402.51,169.62) and (389.37,156.51) .. (389.35,140.31) -- cycle ;
\end{tikzpicture}
 \ \equiv \  Z[\mathfrak{J}] \ .
\end{equation}
A derivative of the generating functional with respect to the source $\mathfrak{J}$ is denoted by a line attached to this blob:
\begin{equation}
\tikzset{every picture/.style={line width=0.75pt}} 
\begin{tikzpicture}[baseline={([yshift=-0.5ex]current bounding box.center)},x=0.75pt,y=0.75pt,yscale=-1,xscale=1]
\draw    (547.33,91) -- (593.33,91) ; 
\draw  [fill={rgb, 255:red, 155; green, 155; blue, 155 }  ,fill opacity=1 ][line width=0.75]  (593.33,91) .. controls (593.33,83.26) and (599.59,76.99) .. (607.33,76.98) .. controls (615.06,76.97) and (621.34,83.23) .. (621.35,90.97) .. controls (621.36,98.71) and (615.09,104.98) .. (607.36,104.99) .. controls (599.62,105) and (593.34,98.74) .. (593.33,91) -- cycle ;
\draw (540,93.4) node [anchor=north west][inner sep=0.75pt]  [font=\scriptsize]  {$\mathnormal{Y}$};
\end{tikzpicture}
 \ \equiv \  \frac{1}{i}\frac{\delta Z[\mathfrak{J}]}{\delta \mathfrak{J} (Y)} \ ,
\end{equation}
where $Y$ is a point in spacetime. Furthermore, the source itself is denoted by a cross:
\begin{equation}
    \times  \ \equiv \  i \mathfrak{J}(Y) \ .
\end{equation}
Here, the cross is understood to be located at the bulk point $Y$.

The Feynman rules of a theory describe the propagators and the interaction vertices used in a graph or diagram. The basic principle of the diagrammatic approach is that in a graph, the edges can either end in a vertex or not.\footnote{Stated in terms of the interacting particles: the particles can choose to interact or not to interact \cite{Cvitanovic:1983eb, Rammer:2007zz}. In the RHS of Eq.~\eqref{eq:Z1SDEbasic}, the first and the second term represent `interact' and `not to interact', respectively.} Let us assume for now that our theory only has one three-point vertex. The functional derivative of the generating functional $Z[\mathfrak{J}]$ can then be written as
\begin{equation}\label{eq:Z1SDEbasic}
    \tikzset{every picture/.style={line width=0.75pt}} 
    \begin{tikzpicture}[x=0.75pt,y=0.75pt,yscale=-1,xscale=1]
    \centering
    \draw    (111,127) -- (171.33,127) ;
    \draw  [fill={rgb, 255:red, 155; green, 155; blue, 155 }  ,fill opacity=1 ][line width=0.75]  (171.33,127) .. controls (171.33,119.26) and (177.59,112.99) .. (185.33,112.98) .. controls (193.06,112.97) and (199.34,119.23) .. (199.35,126.97) .. controls (199.36,134.71) and (193.09,140.98) .. (185.36,140.99) .. controls (177.62,141) and (171.34,134.74) .. (171.33,127) -- cycle ;
    \draw  [fill={rgb, 255:red, 155; green, 155; blue, 155 }  ,fill opacity=1 ] (339.33,127) .. controls (339.33,119.26) and (345.59,112.99) .. (353.33,112.98) .. controls (361.06,112.97) and (367.34,119.23) .. (367.35,126.97) .. controls (367.36,134.71) and (361.09,140.98) .. (353.36,140.99) .. controls (345.62,141) and (339.34,134.74) .. (339.33,127) -- cycle ;
    \draw    (270.33,128) -- (316.33,128) ;
    \draw    (344.33,116) -- (316.33,128) ;
    \draw    (316.33,128) -- (343.33,137) ;
    \draw    (408.33,128) .. controls (427.16,127.59) and (435.1,124) .. (441.54,119.9) .. controls (450.84,113.98) and (457.02,107) .. (488.33,107) ;
    \draw  [fill={rgb, 255:red, 155; green, 155; blue, 155 }  ,fill opacity=1 ] (463.33,126) .. controls (463.33,118.26) and (469.59,111.99) .. (477.33,111.98) .. controls (485.06,111.97) and (491.34,118.23) .. (491.35,125.97) .. controls (491.36,133.71) and (485.09,139.98) .. (477.36,139.99) .. controls (469.62,140) and (463.34,133.74) .. (463.33,126) -- cycle ;
    
    \draw (210,124.4) node [anchor=north west][inner sep=0.75pt]    {$=$};
    \draw (106,129.4) node [anchor=north west][inner sep=0.75pt]  [font=\scriptsize]  {$\mathnormal{Y}$};
    \draw (245,116.4) node [anchor=north west][inner sep=0.75pt]  [font=\normalsize]  {$\frac{1}{2!}$};
    \draw (266,129.4) node [anchor=north west][inner sep=0.75pt]  [font=\scriptsize]  {$\mathnormal{Y}$};
    \draw (379,124.4) node [anchor=north west][inner sep=0.75pt]    {$+$};
    \draw (482,101.4) node [anchor=north west][inner sep=0.75pt]  [font=\scriptsize]  {$\times $};
    \draw (403,128.4) node [anchor=north west][inner sep=0.75pt]  [font=\scriptsize]  {$\mathnormal{Y}$};
    \end{tikzpicture} \ .
\end{equation}
This equation is called the \emph{Schwinger-Dyson} equation for the generating functional $Z[\mathfrak{J}]$ (See the diagrammatic equation in Eq.~(2.12) of \cite{Cvitanovic:1983eb} and Fig.~(9.32) of \cite{Rammer:2007zz}). This is the fundamental diagrammatic equation that allows us to compute the generating functional $Z[\mathfrak{J}]$ using the diagrammatic rules. In other words, it is the functional differential equation for $Z[\mathfrak{J}]$, of which the path integral is then a formal (often ill-defined) functional integral solution.\footnote{The reader interested in a path integral derivation of this equation is referred to Appendix \ref{app:SDEsfromPI}.} The diagrammatic equation \eqref{eq:Z1SDEbasic} can be more explicitly written as
\begin{equation}
    \frac{1}{i}\frac{\delta}{\delta \mathfrak{J}(Y)} Z[\mathfrak{J}] = \int_{Y_0}[-i G(Y|Y_0)] \left[\frac{-i\lambda}{2!} \left(\frac{1}{i} \frac{\delta}{\delta \mathfrak{J}(Y_0)}\right)^2 +i \mathfrak{J}(Y_0) \right] Z[\mathfrak{J}] \ ,
\end{equation}
where $\lambda$ is the three-point coupling constant, and $-iG(Y|Y_0)$ is the propagator of the theory. Both of these are the basic ingredients used to formulate the theory.

The above diagrammatic equation is written down using the basic rule: edges end in vertices or at sources (from the particle point of view: \emph{particles either interact or do not}). The first term on the right-hand side accounts for the only interaction in the theory we are considering, i.e., the three-point interaction. The second term is the disconnected term, which arises only when you turn on sources.

The main point to note here is that, given the propagators and vertices from classical field theory, we can write down the above equation explicitly in perturbation theory. Integrating the resulting equation, we are directly led to the full quantum generating functional in perturbation theory. This is the point of view we will use in what follows.

Perturbatively expanding the Schwinger-Dyson equation results in a sum of all diagrams of the theory: connected and disconnected.  Drawing all these diagrams is clearly not economical, since we will be drawing connected diagrams repeatedly. For economy's sake, we can adapt the discussion in terms of the connected generating functional $W[\mathfrak{J}]$ defined in terms of $Z[\mathfrak{J}]$, by the equation
\begin{equation}
    Z[\mathfrak{J}]=: e^{i W[\mathfrak{J}]} \ .
\end{equation}
This generating functional has a perturbative expansion, which is a sum of purely connected diagrams \cite{Weinberg:1996kr}. The diagrammatic equation for $Z[\mathfrak{J}]$ can now be rewritten in terms of the connected generating functional $W[\mathfrak{J}]$. To this end, we introduce a blob notation for $W[\mathfrak{J}]$:
\begin{equation}
   i W[\mathfrak{J}] \equiv
\tikzset{every picture/.style={line width=0.75pt}} 
\begin{tikzpicture}[baseline={([yshift=-0.5ex]current bounding box.center)},x=0.75pt,y=0.75pt,yscale=-1,xscale=1]

\draw   (305.33,129.96) .. controls (305.33,121.7) and (312.05,115) .. (320.33,115) .. controls (328.62,115) and (335.33,121.7) .. (335.33,129.96) .. controls (335.33,138.22) and (328.62,144.92) .. (320.33,144.92) .. controls (312.05,144.92) and (305.33,138.22) .. (305.33,129.96) -- cycle ;
\draw    (305.61,126.9) -- (320.33,115) ;
\draw    (323.99,115.48) -- (305.37,130.6) ;
\draw    (305.85,133.93) -- (327.22,116.55) ;
\draw    (307.04,137.02) -- (329.84,118.33) ;
\draw    (308.81,139.48) -- (316.38,133.23) -- (331.85,120.44) ;
\draw    (310.98,141.55) -- (333.54,122.86) ;
\draw    (313.37,143.21) -- (334.62,125.48) ;
\draw    (316.35,144.29) -- (335.33,128.57) ;
\draw    (319.46,145) -- (335.09,132.26) ;
\draw    (324.47,144.29) -- (333.54,136.9) ;
\draw    (314.68,116.07) -- (308.12,121.43) ;
\end{tikzpicture}
 \ .
\end{equation}
Then the Schwinger-Dyson equation for the connected generating functional (See the diagrammatic equation Eq.~(2.21) of \cite{Cvitanovic:1983eb} and Ex.~(9.9) of \cite{Rammer:2007zz}) takes the form 
\begin{equation}
    \tikzset{every picture/.style={line width=0.75pt}} 
    \begin{tikzpicture}[x=0.75pt,y=0.75pt,yscale=-1,xscale=1]
    
    \draw    (74.33,128) -- (130.33,128.92) ;
    \draw    (202.33,129) -- (258.33,129.92) ;
    \draw    (314.33,130) -- (343.33,130) ;
    \draw    (343.33,130) -- (365.12,118.43) ;
    \draw    (343.33,130) -- (370.37,140.21) ;
    \draw    (437.33,132) -- (466.33,132) ;
    \draw    (466.33,132) -- (482.85,109.93) ;
    \draw    (466.33,132) -- (484.61,152.9) ;
    \draw   (482.33,105.96) .. controls (482.33,97.7) and (489.05,91) .. (497.33,91) .. controls (505.62,91) and (512.33,97.7) .. (512.33,105.96) .. controls (512.33,114.22) and (505.62,120.92) .. (497.33,120.92) .. controls (489.05,120.92) and (482.33,114.22) .. (482.33,105.96) -- cycle ;
    \draw    (482.61,102.9) -- (497.33,91) ;
    \draw    (500.99,91.48) -- (482.37,106.6) ;
    \draw    (482.85,109.93) -- (504.22,92.55) ;
    \draw    (484.04,113.02) -- (506.84,94.33) ;
    \draw    (485.81,115.48) -- (493.38,109.23) -- (508.85,96.44) ;
    \draw    (487.98,117.55) -- (510.54,98.86) ;
    \draw    (490.37,119.21) -- (511.62,101.48) ;
    \draw    (493.35,120.29) -- (512.33,104.57) ;
    \draw    (496.46,121) -- (512.09,108.26) ;
    \draw    (501.47,120.29) -- (510.54,112.9) ;
    \draw    (491.68,92.07) -- (485.12,97.43) ;
    \draw   (484.33,155.96) .. controls (484.33,147.7) and (491.05,141) .. (499.33,141) .. controls (507.62,141) and (514.33,147.7) .. (514.33,155.96) .. controls (514.33,164.22) and (507.62,170.92) .. (499.33,170.92) .. controls (491.05,170.92) and (484.33,164.22) .. (484.33,155.96) -- cycle ;
    \draw    (484.61,152.9) -- (499.33,141) ;
    \draw    (502.99,141.48) -- (484.37,156.6) ;
    \draw    (484.85,159.93) -- (506.22,142.55) ;
    \draw    (486.04,163.02) -- (508.84,144.33) ;
    \draw    (487.81,165.48) -- (495.38,159.23) -- (510.85,146.44) ;
    \draw    (489.98,167.55) -- (512.54,148.86) ;
    \draw    (492.37,169.21) -- (513.62,151.48) ;
    \draw    (495.35,170.29) -- (514.33,154.57) ;
    \draw    (498.46,171) -- (514.09,158.26) ;
    \draw    (503.47,170.29) -- (512.54,162.9) ;
    \draw    (493.68,142.07) -- (487.12,147.43) ;
    \draw   (130.33,127.96) .. controls (130.33,119.7) and (137.05,113) .. (145.33,113) .. controls (153.62,113) and (160.33,119.7) .. (160.33,127.96) .. controls (160.33,136.22) and (153.62,142.92) .. (145.33,142.92) .. controls (137.05,142.92) and (130.33,136.22) .. (130.33,127.96) -- cycle ;
    \draw    (130.61,124.9) -- (145.33,113) ;
    \draw    (148.99,113.48) -- (130.37,128.6) ;
    \draw    (130.85,131.93) -- (152.22,114.55) ;
    \draw    (132.04,135.02) -- (154.84,116.33) ;
    \draw    (133.81,137.48) -- (141.38,131.23) -- (156.85,118.44) ;
    \draw    (135.98,139.55) -- (158.54,120.86) ;
    \draw    (138.37,141.21) -- (159.62,123.48) ;
    \draw    (141.35,142.29) -- (160.33,126.57) ;
    \draw    (144.46,143) -- (160.09,130.26) ;
    \draw    (149.47,142.29) -- (158.54,134.9) ;
    \draw    (139.68,114.07) -- (133.12,119.43) ;
    \draw   (362.33,126.96) .. controls (362.33,118.7) and (369.05,112) .. (377.33,112) .. controls (385.62,112) and (392.33,118.7) .. (392.33,126.96) .. controls (392.33,135.22) and (385.62,141.92) .. (377.33,141.92) .. controls (369.05,141.92) and (362.33,135.22) .. (362.33,126.96) -- cycle ;
    \draw    (362.61,123.9) -- (377.33,112) ;
    \draw    (380.99,112.48) -- (362.37,127.6) ;
    \draw    (362.85,130.93) -- (384.22,113.55) ;
    \draw    (364.04,134.02) -- (386.84,115.33) ;
    \draw    (365.81,136.48) -- (373.38,130.23) -- (388.85,117.44) ;
    \draw    (367.98,138.55) -- (390.54,119.86) ;
    \draw    (370.37,140.21) -- (391.62,122.48) ;
    \draw    (373.35,141.29) -- (392.33,125.57) ;
    \draw    (376.46,142) -- (392.09,129.26) ;
    \draw    (381.47,141.29) -- (390.54,133.9) ;
    \draw    (371.68,113.07) -- (365.12,118.43) ;
    
    \draw (169,124.4) node [anchor=north west][inner sep=0.75pt]    {$=$};
    \draw (252,124.4) node [anchor=north west][inner sep=0.75pt]  [font=\scriptsize]  {$\times $};
    \draw (277,120.4) node [anchor=north west][inner sep=0.75pt]    {$+$};
    \draw (291,120.4) node [anchor=north west][inner sep=0.75pt]    {$\frac{1}{2!}$};
    \draw (399,120.4) node [anchor=north west][inner sep=0.75pt]    {$+$};
    \draw (417,120.4) node [anchor=north west][inner sep=0.75pt]    {$\frac{1}{2!}$};
    \draw (68,131.4) node [anchor=north west][inner sep=0.75pt]  [font=\footnotesize]  {$Y$};
    \draw (308.33,131.4) node [anchor=north west][inner sep=0.75pt]  [font=\footnotesize]  {$Y$};
    \draw (196,131.4) node [anchor=north west][inner sep=0.75pt]  [font=\footnotesize]  {$Y$};
    \draw (432,133.4) node [anchor=north west][inner sep=0.75pt]  [font=\footnotesize]  {$Y$};

    \end{tikzpicture} \ ,
    \label{eq:SchDysonDiag}
\end{equation}
Note that all vertices and source points that appear in this equation are integrated all over the bulk as always.

The perturbative expansion of these equations clearly does not have any disconnected terms. In what follows, we will use this as our starting point, i.e., we will quantise the classical field theory by proposing Schwinger-Dyson equations for the connected generating functional $W[\mathfrak{J}]$.

With that brief overview of the quantisation procedure using diagrammatics, we now start our discussion of the quantum field theory on the grSK spacetime.  We need to first write down the Schwinger-Dyson equations to quantise the theory. But as we have learnt in the previous section, all the SDEs do is to precisely define the QFT. Perturbatively, all we really require are the propagators and the vertices. Thus, we relegate all the details of the Schwinger-Dyson equations to Appendix \ref{app:grSKExtEFTSDE}, and take a simple-minded view of starting the computation of loops in this section. We already have the rules from Sec \ref{sec:grSKeFTReview}, we will now start computing loops. 
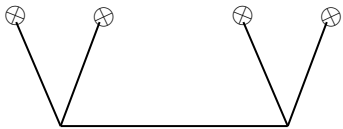
\begin{figure}[H]
    \centering
\tikzset{every picture/.style={line width=0.75pt}} 
\begin{tikzpicture}[x=0.75pt,y=0.75pt,yscale=-0.6,xscale=0.6]

\draw    (288.21,213.37) -- (251.28,126.2) ;
\draw    (288.21,213.37) -- (320.67,126.19) ;
\draw    (477.21,213.37) -- (440.28,126.2) ;
\draw    (477.21,213.37) -- (509.67,126.19) ;
\draw    (288.21,213.37) -- (477.21,213.37) ;

\draw (245.53,135.49) node [anchor=north west][inner sep=0.75pt]  [rotate=-247.04]  {$\otimes $};
\draw (310.14,129.1) node [anchor=north west][inner sep=0.75pt]  [rotate=-289.84]  {$\otimes $};
\draw (434.53,135.49) node [anchor=north west][inner sep=0.75pt]  [rotate=-247.04]  {$\otimes $};
\draw (499.14,129.1) node [anchor=north west][inner sep=0.75pt]  [rotate=-289.84]  {$\otimes $};

\end{tikzpicture}
    \caption{Four-point exchange diagram on the grSK geometry.}
    \label{fig:FourPtExch}
\end{figure}
Before computing loops, we review for the reader's convenience how a tree-level diagram is computed. Let us consider the exchange diagram (see Fig.(\ref{fig:FourPtExch})) in $\phi^3$ theory computed in the grSK geometry. Using the rules in Sec \ref{sec:grSKeFTReview}, we can now compute this diagram, and perform all the monodromy integrals using the results in Appendix \ref{app:MonInt}. 

To neatly arrange the various terms with different powers of the boundary sources, we will employ the useful notation \cite{Jana:2020vyx}:
\begin{equation}
    S_{(n)} = \int \prod_{i=1}^{n} \frac{\diff^d k_i}{(2 \pi)^d}\ (2\pi)^d \delta^{(d)} \left(\sum_{i=1}^{n} k_i\right) \left[\sum_{p=0}^{n}\mathcal{I}_{p,n-p}(k_1, \ldots, k_n)\ \prod_{i=1}^{p} {J}_{\Fb} (k_i) \prod_{j= p+1}^{n} {J}_{\Pb} (k_j)\right] \ .
\label{eq:DefInfluencePhaseNotation}
\end{equation}
Computing the diagrams gives us
\begin{equation}
    \begin{split}
        S^{\rm tree}_{(4)} &= - \int \prod_{i=1}^{4} \frac{\diff^d k_i}{(2\pi)^d}\  (2\pi)^d \delta^{(d)} \left(\sum_{i=1}^{4} k_i\right)\\
        &\times \left(- \frac{\lambda^2_{3\rm B}}{2!}\right)  \oint_{\zeta_1}\oint_{\zeta_2}  \frac{1}{2!}\phi_{(0)} (\zeta_1, k_1) \phi_{(0)} (\zeta_1, k_2) \bbG(\zeta_2|\zeta_1,k_1+k_2) \frac{1}{2!} \phi_{(0)}(\zeta_2, k_3)  \phi_{(0)}(\zeta_2, k_4) \ .
    \end{split}
\end{equation}
Reducing the monodromy integrals to exterior integrals, we obtain four simple terms with neat Bose-Einstein factors:
\begin{equation}
    \begin{split}
        \mathcal{I}_{3,1}^{\rm tr} &= \frac{\lambda_{3 \rm B}^2}{n_{k_4}} \int_{\rm Ext_{1,2}}  \frac{1}{2!}\Gin(\zeta_1,k_1) \Gin(\zeta_1,k_2) \ \bbGR(\zeta_2|\zeta_1, k_1+k_2) \ \Gin(\zeta_2,k_3) G^{\rm out}(\zeta_2,k_4) \ ,\\
        \mathcal{I}_{2,2}^{\rm tr} &= -\lambda_{3\rm B}^2 \int_{\rm Ext_{1,2}} \\
        &\hspace{1cm}\Bigg\{ \frac{1}{n_{k_3+k_4}}    \frac{1}{2!} \Gin(\zeta_1,k_1) \Gin(\zeta_1,k_2) \ \bbGR(\zeta_2|\zeta_1, k_1+k_2)\frac{1}{2!}G^{\rm out}(\zeta_2,k_3) G^{\rm out}(\zeta_2,k_4)\\
        &\hspace{0.6cm}+\frac{1+n_{k_1+k_3}}{(1+n_{k_1})n_{k_4}} \Gin(\zeta_1,k_1)G^{\rm out}(\zeta_1,k_3) \ \bbGR(\zeta_2|\zeta_1, k_1+k_3)\  \Gin(\zeta_2,k_2)  G^{\rm out}(\zeta_2,k_4)\Bigg\}\ ,\\
        \mathcal{I}_{1,3}^{\rm tr} &= \frac{\lambda_{3\rm B}^2}{n_{k_2+k_3+k_4}}\int_{\rm Ext_{1,2}}  \Gin(\zeta_1,k_1)  G^{\rm out}(\zeta_1,k_2)  \bbGR(\zeta_2|\zeta_1, k_1+k_2)   \frac{1}{2!} G^{\rm out}(\zeta_2,k_3)  G^{\rm out}(\zeta_2,k_4) \ .
    \end{split}
\end{equation}
This set of terms is exactly what we obtain from the following diagrams drawn in the exterior field theory. This is what was first realised in \cite{Loganayagam:2024mnj}, and is an example of the statement that the exterior field theory is equivalent to the grSK field theory at tree level.
\begin{table}
    \centering
    \begin{tabular}{cc}

\tikzset{every picture/.style={line width=0.75pt}} 

\begin{tikzpicture}[x=0.75pt,y=0.75pt,yscale=-0.7,xscale=0.7]

\draw    (270.89,117.79) -- (288.15,162.17) ;
\draw [shift={(289.24,164.96)}, rotate = 248.75] [fill={rgb, 255:red, 0; green, 0; blue, 0 }  ][line width=0.08]  [draw opacity=0] (8.93,-4.29) -- (0,0) -- (8.93,4.29) -- cycle    ;
\draw    (288.72,163.53) -- (304.95,200.88) ;
\draw [shift={(288.72,163.53)}, rotate = 246.51] [color={rgb, 255:red, 0; green, 0; blue, 0 }  ][line width=0.75]    (0,5.59) -- (0,-5.59)   ;

\draw    (336.51,118.04) -- (319.12,162.37) ;
\draw [shift={(318.03,165.16)}, rotate = 291.42] [fill={rgb, 255:red, 0; green, 0; blue, 0 }  ][line width=0.08]  [draw opacity=0] (8.93,-4.29) -- (0,0) -- (8.93,4.29) -- cycle    ;
\draw    (318.61,163.75) -- (304.95,200.88) ;
\draw [shift={(318.61,163.75)}, rotate = 290.2] [color={rgb, 255:red, 0; green, 0; blue, 0 }  ][line width=0.75]    (0,5.59) -- (0,-5.59)   ;

\draw    (418.89,117.79) -- (436.15,162.17) ;
\draw [shift={(437.24,164.96)}, rotate = 248.75] [fill={rgb, 255:red, 0; green, 0; blue, 0 }  ][line width=0.08]  [draw opacity=0] (8.93,-4.29) -- (0,0) -- (8.93,4.29) -- cycle    ;
\draw    (436.72,163.53) -- (452.95,200.88) ;
\draw [shift={(436.72,163.53)}, rotate = 246.51] [color={rgb, 255:red, 0; green, 0; blue, 0 }  ][line width=0.75]    (0,5.59) -- (0,-5.59)   ;

\draw    (485.09,116.63) -- (470.5,155) ;
\draw [shift={(470.5,155)}, rotate = 290.82] [color={rgb, 255:red, 0; green, 0; blue, 0 }  ][line width=0.75]    (0,5.59) -- (0,-5.59)   ;
\draw    (469.43,157.8) -- (452.95,200.88) ;
\draw [shift={(470.5,155)}, rotate = 110.93] [fill={rgb, 255:red, 0; green, 0; blue, 0 }  ][line width=0.08]  [draw opacity=0] (8.93,-4.29) -- (0,0) -- (8.93,4.29) -- cycle    ;
\draw    (304.95,200.88) -- (378.73,200.03) ;
\draw [shift={(381.73,200)}, rotate = 179.34] [fill={rgb, 255:red, 0; green, 0; blue, 0 }  ][line width=0.08]  [draw opacity=0] (8.93,-4.29) -- (0,0) -- (8.93,4.29) -- cycle    ;
\draw    (381.73,200) -- (452.95,200.88) ;
\draw [shift={(381.73,200)}, rotate = 180.71] [color={rgb, 255:red, 0; green, 0; blue, 0 }  ][line width=0.75]    (0,5.59) -- (0,-5.59)   ;

\draw (411.23,124.27) node [anchor=north west][inner sep=0.75pt]  [rotate=-251.04]  {$\otimes $};
\draw (297.83,191.6) node [anchor=north west][inner sep=0.75pt]  [font=\large]  {$\bullet $};
\draw (326.48,117.61) node [anchor=north west][inner sep=0.75pt]  [rotate=-293.71]  {$\otimes $};
\draw (263.23,124.27) node [anchor=north west][inner sep=0.75pt]  [rotate=-251.04]  {$\otimes $};
\draw (474.48,117.61) node [anchor=north west][inner sep=0.75pt]  [rotate=-293.71]  {$\otimes $};
\draw (445.83,191.6) node [anchor=north west][inner sep=0.75pt]  [font=\large]  {$\bullet $};

\end{tikzpicture} & 

\tikzset{every picture/.style={line width=0.75pt}} 

\begin{tikzpicture}[x=0.75pt,y=0.75pt,yscale=-0.7,xscale=0.7]

\draw    (266.89,90.79) -- (284.15,135.17) ;
\draw [shift={(285.24,137.96)}, rotate = 248.75] [fill={rgb, 255:red, 0; green, 0; blue, 0 }  ][line width=0.08]  [draw opacity=0] (8.93,-4.29) -- (0,0) -- (8.93,4.29) -- cycle    ;
\draw    (284.72,136.53) -- (300.95,173.88) ;
\draw [shift={(284.72,136.53)}, rotate = 246.51] [color={rgb, 255:red, 0; green, 0; blue, 0 }  ][line width=0.75]    (0,5.59) -- (0,-5.59)   ;

\draw    (332.51,91.04) -- (315.12,135.37) ;
\draw [shift={(314.03,138.16)}, rotate = 291.42] [fill={rgb, 255:red, 0; green, 0; blue, 0 }  ][line width=0.08]  [draw opacity=0] (8.93,-4.29) -- (0,0) -- (8.93,4.29) -- cycle    ;
\draw    (314.61,136.75) -- (300.95,173.88) ;
\draw [shift={(314.61,136.75)}, rotate = 290.2] [color={rgb, 255:red, 0; green, 0; blue, 0 }  ][line width=0.75]    (0,5.59) -- (0,-5.59)   ;

\draw    (414.89,90.79) -- (429.5,127) ;
\draw [shift={(429.5,127)}, rotate = 248.03] [color={rgb, 255:red, 0; green, 0; blue, 0 }  ][line width=0.75]    (0,5.59) -- (0,-5.59)   ;
\draw    (430.65,129.77) -- (449.48,175.31) ;
\draw [shift={(429.5,127)}, rotate = 67.54] [fill={rgb, 255:red, 0; green, 0; blue, 0 }  ][line width=0.08]  [draw opacity=0] (8.93,-4.29) -- (0,0) -- (8.93,4.29) -- cycle    ;
\draw    (481.09,89.63) -- (466.5,128) ;
\draw [shift={(466.5,128)}, rotate = 290.82] [color={rgb, 255:red, 0; green, 0; blue, 0 }  ][line width=0.75]    (0,5.59) -- (0,-5.59)   ;
\draw    (465.43,130.8) -- (448.95,173.88) ;
\draw [shift={(466.5,128)}, rotate = 110.93] [fill={rgb, 255:red, 0; green, 0; blue, 0 }  ][line width=0.08]  [draw opacity=0] (8.93,-4.29) -- (0,0) -- (8.93,4.29) -- cycle    ;
\draw    (300.95,173.88) -- (374.73,173.03) ;
\draw [shift={(377.73,173)}, rotate = 179.34] [fill={rgb, 255:red, 0; green, 0; blue, 0 }  ][line width=0.08]  [draw opacity=0] (8.93,-4.29) -- (0,0) -- (8.93,4.29) -- cycle    ;
\draw    (377.73,173) -- (448.95,173.88) ;
\draw [shift={(377.73,173)}, rotate = 180.71] [color={rgb, 255:red, 0; green, 0; blue, 0 }  ][line width=0.75]    (0,5.59) -- (0,-5.59)   ;

\draw (470.48,90.61) node [anchor=north west][inner sep=0.75pt]  [rotate=-293.71]  {$\otimes $};
\draw (441.83,166.6) node [anchor=north west][inner sep=0.75pt]  [font=\large]  {$\bullet $};
\draw (293.83,166.6) node [anchor=north west][inner sep=0.75pt]  [font=\large]  {$\bullet $};
\draw (322.48,90.61) node [anchor=north west][inner sep=0.75pt]  [rotate=-293.71]  {$\otimes $};
\draw (259.23,97.27) node [anchor=north west][inner sep=0.75pt]  [rotate=-251.04]  {$\otimes $};
\draw (407.23,97.27) node [anchor=north west][inner sep=0.75pt]  [rotate=-251.04]  {$\otimes $};

\end{tikzpicture}
 \\

\tikzset{every picture/.style={line width=0.75pt}} 

\begin{tikzpicture}[x=0.75pt,y=0.75pt,yscale=-0.7,xscale=0.7]

\draw    (104.89,123.79) -- (122.15,168.17) ;
\draw [shift={(123.24,170.96)}, rotate = 248.75] [fill={rgb, 255:red, 0; green, 0; blue, 0 }  ][line width=0.08]  [draw opacity=0] (8.93,-4.29) -- (0,0) -- (8.93,4.29) -- cycle    ;
\draw    (122.72,169.53) -- (138.95,206.88) ;
\draw [shift={(122.72,169.53)}, rotate = 246.51] [color={rgb, 255:red, 0; green, 0; blue, 0 }  ][line width=0.75]    (0,5.59) -- (0,-5.59)   ;

\draw    (171.09,122.63) -- (156.5,161) ;
\draw [shift={(156.5,161)}, rotate = 290.82] [color={rgb, 255:red, 0; green, 0; blue, 0 }  ][line width=0.75]    (0,5.59) -- (0,-5.59)   ;
\draw    (155.43,163.8) -- (138.95,206.88) ;
\draw [shift={(156.5,161)}, rotate = 110.93] [fill={rgb, 255:red, 0; green, 0; blue, 0 }  ][line width=0.08]  [draw opacity=0] (8.93,-4.29) -- (0,0) -- (8.93,4.29) -- cycle    ;
\draw    (251.89,123.79) -- (269.15,168.17) ;
\draw [shift={(270.24,170.96)}, rotate = 248.75] [fill={rgb, 255:red, 0; green, 0; blue, 0 }  ][line width=0.08]  [draw opacity=0] (8.93,-4.29) -- (0,0) -- (8.93,4.29) -- cycle    ;
\draw    (269.72,169.53) -- (285.95,206.88) ;
\draw [shift={(269.72,169.53)}, rotate = 246.51] [color={rgb, 255:red, 0; green, 0; blue, 0 }  ][line width=0.75]    (0,5.59) -- (0,-5.59)   ;

\draw    (318.09,122.63) -- (303.5,161) ;
\draw [shift={(303.5,161)}, rotate = 290.82] [color={rgb, 255:red, 0; green, 0; blue, 0 }  ][line width=0.75]    (0,5.59) -- (0,-5.59)   ;
\draw    (302.43,163.8) -- (285.95,206.88) ;
\draw [shift={(303.5,161)}, rotate = 110.93] [fill={rgb, 255:red, 0; green, 0; blue, 0 }  ][line width=0.08]  [draw opacity=0] (8.93,-4.29) -- (0,0) -- (8.93,4.29) -- cycle    ;

\draw    (137.95,206.88) -- (211.73,206.03) ;
\draw [shift={(214.73,206)}, rotate = 179.34] [fill={rgb, 255:red, 0; green, 0; blue, 0 }  ][line width=0.08]  [draw opacity=0] (8.93,-4.29) -- (0,0) -- (8.93,4.29) -- cycle    ;
\draw    (214.73,206) -- (285.95,206.88) ;
\draw [shift={(214.73,206)}, rotate = 180.71] [color={rgb, 255:red, 0; green, 0; blue, 0 }  ][line width=0.75]    (0,5.59) -- (0,-5.59)   ;

\draw (160.48,123.61) node [anchor=north west][inner sep=0.75pt]  [rotate=-293.71]  {$\otimes $};
\draw (131.83,199.6) node [anchor=north west][inner sep=0.75pt]  [font=\large]  {$\bullet $};
\draw (97.23,130.27) node [anchor=north west][inner sep=0.75pt]  [rotate=-251.04]  {$\otimes $};
\draw (244.23,130.27) node [anchor=north west][inner sep=0.75pt]  [rotate=-251.04]  {$\otimes $};
\draw (307.48,123.61) node [anchor=north west][inner sep=0.75pt]  [rotate=-293.71]  {$\otimes $};
\draw (278.83,199.6) node [anchor=north west][inner sep=0.75pt]  [font=\large]  {$\bullet $};

\end{tikzpicture}
  & 

\tikzset{every picture/.style={line width=0.75pt}} 

\begin{tikzpicture}[x=0.75pt,y=0.75pt,yscale=-0.7,xscale=0.7]

\draw    (333.89,109.79) -- (348.5,146) ;
\draw [shift={(348.5,146)}, rotate = 248.03] [color={rgb, 255:red, 0; green, 0; blue, 0 }  ][line width=0.75]    (0,5.59) -- (0,-5.59)   ;
\draw    (349.65,148.77) -- (368.48,194.31) ;
\draw [shift={(348.5,146)}, rotate = 67.54] [fill={rgb, 255:red, 0; green, 0; blue, 0 }  ][line width=0.08]  [draw opacity=0] (8.93,-4.29) -- (0,0) -- (8.93,4.29) -- cycle    ;
\draw    (400.09,108.63) -- (385.5,147) ;
\draw [shift={(385.5,147)}, rotate = 290.82] [color={rgb, 255:red, 0; green, 0; blue, 0 }  ][line width=0.75]    (0,5.59) -- (0,-5.59)   ;
\draw    (384.43,149.8) -- (367.95,192.88) ;
\draw [shift={(385.5,147)}, rotate = 110.93] [fill={rgb, 255:red, 0; green, 0; blue, 0 }  ][line width=0.08]  [draw opacity=0] (8.93,-4.29) -- (0,0) -- (8.93,4.29) -- cycle    ;

\draw    (185.89,109.79) -- (203.15,154.17) ;
\draw [shift={(204.24,156.96)}, rotate = 248.75] [fill={rgb, 255:red, 0; green, 0; blue, 0 }  ][line width=0.08]  [draw opacity=0] (8.93,-4.29) -- (0,0) -- (8.93,4.29) -- cycle    ;
\draw    (203.72,155.53) -- (219.95,192.88) ;
\draw [shift={(203.72,155.53)}, rotate = 246.51] [color={rgb, 255:red, 0; green, 0; blue, 0 }  ][line width=0.75]    (0,5.59) -- (0,-5.59)   ;

\draw    (252.09,108.63) -- (237.5,147) ;
\draw [shift={(237.5,147)}, rotate = 290.82] [color={rgb, 255:red, 0; green, 0; blue, 0 }  ][line width=0.75]    (0,5.59) -- (0,-5.59)   ;
\draw    (236.43,149.8) -- (219.95,192.88) ;
\draw [shift={(237.5,147)}, rotate = 110.93] [fill={rgb, 255:red, 0; green, 0; blue, 0 }  ][line width=0.08]  [draw opacity=0] (8.93,-4.29) -- (0,0) -- (8.93,4.29) -- cycle    ;

\draw    (295.73,192) -- (367.95,192.88) ;
\draw [shift={(295.73,192)}, rotate = 180.7] [color={rgb, 255:red, 0; green, 0; blue, 0 }  ][line width=0.75]    (0,5.59) -- (0,-5.59)   ;
\draw    (219.95,192.88) -- (293.73,192.03) ;
\draw [shift={(296.73,192)}, rotate = 179.34] [fill={rgb, 255:red, 0; green, 0; blue, 0 }  ][line width=0.08]  [draw opacity=0] (8.93,-4.29) -- (0,0) -- (8.93,4.29) -- cycle    ;

\draw (326.23,116.27) node [anchor=north west][inner sep=0.75pt]  [rotate=-251.04]  {$\otimes $};
\draw (360.83,185.6) node [anchor=north west][inner sep=0.75pt]  [font=\large]  {$\bullet $};
\draw (389.48,109.61) node [anchor=north west][inner sep=0.75pt]  [rotate=-293.71]  {$\otimes $};
\draw (212.83,185.6) node [anchor=north west][inner sep=0.75pt]  [font=\large]  {$\bullet $};
\draw (241.48,109.61) node [anchor=north west][inner sep=0.75pt]  [rotate=-293.71]  {$\otimes $};
\draw (178.23,116.27) node [anchor=north west][inner sep=0.75pt]  [rotate=-251.04]  {$\otimes $};

\end{tikzpicture}

    \end{tabular}
    \caption{The exchange diagrams in $\phi^3$ theory arising from the exterior EFT rules.}
    \label{tab:placeholder}
\end{table}

We now follow the same logic as above and compute the simplest two-point loop diagram in $\phi^3$ theory. The grSK loop is of the form
\begin{equation}
    \tikzset{every picture/.style={line width=0.75pt}} 
\begin{tikzpicture}[x=0.75pt,y=0.75pt,yscale=-1,xscale=1, baseline={([yshift=-0.5ex]current bounding box.center)}]

\draw    (138,140) -- (198,140) ;
\draw   (198,140) .. controls (198,126.19) and (209.19,115) .. (223,115) .. controls (236.81,115) and (248,126.19) .. (248,140) .. controls (248,153.81) and (236.81,165) .. (223,165) .. controls (209.19,165) and (198,153.81) .. (198,140) -- cycle ;
\draw    (248,140) -- (308,140) ;

\draw (125,134.4) node [anchor=north west][inner sep=0.75pt]    {$\otimes $};
\draw (303,134.4) node [anchor=north west][inner sep=0.75pt]    {$\otimes $};
\end{tikzpicture} \ .
\end{equation}
As in the tree-level case, we expect this loop to have the exterior field theory descendants given in table \ref{tab:DiagSelfEnergyPhi3}. As we will see in the next section, this expectation is indeed borne out. Furthermore, we will also see that of these diagrams vanish due to SK collapse and KMS conditions.
\begin{table}[H]
    \centering
    \begin{tabular}{cc}
\tikzset{every picture/.style={line width=0.75pt}} 

\begin{tikzpicture}[x=0.75pt,y=0.75pt,yscale=-0.75,xscale=0.75]

\draw    (166.33,195) -- (195.3,195.37) ;
\draw [shift={(198.3,195.41)}, rotate = 180.74] [fill={rgb, 255:red, 0; green, 0; blue, 0 }  ][line width=0.08]  [draw opacity=0] (8.93,-4.29) -- (0,0) -- (8.93,4.29) -- cycle    ;
\draw    (197.3,195.41) -- (227.33,195.67) ;
\draw [shift={(197.3,195.41)}, rotate = 180.49] [color={rgb, 255:red, 0; green, 0; blue, 0 }  ][line width=0.75]    (0,5.59) -- (0,-5.59)   ;
\draw    (286.33,195.55) -- (315.3,195.94) ;
\draw [shift={(318.3,195.98)}, rotate = 180.77] [fill={rgb, 255:red, 0; green, 0; blue, 0 }  ][line width=0.08]  [draw opacity=0] (8.93,-4.29) -- (0,0) -- (8.93,4.29) -- cycle    ;
\draw    (317.3,195.98) -- (347.33,196.25) ;
\draw [shift={(317.3,195.98)}, rotate = 180.51] [color={rgb, 255:red, 0; green, 0; blue, 0 }  ][line width=0.75]    (0,5.59) -- (0,-5.59)   ;
\draw  [draw opacity=0] (258,225.98) .. controls (257.67,225.99) and (257.33,226) .. (257,226) .. controls (240.43,226) and (227,212.57) .. (227,196) .. controls (227,179.43) and (240.43,166) .. (257,166) -- (257,196) -- cycle ; \draw    (257,226) .. controls (240.43,226) and (227,212.57) .. (227,196) .. controls (227,180.43) and (238.87,167.62) .. (254.05,166.14) ; \draw [shift={(257,166)}, rotate = 168.44] [fill={rgb, 255:red, 0; green, 0; blue, 0 }  ][line width=0.08]  [draw opacity=0] (8.93,-4.29) -- (0,0) -- (8.93,4.29) -- cycle    ; \draw [shift={(258,225.98)}, rotate = 189.13] [fill={rgb, 255:red, 0; green, 0; blue, 0 }  ][line width=0.08]  [draw opacity=0] (8.93,-4.29) -- (0,0) -- (8.93,4.29) -- cycle    ;
\draw  [draw opacity=0] (255.88,166.02) .. controls (256.22,166.01) and (256.55,166) .. (256.88,166) .. controls (273.45,165.94) and (286.94,179.32) .. (287,195.88) .. controls (287.06,212.45) and (273.68,225.94) .. (257.12,226) -- (257,196) -- cycle ; \draw    (255.88,166.02) .. controls (256.22,166.01) and (256.55,166) .. (256.88,166) .. controls (273.45,165.94) and (286.94,179.32) .. (287,195.88) .. controls (287.06,212.45) and (273.68,225.94) .. (257.12,226) ; \draw [shift={(257.12,226)}, rotate = 348.22] [color={rgb, 255:red, 0; green, 0; blue, 0 }  ][line width=0.75]    (0,5.59) -- (0,-5.59)   ; \draw [shift={(255.88,166.02)}, rotate = 188.91] [color={rgb, 255:red, 0; green, 0; blue, 0 }  ][line width=0.75]    (0,5.59) -- (0,-5.59)   ;

\draw (152,185.4) node [anchor=north west][inner sep=0.75pt]    {$\otimes $};
\draw (341,186.9) node [anchor=north west][inner sep=0.75pt]    {$\otimes $};

\draw [shift={(287,195.88)}, rotate = 210.82] [color={rgb, 255:red, 0; green, 0; blue, 0 }  ][fill={rgb, 255:red, 0; green, 0; blue, 0 }  ][line width=0.75]      (0, 0) circle [x radius= 2.2, y radius= 2.2]   ;

\draw [shift={(226,195.88)}, rotate = 210.82] [color={rgb, 255:red, 0; green, 0; blue, 0 }  ][fill={rgb, 255:red, 0; green, 0; blue, 0 }  ][line width=0.75]      (0, 0) circle [x radius= 2.2, y radius= 2.2]   ;
\end{tikzpicture}  &  

\tikzset{every picture/.style={line width=0.75pt}} 

\begin{tikzpicture}[x=0.75pt,y=0.75pt,yscale=-0.75,xscale=0.75]

\draw    (166.33,195) -- (195.3,195.37) ;
\draw [shift={(198.3,195.41)}, rotate = 180.74] [fill={rgb, 255:red, 0; green, 0; blue, 0 }  ][line width=0.08]  [draw opacity=0] (8.93,-4.29) -- (0,0) -- (8.93,4.29) -- cycle    ;
\draw    (197.3,195.41) -- (227.33,195.67) ;
\draw [shift={(197.3,195.41)}, rotate = 180.49] [color={rgb, 255:red, 0; green, 0; blue, 0 }  ][line width=0.75]    (0,5.59) -- (0,-5.59)   ;
\draw    (286.33,195.55) -- (315.3,195.94) ;
\draw [shift={(318.3,195.98)}, rotate = 180.77] [fill={rgb, 255:red, 0; green, 0; blue, 0 }  ][line width=0.08]  [draw opacity=0] (8.93,-4.29) -- (0,0) -- (8.93,4.29) -- cycle    ;
\draw    (317.3,195.98) -- (347.33,196.25) ;
\draw [shift={(317.3,195.98)}, rotate = 180.51] [color={rgb, 255:red, 0; green, 0; blue, 0 }  ][line width=0.75]    (0,5.59) -- (0,-5.59)   ;
\draw  [draw opacity=0] (258,225.98) .. controls (257.67,225.99) and (257.33,226) .. (257,226) .. controls (240.43,226) and (227,212.57) .. (227,196) .. controls (227,179.43) and (240.43,166) .. (257,166) -- (257,196) -- cycle ; \draw    (258,225.98) .. controls (257.67,225.99) and (257.33,226) .. (257,226) .. controls (240.43,226) and (227,212.57) .. (227,196) .. controls (227,180.43) and (238.87,167.62) .. (254.05,166.14) ; \draw [shift={(257,166)}, rotate = 168.44] [fill={rgb, 255:red, 0; green, 0; blue, 0 }  ][line width=0.08]  [draw opacity=0] (8.93,-4.29) -- (0,0) -- (8.93,4.29) -- cycle    ; \draw [shift={(258,225.98)}, rotate = 9.13] [color={rgb, 255:red, 0; green, 0; blue, 0 }  ][line width=0.75]    (0,5.59) -- (0,-5.59)   ;
\draw  [draw opacity=0] (255.88,166.02) .. controls (256.22,166.01) and (256.55,166) .. (256.88,166) .. controls (273.45,165.94) and (286.94,179.32) .. (287,195.88) .. controls (287.06,212.45) and (273.68,225.94) .. (257.12,226) -- (257,196) -- cycle ; \draw    (255.88,166.02) .. controls (256.22,166.01) and (256.55,166) .. (256.88,166) .. controls (273.45,165.94) and (286.94,179.32) .. (287,195.88) .. controls (287.06,211.46) and (275.24,224.31) .. (260.06,225.85) ; \draw [shift={(257.12,226)}, rotate = 348.22] [fill={rgb, 255:red, 0; green, 0; blue, 0 }  ][line width=0.08]  [draw opacity=0] (8.93,-4.29) -- (0,0) -- (8.93,4.29) -- cycle    ; \draw [shift={(255.88,166.02)}, rotate = 188.91] [color={rgb, 255:red, 0; green, 0; blue, 0 }  ][line width=0.75]    (0,5.59) -- (0,-5.59)   ;

\draw (152,185.4) node [anchor=north west][inner sep=0.75pt]    {$\otimes $};
\draw (341,186.9) node [anchor=north west][inner sep=0.75pt]    {$\otimes $};

\draw [shift={(287,195.88)}, rotate = 210.82] [color={rgb, 255:red, 0; green, 0; blue, 0 }  ][fill={rgb, 255:red, 0; green, 0; blue, 0 }  ][line width=0.75]      (0, 0) circle [x radius= 2.2, y radius= 2.2]   ;

\draw [shift={(226,195.88)}, rotate = 210.82] [color={rgb, 255:red, 0; green, 0; blue, 0 }  ][fill={rgb, 255:red, 0; green, 0; blue, 0 }  ][line width=0.75]      (0, 0) circle [x radius= 2.2, y radius= 2.2]   ;

\end{tikzpicture}\\

\tikzset{every picture/.style={line width=0.75pt}} 
\begin{tikzpicture}[x=0.75pt,y=0.75pt,yscale=-0.75,xscale=0.75]

\draw    (126.33,208) -- (155.3,208.37) ;
\draw [shift={(158.3,208.41)}, rotate = 180.74] [fill={rgb, 255:red, 0; green, 0; blue, 0 }  ][line width=0.08]  [draw opacity=0] (8.93,-4.29) -- (0,0) -- (8.93,4.29) -- cycle    ;
\draw    (157.3,208.41) -- (187.33,208.67) ;
\draw [shift={(157.3,208.41)}, rotate = 180.49] [color={rgb, 255:red, 0; green, 0; blue, 0 }  ][line width=0.75]    (0,5.59) -- (0,-5.59)   ;
\draw    (246.33,208.55) -- (278.3,208.98) ;
\draw [shift={(278.3,208.98)}, rotate = 180.77] [color={rgb, 255:red, 0; green, 0; blue, 0 }  ][line width=0.75]    (0,5.59) -- (0,-5.59)   ;
\draw    (280.3,209.01) -- (307.33,209.25) ;
\draw [shift={(277.3,208.98)}, rotate = 0.51] [fill={rgb, 255:red, 0; green, 0; blue, 0 }  ][line width=0.08]  [draw opacity=0] (8.93,-4.29) -- (0,0) -- (8.93,4.29) -- cycle    ;
\draw  [draw opacity=0] (218,238.98) .. controls (217.67,238.99) and (217.33,239) .. (217,239) .. controls (200.43,239) and (187,225.57) .. (187,209) .. controls (187,192.43) and (200.43,179) .. (217,179) -- (217,209) -- cycle ; \draw    (218,238.98) .. controls (217.67,238.99) and (217.33,239) .. (217,239) .. controls (200.43,239) and (187,225.57) .. (187,209) .. controls (187,193.43) and (198.87,180.62) .. (214.05,179.14) ; \draw [shift={(217,179)}, rotate = 168.44] [fill={rgb, 255:red, 0; green, 0; blue, 0 }  ][line width=0.08]  [draw opacity=0] (8.93,-4.29) -- (0,0) -- (8.93,4.29) -- cycle    ; \draw [shift={(218,238.98)}, rotate = 9.13] [color={rgb, 255:red, 0; green, 0; blue, 0 }  ][line width=0.75]    (0,5.59) -- (0,-5.59)   ;
\draw  [draw opacity=0] (215.88,179.02) .. controls (216.22,179.01) and (216.55,179) .. (216.88,179) .. controls (233.45,178.94) and (246.94,192.32) .. (247,208.88) .. controls (247.06,225.45) and (233.68,238.94) .. (217.12,239) -- (217,209) -- cycle ; \draw    (215.88,179.02) .. controls (216.22,179.01) and (216.55,179) .. (216.88,179) .. controls (233.45,178.94) and (246.94,192.32) .. (247,208.88) .. controls (247.06,224.46) and (235.24,237.31) .. (220.06,238.85) ; \draw [shift={(217.12,239)}, rotate = 348.22] [fill={rgb, 255:red, 0; green, 0; blue, 0 }  ][line width=0.08]  [draw opacity=0] (8.93,-4.29) -- (0,0) -- (8.93,4.29) -- cycle    ; \draw [shift={(215.88,179.02)}, rotate = 188.91] [color={rgb, 255:red, 0; green, 0; blue, 0 }  ][line width=0.75]    (0,5.59) -- (0,-5.59)   ;

\draw (112,198.4) node [anchor=north west][inner sep=0.75pt]    {$\otimes $};
\draw (301,199.9) node [anchor=north west][inner sep=0.75pt]    {$\otimes $};

\draw [shift={(247,208.88)}, rotate = 210.82] [color={rgb, 255:red, 0; green, 0; blue, 0 }  ][fill={rgb, 255:red, 0; green, 0; blue, 0 }  ][line width=0.75]      (0, 0) circle [x radius= 2.2, y radius= 2.2]   ;

\draw [shift={(186.99,208.88)}, rotate = 210.82] [color={rgb, 255:red, 0; green, 0; blue, 0 }  ][fill={rgb, 255:red, 0; green, 0; blue, 0 }  ][line width=0.75]      (0, 0) circle [x radius= 2.2, y radius= 2.2]   ;

\draw [shift={(247,208.88)}, rotate = 210.82] [color={rgb, 255:red, 0; green, 0; blue, 0 }  ][fill={rgb, 255:red, 0; green, 0; blue, 0 }  ][line width=0.75]      (0, 0) circle [x radius= 2.2, y radius= 2.2]   ;

\draw [shift={(186.99,208.88)}, rotate = 210.82] [color={rgb, 255:red, 0; green, 0; blue, 0 }  ][fill={rgb, 255:red, 0; green, 0; blue, 0 }  ][line width=0.75]      (0, 0) circle [x radius= 2.2, y radius= 2.2]   ;
\end{tikzpicture}
  & 

\tikzset{every picture/.style={line width=0.75pt}} 

\begin{tikzpicture}[x=0.75pt,y=0.75pt,yscale=-0.75,xscale=0.75]

\draw    (126.33,208) -- (158.3,208.41) ;
\draw [shift={(158.3,208.41)}, rotate = 180.74] [color={rgb, 255:red, 0; green, 0; blue, 0 }  ][line width=0.75]    (0,5.59) -- (0,-5.59)   ;
\draw    (160.3,208.44) -- (187.33,208.67) ;
\draw [shift={(157.3,208.41)}, rotate = 0.49] [fill={rgb, 255:red, 0; green, 0; blue, 0 }  ][line width=0.08]  [draw opacity=0] (8.93,-4.29) -- (0,0) -- (8.93,4.29) -- cycle    ;
\draw    (246.33,208.55) -- (275.3,208.94) ;
\draw [shift={(278.3,208.98)}, rotate = 180.77] [fill={rgb, 255:red, 0; green, 0; blue, 0 }  ][line width=0.08]  [draw opacity=0] (8.93,-4.29) -- (0,0) -- (8.93,4.29) -- cycle    ;
\draw    (277.3,208.98) -- (307.33,209.25) ;
\draw [shift={(277.3,208.98)}, rotate = 180.51] [color={rgb, 255:red, 0; green, 0; blue, 0 }  ][line width=0.75]    (0,5.59) -- (0,-5.59)   ;
\draw  [draw opacity=0] (218,238.98) .. controls (217.67,238.99) and (217.33,239) .. (217,239) .. controls (200.43,239) and (187,225.57) .. (187,209) .. controls (187,192.43) and (200.43,179) .. (217,179) -- (217,209) -- cycle ; \draw    (218,238.98) .. controls (217.67,238.99) and (217.33,239) .. (217,239) .. controls (200.43,239) and (187,225.57) .. (187,209) .. controls (187,193.43) and (198.87,180.62) .. (214.05,179.14) ; \draw [shift={(217,179)}, rotate = 168.44] [fill={rgb, 255:red, 0; green, 0; blue, 0 }  ][line width=0.08]  [draw opacity=0] (8.93,-4.29) -- (0,0) -- (8.93,4.29) -- cycle    ; \draw [shift={(218,238.98)}, rotate = 9.13] [color={rgb, 255:red, 0; green, 0; blue, 0 }  ][line width=0.75]    (0,5.59) -- (0,-5.59)   ;
\draw  [draw opacity=0] (215.88,179.02) .. controls (216.22,179.01) and (216.55,179) .. (216.88,179) .. controls (233.45,178.94) and (246.94,192.32) .. (247,208.88) .. controls (247.06,225.45) and (233.68,238.94) .. (217.12,239) -- (217,209) -- cycle ; \draw    (215.88,179.02) .. controls (216.22,179.01) and (216.55,179) .. (216.88,179) .. controls (233.45,178.94) and (246.94,192.32) .. (247,208.88) .. controls (247.06,224.46) and (235.24,237.31) .. (220.06,238.85) ; \draw [shift={(217.12,239)}, rotate = 348.22] [fill={rgb, 255:red, 0; green, 0; blue, 0 }  ][line width=0.08]  [draw opacity=0] (8.93,-4.29) -- (0,0) -- (8.93,4.29) -- cycle    ; \draw [shift={(215.88,179.02)}, rotate = 188.91] [color={rgb, 255:red, 0; green, 0; blue, 0 }  ][line width=0.75]    (0,5.59) -- (0,-5.59)   ;

\draw (112,198.4) node [anchor=north west][inner sep=0.75pt]    {$\otimes $};
\draw (301,199.9) node [anchor=north west][inner sep=0.75pt]    {$\otimes $};

\draw [shift={(247,208.88)}, rotate = 210.82] [color={rgb, 255:red, 0; green, 0; blue, 0 }  ][fill={rgb, 255:red, 0; green, 0; blue, 0 }  ][line width=0.75]      (0, 0) circle [x radius= 2.2, y radius= 2.2]   ;

\draw [shift={(186.99,208.88)}, rotate = 210.82] [color={rgb, 255:red, 0; green, 0; blue, 0 }  ][fill={rgb, 255:red, 0; green, 0; blue, 0 }  ][line width=0.75]      (0, 0) circle [x radius= 2.2, y radius= 2.2]   ;

\end{tikzpicture}
    \end{tabular}
    \caption{The diagrams arising from the EFT SDEs with two boundary sources at one-loop order.}
    \label{tab:DiagSelfEnergyPhi3}
\end{table}

\section{Equivalence of grSK QFT and exterior QFT}\label{sec:Equivalence}

We now have two distinct proposals to calculate the generating functional of boundary correlators. In this section, we intend to show that the two proposals are identical perturbatively. To verify this claim, we will now use the perturbative expansion of both the SDEs. At tree-level, we already know this by definition, since the exterior EFT was derived from grSK for tree-level. Thus, we will start the check with one-loop computations.

\begin{figure}[H]
    \centering
    \tikzset{every picture/.style={line width=0.75pt}} 
\begin{tikzpicture}[x=0.75pt,y=0.75pt,yscale=-1,xscale=1]

\draw    (138,140) -- (198,140) ;
\draw   (198,140) .. controls (198,126.19) and (209.19,115) .. (223,115) .. controls (236.81,115) and (248,126.19) .. (248,140) .. controls (248,153.81) and (236.81,165) .. (223,165) .. controls (209.19,165) and (198,153.81) .. (198,140) -- cycle ;
\draw    (248,140) -- (308,140) ;

\draw (125,134.4) node [anchor=north west][inner sep=0.75pt]    {$\otimes $};
\draw (303,134.4) node [anchor=north west][inner sep=0.75pt]    {$\otimes $};
\end{tikzpicture}
    \caption{One-loop self-energy diagram contributing to the two-point correlator.}
    \label{fig:OneLoopMassRen}
\end{figure}
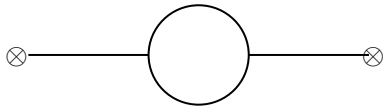

We start with the one-loop mass renormalisation in $\phi^3$ theory. Starting with the QFT defined on the grSK spacetime, the only diagram that contributes is given in Fig.~(\ref{fig:OneLoopMassRen}). Evaluating this diagram with the Feynman rules given in Sec.~(\ref{sec:ClassicalFieldTheorygrSK}), we get
\begin{equation}
    \begin{split}
        S_{(2)}^{1\rm-loop} &= -\frac{i}{2!} \left(-i \lambda_{3 \rm B}\right)^2\int_{k_{1,2}}\oint_{\zeta_1,\zeta_2} \phi_{(0)} (\zeta_1,k_1) \phi_{(0)} (\zeta_2,k_2)\\
        &\hspace{4cm}\times\int_p \left[-i \bbG(\zeta_2|\zeta_1, p)\right] \left[-i \bbG(\zeta_2|\zeta_1, k_1-p)\right]\\
        &= -i \frac{\lambda_{3 \rm B}^2}{2!} \int_{k_{1,2}}\oint_{\zeta_1,\zeta_2} \phi_{(0)} (\zeta_1,k_1) \phi_{(0)} (\zeta_2,k_2) \int_p \bbG(\zeta_2|\zeta_1, p)  \bbG(\zeta_2|\zeta_1, k_1-p) \ .
    \end{split}
\end{equation}
Here, we include a factor of $\frac{1}{2!}$ because of the $\mathbb{Z}_2$ symmetry of the diagram. We can now simplify this expression greatly. As we learnt in Eq.~\eqref{eq:LeadingOrderSolutionPF}, the leading order solution $\phi_{(0)}$ can be written in terms of the ingoing and the outgoing boundary-to-bulk propagators, and the appropriate boundary sources.

We can now use the explicit form of the leading-order solution $\phi_{(0)}(\zeta,k)$ given in Eq.~\eqref{eq:LeadingOrderSolutionPF} to read off the terms with different products of boundary sources. This gives us
\begin{equation}
    \begin{split}
        \mathcal{I}^{\rm 1loop}_{2,0} &= -i \frac{\lambda_{3 \rm B}^2}{2!} \oint_{\zeta_1,\zeta_2} \Gin (\zeta_1,k_1) \Gin(\zeta_2,k_2) \int_p \bbG(\zeta_2|\zeta_1, p)  \bbG(\zeta_2|\zeta_1, k_1-p)\ ,\\
        \mathcal{I}^{\rm 1loop}_{1,1} &= i \frac{\lambda_{3 \rm B}^2}{2!} \oint_{\zeta_1,\zeta_2} \Gin (\zeta_1,k_1) e^{\beta k_2^0} \Gout(\zeta_2,k_2) \int_p \bbG(\zeta_2|\zeta_1, p)  \bbG(\zeta_2|\zeta_1, k_1-p)\ ,\\
        \mathcal{I}^{\rm 1loop}_{0,2} &= -i \frac{\lambda_{3 \rm B}^2}{2!} \oint_{\zeta_1,\zeta_2} e^{\beta k_1^0} \Gout (\zeta_1,k_1) e^{\beta k_2^0} \Gout(\zeta_2,k_2) \int_p \bbG(\zeta_2|\zeta_1, p)  \bbG(\zeta_2|\zeta_1, k_1-p)\ .
    \end{split}
\end{equation}
This leaves us with the monodromy integrals over the grSK contour. As the reader can already see, this allows us to recast the grSK answer (which is a single term) into many terms that are all evaluated on a single copy of the exterior. Let us see how this is done. The important point to realise here is that the integrand in all of the above equations is of the form
\begin{equation}
    \begin{split}
        &\oint_{\zeta_1} \oint_{\zeta_2}e^{\beta \Cnst_1 (1-\zeta_1)}e^{\beta \Cnst_2 (1-\zeta_2)}\ \bbG(\zeta_2|\zeta_1, p_1)\bbG(\zeta_2|\zeta_1, p_2) \mathscr{F}(\zeta_1, \zeta_2)\ ,
    \end{split}
\end{equation}
where $ \mathscr{F}(\zeta_1, \zeta_2)$ is an analytic function of $\zeta_1$ and $\zeta_2$. All the various factors of the ingoing propagators and their complex conjugates are sitting inside this function. Since all the propagators are free of poles at the horizon, we can reduce the above monodromy integral to an integral over just one branch of the contour:
\begin{equation}
    \begin{split}
        &\oint_{\zeta_1} \oint_{\zeta_2}e^{\beta \Cnst_1 (1-\zeta_1)}e^{\beta \Cnst_2 (1-\zeta_2)}\ \bbG(\zeta_2|\zeta_1, p_1)\bbG(\zeta_2|\zeta_1, p_2) \mathscr{F}(\zeta_1, \zeta_2)\\
        &\hspace{2cm}= \int_{\rm Ext_1} \int_{\rm Ext_2}e^{\beta \Cnst_1 (1-\zeta_1)}e^{\beta \Cnst_2 (1-\zeta_2)}\ \bbG_{\rm DD}(\zeta_2|\zeta_1, p_1, p_2) \mathscr{F}(\zeta_1, \zeta_2)\ ,
    \end{split}
\end{equation}
where $\bbG_{\rm DD}(\zeta_2|\zeta_1, p_1)$ is the function that encapsulates the double discontinuity, and is given by
\begin{equation}
\begin{split}
&\bbG_{\rm DD}(\z_2,\z_1|,p_1,p_2) \equiv  \bbG(\z_2|\z_1,p_1) \bbG(\z_2|\z_1,p_2)-e^{-\beta \kappa_1}\bbG(\z_2|\z_1+1,p_1) \bbG(\z_2|\z_1+1,p_2)\\
&\hspace{3.7cm}-e^{-\beta \kappa_2}\bbG(\z_2+1|\z_1,p_1) \bbG(\z_2+1|\z_1,p_2)\\
&\hspace{3.7cm}+e^{-\beta(\kappa_1 +\kappa_2)}\bbG(\z_2+1|\z_1+1,p_1) \bbG(\z_2+1|\z_1+1,p_2) \ .
\end{split}
\end{equation}
We now simply have to evaluate this double discontinuity function. We leave the details to Appendix \ref{app:MonInt}. In this appendix, we provide a conjecture for doing such monodromy integrals for an arbitrary loop diagram.

Using the result in Eq.~\eqref{eq:DoubleDisc}, we can now perform all the monodromy integrals. For ease of discussion, we focus on the term with one past and one future source (the reader can check that the two other terms vanish):
\begin{equation}
    \begin{split}
        \mathcal{I}^{\rm 1loop}_{1,1}&= i \frac{\lambda_{3 \rm B}^2}{2!} \int_{\rm Ext_{1,2}} \Gin (\zeta_1,k_1) \Gout(\zeta_2,k_2)\int_p \Bigg[ \frac{\nbe_p (1+\nbe_{p-k_1})}{\nbe_{-k_1} \nbe_{k_1}}\bbGR(\zeta_2|\zeta_1, p)  \bbGR(\zeta_2|\zeta_1, k_1-p)\\
        &+ \frac{\nbe_{k_1-p}}{\nbe_{-p}} \bbGR(\zeta_2|\zeta_1, p)  \bbGA(\zeta_2|\zeta_1, k_1-p) + \frac{\nbe_{p}}{\nbe_{-k_1+p}}\bbGA(\zeta_2|\zeta_1, p)  \bbGR(\zeta_2|\zeta_1, k_1-p) \Bigg]\\
        &= i \ \lambda_{3 \rm B}^2 \int_{\rm Ext_{1,2}} \Gin (\zeta_1,k_1) \Gout(\zeta_2,k_2)\bbGR(\zeta_2|\zeta_1, p)\\
        &\hspace{2cm} \times \int_p \Bigg[\frac{1}{2!} \frac{\nbe_p \nbe_{k_1-p}}{(1+\nbe_{k_1)} \nbe_{k_1}} \bbGR(\zeta_2|\zeta_1, k_1-p)+ \frac{\nbe_{k_1-p}}{\nbe_{-p}}   \bbGA(\zeta_2|\zeta_1, k_1-p) \Bigg]\ .
    \end{split}
    \label{eq:I11grSK}
\end{equation}

We will now repeat the computation of the one-loop self-energy. This time, we will use the exterior QFT rules instead, i.e., we will use the diagrams and the rules that arise out of the perturbative expansion of the exterior QFT SDEs. The diagrams for all the terms in the mass renormalisation at one-loop are given in table \ref{tab:DiagSelfEnergyPhi3}.

Once again, we will focus on the terms with one past source and one future source. The two diagrams that contribute are the ones in the first row of the table. Evaluating these with the rules in Sec.~(\ref{sec:EFTwithFeynmanRules}), we obtain
\begin{equation}
    \begin{split}
        \mathcal{I}^{\rm 1loop}_{1,1} &= -i \lambda_{3 \rm B}^2 \int_{\rm Ext_{1,2}} \frac{1}{1+\nbe_{k_1}}\Gin (\zeta_1,k_1) \Gout(\zeta_2,k_2)\bbGR(\zeta_2|\zeta_1, p)\\
        &\qquad\qquad\times \int_p \left[\frac{1}{2} \frac{ \nbe_{-p}\nbe_{-k_1+p}}{\nbe_{k_2}}   \bbGR(\zeta_2|\zeta_1, k_1-p) + \frac{\nbe_{-k_1}\nbe_{k_1-p}}{\nbe_{-p}} \bbGA(\zeta_2|\zeta_1, k_1-p) \right] \ .
    \end{split}
\end{equation}
This expression can be gleaned to be the same as the one in Eq.~\eqref{eq:I11grSK} by using the identity
\begin{equation}
    \frac{\nbe_p \nbe_{k_1-p}}{\nbe_{k_1}} = - \frac{\nbe_{-p} \nbe_{-k_1+p}}{\nbe_{-k_1}} \ .
\end{equation}

We have checked that the grSK SDEs and the exterior QFT SDEs produce the same answer with all the correct signs and symmetry factors for many diagrams until four loops. Among the many diagrams are the $\phi^3$ triangle diagram for the three-point function, the two-loop sunset diagram in $\phi^3$, the two-loop non-planar vertex correction in $\phi^3$, the three- and four-loop melon diagrams for two-point functions in $\phi^5$ and $\phi^6$ theories, respectively.

\section{Microscopic Unitarity and Thermality}\label{sec:MicUniTherm}

We have seen that the QFT defined on the grSK contour, as well as the QFT defined on the black hole exterior, both give the same answers. Our next goal is to check whether these answers are consistent with microscopic unitarity and thermality of the boundary theory. In other words, we would like to ask if the boundary correlators satisfy the SK collapse and the KMS conditions. Here, we choose to perform these checks on the answers coming from the exterior QFT. There is not so much of a choice since even in the case of the grSK results, we would have to perform the monodromy integrals to check SK collapse and KMS conditions, therefore reducing it to the exterior field theory results. 

It is worth emphasising, however, that these two perspectives explain unitarity and thermality through fundamentally different mechanisms, even at the tree level \cite{Loganayagam:2024mnj, Sharma:2025hbk}. In the grSK picture, these consistency checks follow from the analyticity of the propagators in the radial integrals. In contrast, in the exterior field theory, they follow from the causality of propagators and the structure of the interaction vertices --- specifically, from the absence of vertices surrounded by either all semi-capacitors or all semi-diodes.

We now move towards performing these consistency checks perturbatively at the loop level. Recall that, in the past-future basis, the SK collapse and KMS conditions correspond to the statement that all correlators with purely future sources and purely past sources vanish, respectively. But from the exterior QFT perspective, it may seem that loop corrections could generate diagrams contributing to such correlators. The key point of this section is to show that, under certain assumptions, these loop diagrams vanish trivially. The main statement we will use is the fact that the past–future basis is a causal basis. This suggests that any diagram in which all \emph{causal arrows} (i.e. the directions associated with semi-diode or semi-capacitor lines, not momenta) point entirely inward or entirely outward necessarily vanishes. Provided this is granted, we can give a general argument for SK collapse and KMS.

Once again, for illustration, we will use the example of the one-loop self-energy diagram in $\phi^3$ theory, but our arguments will mostly generalise. We will demonstrate this fact by drawing some more diagrams at the end of this section. 

Consider for now the purely future term, i.e., the first diagram in the second row of table \ref{tab:DiagSelfEnergyPhi3}:
\begin{equation}
\begin{split}
\mathcal{I}_{2,0}^{\rm 1 loop}
&= -i \lambda_{3\rm B}^2 \int_{\rm Ext_{1,2}}  \Gin(\z_1,k_1) \Gin(\z_2,k_2) \left[\int_{p}  \bbGR(\z_2|\z_1,p) \bbGR(\z_1|\z_2, p-k_1)\right]
\end{split}
\end{equation}
To satisfy the SK collapse rule, this term has to now vanish. The main point to notice here is that $\bbGR(\z_2|\z_1,p)$  is analytic in the upper half plane (UHP) of $p^0$. Since we have a product of these retarded bulk-to-bulk propagators, we can say the following:
\begin{equation}
\begin{split}
&\bbGR(\z_2|\z_1,p) \text{ is analytic when } \text{Im}[p^0] \ > \ 0 \ ,\\ 
&\bbGR(\z_1|\z_2, p-k_1) \text{ is analytic when } \text{Im}[p^0] \ > \ \text{Im}[k^0_1] \ .
\end{split}
\end{equation}

We will now assume that the exterior momenta $k_1$ and $k_2$ are both real, i.e., $\text{Im}(k^0)=0$. In this case, both propagators appearing in the product above are analytic in the upper half-plane (UHP) of $p^0$. We can therefore choose to close the contour of integration in the UHP of $p^0$ using a semicircular arc at infinity. The original integration over real loop momentum can then be reinterpreted as an integral over the real axis closed in the UHP. Since the integrand is analytic there, the closed contour integral vanishes.

Consequently, the diagram under consideration also vanishes,
\begin{equation}
    \mathcal{I}_{2,0}^{\rm 1 loop} = \ 0 \ ,
\end{equation}
and similarly the diagram with purely past sources also vanishes, i.e., $\mathcal{I}_{0,2}^{\rm 1 loop} = 0 $. Here we have crucially assumed that the falloff of $\bbGR$ at large $|p^0|$ is sufficiently rapid so that the contribution from the semicircular arc vanishes.

As the reader might have guessed by now, this argument is very general: it applies whenever the loop integral contains more than one competing retarded bulk–to–bulk propagator. At the end, the lesson can be summarised as follows:
\begin{center}
    \emph{The arrows in the past–future diagrammatics should be regarded as causal arrows, and any diagram without a flow of `causal time' vanishes.}
\end{center} 
As usual, causality is manifesting here in the form of analyticity in the frequency.

Finally, even if one draws diagrams (following the Feynman rules of exterior field theory) for purely future or purely past correlators, the result always contains a loop of arrows pointing in the same causal direction. By the contour argument above, such loops necessarily vanish, 
\begin{equation}
    \int_p \bbGR(\z_2|\z_1,p) \bbGR(\z_3|\z_2,k_1+p) \ldots \bbGR(\z_n|\z_{n-1},k_{n-2}+p) \bbGR(\z_1|\z_n,k_{n-1}+p) = 0 \ ,
\end{equation}
Here, the $k_i$s are all sums of various external momenta, which are all real. Diagrammatically, this means that
\begin{equation}
\tikzset{every picture/.style={line width=0.75pt}} 
\begin{tikzpicture}[x=0.75pt,y=0.75pt,yscale=-1,xscale=1]

\draw    (416.99,128.59) -- (435,104.62) ;
\draw [shift={(436.8,102.22)}, rotate = 126.9] [fill={rgb, 255:red, 0; green, 0; blue, 0 }  ][line width=0.08]  [draw opacity=0] (8.93,-4.29) -- (0,0) -- (8.93,4.29) -- cycle    ;
\draw    (436.18,103.01) -- (451.33,81.51) ;
\draw [shift={(436.18,103.01)}, rotate = 125.17] [color={rgb, 255:red, 0; green, 0; blue, 0 }  ][line width=0.75]    (0,5.59) -- (0,-5.59)   ;
\draw    (451.33,81.51) -- (480.31,81.4) ;
\draw [shift={(483.31,81.39)}, rotate = 179.78] [fill={rgb, 255:red, 0; green, 0; blue, 0 }  ][line width=0.08]  [draw opacity=0] (8.93,-4.29) -- (0,0) -- (8.93,4.29) -- cycle    ;
\draw    (482.31,81.4) -- (512.33,81.16) ;
\draw [shift={(482.31,81.4)}, rotate = 179.53] [color={rgb, 255:red, 0; green, 0; blue, 0 }  ][line width=0.75]    (0,5.59) -- (0,-5.59)   ;

\draw    (512.33,81.16) -- (531.77,102.24) ;
\draw [shift={(533.8,104.44)}, rotate = 227.33] [fill={rgb, 255:red, 0; green, 0; blue, 0 }  ][line width=0.08]  [draw opacity=0] (8.93,-4.29) -- (0,0) -- (8.93,4.29) -- cycle    ;
\draw    (533.11,103.73) -- (554.7,126.61) ;
\draw [shift={(533.11,103.73)}, rotate = 226.66] [color={rgb, 255:red, 0; green, 0; blue, 0 }  ][line width=0.75]    (0,5.59) -- (0,-5.59)   ;
\draw    (554.7,126.61) -- (537.68,150.06) ;
\draw [shift={(535.92,152.49)}, rotate = 305.96] [fill={rgb, 255:red, 0; green, 0; blue, 0 }  ][line width=0.08]  [draw opacity=0] (8.93,-4.29) -- (0,0) -- (8.93,4.29) -- cycle    ;
\draw    (536.5,151.67) -- (518.97,176.06) ;
\draw [shift={(536.5,151.67)}, rotate = 305.71] [color={rgb, 255:red, 0; green, 0; blue, 0 }  ][line width=0.75]    (0,5.59) -- (0,-5.59)   ;

\draw    (452.67,178.07) -- (435.77,154.54) ;
\draw [shift={(434.02,152.1)}, rotate = 54.32] [fill={rgb, 255:red, 0; green, 0; blue, 0 }  ][line width=0.08]  [draw opacity=0] (8.93,-4.29) -- (0,0) -- (8.93,4.29) -- cycle    ;
\draw    (434.62,152.91) -- (416.99,128.59) ;
\draw [shift={(434.62,152.91)}, rotate = 54.07] [color={rgb, 255:red, 0; green, 0; blue, 0 }  ][line width=0.75]    (0,5.59) -- (0,-5.59)   ;

\draw (442,57.4) node [anchor=north west][inner sep=0.75pt]    {$\zeta _{1}$};
\draw (506,58.4) node [anchor=north west][inner sep=0.75pt]    {$\zeta _{2}$};
\draw (556,115.4) node [anchor=north west][inner sep=0.75pt]    {$\zeta _{3}$};
\draw (520.33,179.83) node [anchor=north west][inner sep=0.75pt]    {$\zeta _{4}$};
\draw (398,115.4) node [anchor=north west][inner sep=0.75pt]    {$\zeta _{n}$};
\draw (432,179.4) node [anchor=north west][inner sep=0.75pt]    {$\zeta _{n-1}$};
\draw (469,163.4) node [anchor=north west][inner sep=0.75pt]  [font=\large]  {$\dotsc $};
\draw (606,119.4) node [anchor=north west][inner sep=0.75pt]    {$= \ \ \ 0 \ .$};
\draw [shift={(451.33,81.51)}, rotate = 210.82] [color={rgb, 255:red, 0; green, 0; blue, 0 }  ][fill={rgb, 255:red, 0; green, 0; blue, 0 }  ][line width=0.75]      (0, 0) circle [x radius= 2.2, y radius= 2.2]   ;

\draw [shift={(416.99,128.59)}, rotate = 210.82] [color={rgb, 255:red, 0; green, 0; blue, 0 }  ][fill={rgb, 255:red, 0; green, 0; blue, 0 }  ][line width=0.75]      (0, 0) circle [x radius= 2.2, y radius= 2.2]   ;

\draw [shift={(554.7,126.61)}, rotate = 210.82] [color={rgb, 255:red, 0; green, 0; blue, 0 }  ][fill={rgb, 255:red, 0; green, 0; blue, 0 }  ][line width=0.75]      (0, 0) circle [x radius= 2.2, y radius= 2.2]   ;

\draw [shift={(512.33,81.16)}, rotate = 210.82] [color={rgb, 255:red, 0; green, 0; blue, 0 }  ][fill={rgb, 255:red, 0; green, 0; blue, 0 }  ][line width=0.75]      (0, 0) circle [x radius= 2.2, y radius= 2.2]   ;
\end{tikzpicture}
\end{equation}

It is easy to convince oneself that at one loop, any diagram contributing to purely future or purely past correlators must have a \emph{causal loop} of the type described above. To see this, let us take the example of a pentagon diagram. Consider the case where all external arrows are incoming, i.e., the diagrams with all future sources. We now have to show that the only diagrams possible are ones with a directed cycle of the diode arrows (or ones with a causal loop). There should be no other diagram possible. This, of course, means that the two outgoing semi-diode vertices cannot be used. Let us see why this is the case by contradiction. Let us assume that the two outgoing semi–diode vertices are used at a point $\zeta_1$, as shown below:
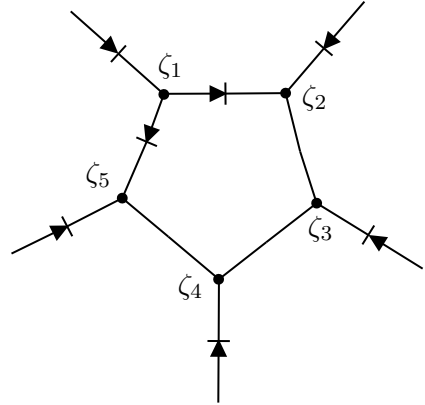
\begin{equation}
\tikzset{every picture/.style={line width=0.75pt}} 
\begin{tikzpicture}[x=0.75pt,y=0.75pt,yscale=-1,xscale=1]

\draw    (450.79,153.98) -- (462.96,125.63) ;
\draw [shift={(462.96,125.63)}, rotate = 113.23] [color={rgb, 255:red, 0; green, 0; blue, 0 }  ][line width=0.75]    (0,5.59) -- (0,-5.59)   ;
\draw    (463.6,123.73) -- (471.54,101.82) ;
\draw [shift={(462.58,126.55)}, rotate = 289.92] [fill={rgb, 255:red, 0; green, 0; blue, 0 }  ][line width=0.08]  [draw opacity=0] (8.93,-4.29) -- (0,0) -- (8.93,4.29) -- cycle    ;
\draw    (471.33,101.51) -- (500.31,101.4) ;
\draw [shift={(503.31,101.39)}, rotate = 179.78] [fill={rgb, 255:red, 0; green, 0; blue, 0 }  ][line width=0.08]  [draw opacity=0] (8.93,-4.29) -- (0,0) -- (8.93,4.29) -- cycle    ;
\draw    (502.31,101.4) -- (532.33,101.16) ;
\draw [shift={(502.31,101.4)}, rotate = 179.53] [color={rgb, 255:red, 0; green, 0; blue, 0 }  ][line width=0.75]    (0,5.59) -- (0,-5.59)   ;

\draw    (532.33,101.16) -- (539.57,130.51) ;
\draw    (539.57,130.51) -- (547.83,156.51) ;
\draw    (547.83,156.51) -- (522.64,176.19) ;
\draw    (522.64,176.19) -- (499.05,194.79) ;
\draw    (499.05,194.79) -- (473.75,173.33) ;
\draw    (473.75,173.33) -- (450.79,153.98) ;
\draw    (600.73,189.45) -- (576.16,174.09) ;
\draw [shift={(573.62,172.5)}, rotate = 32.01] [fill={rgb, 255:red, 0; green, 0; blue, 0 }  ][line width=0.08]  [draw opacity=0] (8.93,-4.29) -- (0,0) -- (8.93,4.29) -- cycle    ;
\draw    (573.62,172.5) -- (547.83,156.51) ;
\draw [shift={(573.62,172.5)}, rotate = 31.81] [color={rgb, 255:red, 0; green, 0; blue, 0 }  ][line width=0.75]    (0,5.59) -- (0,-5.59)   ;
\draw    (498.73,256.83) -- (498.88,227.86) ;
\draw [shift={(498.9,224.86)}, rotate = 90.3] [fill={rgb, 255:red, 0; green, 0; blue, 0 }  ][line width=0.08]  [draw opacity=0] (8.93,-4.29) -- (0,0) -- (8.93,4.29) -- cycle    ;
\draw    (498.91,225.86) -- (499.05,194.79) ;
\draw [shift={(498.91,225.86)}, rotate = 90.27] [color={rgb, 255:red, 0; green, 0; blue, 0 }  ][line width=0.75]    (0,5.59) -- (0,-5.59)   ;
\draw    (395.5,181.87) -- (421.49,169.06) ;
\draw [shift={(424.18,167.74)}, rotate = 153.76] [fill={rgb, 255:red, 0; green, 0; blue, 0 }  ][line width=0.08]  [draw opacity=0] (8.93,-4.29) -- (0,0) -- (8.93,4.29) -- cycle    ;
\draw    (423.29,168.19) -- (450.79,153.98) ;
\draw [shift={(423.29,168.19)}, rotate = 152.67] [color={rgb, 255:red, 0; green, 0; blue, 0 }  ][line width=0.75]    (0,5.59) -- (0,-5.59)   ;
\draw    (571.65,55.15) -- (552.78,77.14) ;
\draw [shift={(550.83,79.41)}, rotate = 310.63] [fill={rgb, 255:red, 0; green, 0; blue, 0 }  ][line width=0.08]  [draw opacity=0] (8.93,-4.29) -- (0,0) -- (8.93,4.29) -- cycle    ;
\draw    (551.47,78.64) -- (532.02,101.52) ;
\draw [shift={(551.47,78.64)}, rotate = 310.38] [color={rgb, 255:red, 0; green, 0; blue, 0 }  ][line width=0.75]    (0,5.59) -- (0,-5.59)   ;

\draw    (425.07,60.04) -- (446.66,79.36) ;
\draw [shift={(448.89,81.36)}, rotate = 221.83] [fill={rgb, 255:red, 0; green, 0; blue, 0 }  ][line width=0.08]  [draw opacity=0] (8.93,-4.29) -- (0,0) -- (8.93,4.29) -- cycle    ;
\draw    (448.89,81.36) -- (471.33,101.51) ;
\draw [shift={(448.89,81.36)}, rotate = 221.92] [color={rgb, 255:red, 0; green, 0; blue, 0 }  ][line width=0.75]    (0,5.59) -- (0,-5.59)   ;

\draw (467,79.4) node [anchor=north west][inner sep=0.75pt]    {$\zeta _{1}$};
\draw (539,95.4) node [anchor=north west][inner sep=0.75pt]    {$\zeta _{2}$};
\draw (542,161.4) node [anchor=north west][inner sep=0.75pt]    {$\zeta _{3}$};
\draw (431,134.4) node [anchor=north west][inner sep=0.75pt]    {$\zeta _{5}$};
\draw (477,190.4) node [anchor=north west][inner sep=0.75pt]    {$\zeta _{4}$};

\draw [shift={(532.33,101.16)}, rotate = 210.82] [color={rgb, 255:red, 0; green, 0; blue, 0 }  ][fill={rgb, 255:red, 0; green, 0; blue, 0 }  ][line width=0.75]      (0, 0) circle [x radius= 2.2, y radius= 2.2]   ;

\draw [shift={(471.54,101.82)}, rotate = 210.82] [color={rgb, 255:red, 0; green, 0; blue, 0 }  ][fill={rgb, 255:red, 0; green, 0; blue, 0 }  ][line width=0.75]      (0, 0) circle [x radius= 2.2, y radius= 2.2]   ;

\draw [shift={(450.79,153.98)}, rotate = 210.82] [color={rgb, 255:red, 0; green, 0; blue, 0 }  ][fill={rgb, 255:red, 0; green, 0; blue, 0 }  ][line width=0.75]      (0, 0) circle [x radius= 2.2, y radius= 2.2]   ;

\draw [shift={(547.83,156.51)}, rotate = 210.82] [color={rgb, 255:red, 0; green, 0; blue, 0 }  ][fill={rgb, 255:red, 0; green, 0; blue, 0 }  ][line width=0.75]      (0, 0) circle [x radius= 2.2, y radius= 2.2]   ;

\draw [shift={(498.88,194.79)}, rotate = 210.82] [color={rgb, 255:red, 0; green, 0; blue, 0 }  ][fill={rgb, 255:red, 0; green, 0; blue, 0 }  ][line width=0.75]      (0, 0) circle [x radius= 2.2, y radius= 2.2]   ;

\end{tikzpicture} \ .
\end{equation}
At each vertex, there are two possible choices of arrows: one with two semi-diodes and one semi-capacitor, and the other with two sem-capacitors and one semi-diode. This is because vertices with all legs of the same type (either all semi-diodes or all semi-capacitors) vanish. Then, once the choice at $\zeta_1$ is fixed --- here taken to be the two–outgoing semi–diode vertex — the configuration at $\zeta_2$ is forced since two of its legs are already occupied by semi–capacitors. The same argument will then fix the choice at $\zeta_3$ and so on, until we reach $\zeta_5$, whence we will get a diagram like
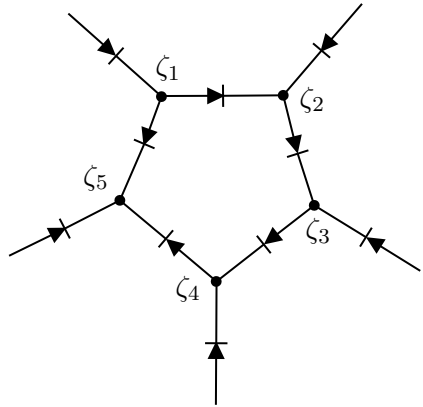
\begin{equation}
\tikzset{every picture/.style={line width=0.75pt}} 
\begin{tikzpicture}[x=0.75pt,y=0.75pt,yscale=-1,xscale=1]

\draw    (450.79,153.98) -- (462.96,125.63) ;
\draw [shift={(462.96,125.63)}, rotate = 113.23] [color={rgb, 255:red, 0; green, 0; blue, 0 }  ][line width=0.75]    (0,5.59) -- (0,-5.59)   ;
\draw    (463.6,123.73) -- (471.54,101.82) ;
\draw [shift={(462.58,126.55)}, rotate = 289.92] [fill={rgb, 255:red, 0; green, 0; blue, 0 }  ][line width=0.08]  [draw opacity=0] (8.93,-4.29) -- (0,0) -- (8.93,4.29) -- cycle    ;
\draw    (471.33,101.51) -- (500.31,101.4) ;
\draw [shift={(503.31,101.39)}, rotate = 179.78] [fill={rgb, 255:red, 0; green, 0; blue, 0 }  ][line width=0.08]  [draw opacity=0] (8.93,-4.29) -- (0,0) -- (8.93,4.29) -- cycle    ;
\draw    (502.31,101.4) -- (532.33,101.16) ;
\draw [shift={(502.31,101.4)}, rotate = 179.53] [color={rgb, 255:red, 0; green, 0; blue, 0 }  ][line width=0.75]    (0,5.59) -- (0,-5.59)   ;

\draw    (532.33,101.16) -- (538.85,127.6) ;
\draw [shift={(539.57,130.51)}, rotate = 256.16] [fill={rgb, 255:red, 0; green, 0; blue, 0 }  ][line width=0.08]  [draw opacity=0] (8.93,-4.29) -- (0,0) -- (8.93,4.29) -- cycle    ;
\draw    (539.57,130.51) -- (547.83,156.51) ;
\draw [shift={(539.57,130.51)}, rotate = 252.36] [color={rgb, 255:red, 0; green, 0; blue, 0 }  ][line width=0.75]    (0,5.59) -- (0,-5.59)   ;
\draw    (547.83,156.51) -- (525,174.35) ;
\draw [shift={(522.64,176.19)}, rotate = 322] [fill={rgb, 255:red, 0; green, 0; blue, 0 }  ][line width=0.08]  [draw opacity=0] (8.93,-4.29) -- (0,0) -- (8.93,4.29) -- cycle    ;
\draw    (522.64,176.19) -- (499.05,194.79) ;
\draw [shift={(522.64,176.19)}, rotate = 321.75] [color={rgb, 255:red, 0; green, 0; blue, 0 }  ][line width=0.75]    (0,5.59) -- (0,-5.59)   ;
\draw    (499.05,194.79) -- (476.04,175.27) ;
\draw [shift={(473.75,173.33)}, rotate = 40.29] [fill={rgb, 255:red, 0; green, 0; blue, 0 }  ][line width=0.08]  [draw opacity=0] (8.93,-4.29) -- (0,0) -- (8.93,4.29) -- cycle    ;
\draw    (473.75,173.33) -- (450.79,153.98) ;
\draw [shift={(473.75,173.33)}, rotate = 40.14] [color={rgb, 255:red, 0; green, 0; blue, 0 }  ][line width=0.75]    (0,5.59) -- (0,-5.59)   ;
\draw    (600.73,189.45) -- (576.16,174.09) ;
\draw [shift={(573.62,172.5)}, rotate = 32.01] [fill={rgb, 255:red, 0; green, 0; blue, 0 }  ][line width=0.08]  [draw opacity=0] (8.93,-4.29) -- (0,0) -- (8.93,4.29) -- cycle    ;
\draw    (573.62,172.5) -- (547.83,156.51) ;
\draw [shift={(573.62,172.5)}, rotate = 31.81] [color={rgb, 255:red, 0; green, 0; blue, 0 }  ][line width=0.75]    (0,5.59) -- (0,-5.59)   ;
\draw    (498.73,256.83) -- (498.88,227.86) ;
\draw [shift={(498.9,224.86)}, rotate = 90.3] [fill={rgb, 255:red, 0; green, 0; blue, 0 }  ][line width=0.08]  [draw opacity=0] (8.93,-4.29) -- (0,0) -- (8.93,4.29) -- cycle    ;
\draw    (498.91,225.86) -- (499.05,194.79) ;
\draw [shift={(498.91,225.86)}, rotate = 90.27] [color={rgb, 255:red, 0; green, 0; blue, 0 }  ][line width=0.75]    (0,5.59) -- (0,-5.59)   ;
\draw    (395.5,181.87) -- (421.49,169.06) ;
\draw [shift={(424.18,167.74)}, rotate = 153.76] [fill={rgb, 255:red, 0; green, 0; blue, 0 }  ][line width=0.08]  [draw opacity=0] (8.93,-4.29) -- (0,0) -- (8.93,4.29) -- cycle    ;
\draw    (423.29,168.19) -- (450.79,153.98) ;
\draw [shift={(423.29,168.19)}, rotate = 152.67] [color={rgb, 255:red, 0; green, 0; blue, 0 }  ][line width=0.75]    (0,5.59) -- (0,-5.59)   ;
\draw    (571.65,55.15) -- (552.78,77.14) ;
\draw [shift={(550.83,79.41)}, rotate = 310.63] [fill={rgb, 255:red, 0; green, 0; blue, 0 }  ][line width=0.08]  [draw opacity=0] (8.93,-4.29) -- (0,0) -- (8.93,4.29) -- cycle    ;
\draw    (551.47,78.64) -- (532.02,101.52) ;
\draw [shift={(551.47,78.64)}, rotate = 310.38] [color={rgb, 255:red, 0; green, 0; blue, 0 }  ][line width=0.75]    (0,5.59) -- (0,-5.59)   ;

\draw    (425.07,60.04) -- (446.66,79.36) ;
\draw [shift={(448.89,81.36)}, rotate = 221.83] [fill={rgb, 255:red, 0; green, 0; blue, 0 }  ][line width=0.08]  [draw opacity=0] (8.93,-4.29) -- (0,0) -- (8.93,4.29) -- cycle    ;
\draw    (448.89,81.36) -- (471.33,101.51) ;
\draw [shift={(448.89,81.36)}, rotate = 221.92] [color={rgb, 255:red, 0; green, 0; blue, 0 }  ][line width=0.75]    (0,5.59) -- (0,-5.59)   ;

\draw (467,79.4) node [anchor=north west][inner sep=0.75pt]    {$\zeta _{1}$};
\draw (539,95.4) node [anchor=north west][inner sep=0.75pt]    {$\zeta _{2}$};
\draw (542,161.4) node [anchor=north west][inner sep=0.75pt]    {$\zeta _{3}$};
\draw (431,134.4) node [anchor=north west][inner sep=0.75pt]    {$\zeta _{5}$};
\draw (477,190.4) node [anchor=north west][inner sep=0.75pt]    {$\zeta _{4}$};
\draw (677,190.4) node [anchor=north west][inner sep=0.75pt]    {$\ .$};

\draw [shift={(532.33,101.16)}, rotate = 210.82] [color={rgb, 255:red, 0; green, 0; blue, 0 }  ][fill={rgb, 255:red, 0; green, 0; blue, 0 }  ][line width=0.75]      (0, 0) circle [x radius= 2.2, y radius= 2.2]   ;

\draw [shift={(471.54,101.82)}, rotate = 210.82] [color={rgb, 255:red, 0; green, 0; blue, 0 }  ][fill={rgb, 255:red, 0; green, 0; blue, 0 }  ][line width=0.75]      (0, 0) circle [x radius= 2.2, y radius= 2.2]   ;

\draw [shift={(450.79,153.98)}, rotate = 210.82] [color={rgb, 255:red, 0; green, 0; blue, 0 }  ][fill={rgb, 255:red, 0; green, 0; blue, 0 }  ][line width=0.75]      (0, 0) circle [x radius= 2.2, y radius= 2.2]   ;

\draw [shift={(547.83,156.51)}, rotate = 210.82] [color={rgb, 255:red, 0; green, 0; blue, 0 }  ][fill={rgb, 255:red, 0; green, 0; blue, 0 }  ][line width=0.75]      (0, 0) circle [x radius= 2.2, y radius= 2.2]   ;

\draw [shift={(498.88,194.79)}, rotate = 210.82] [color={rgb, 255:red, 0; green, 0; blue, 0 }  ][fill={rgb, 255:red, 0; green, 0; blue, 0 }  ][line width=0.75]      (0, 0) circle [x radius= 2.2, y radius= 2.2]   ;

\end{tikzpicture}
\end{equation}
Clearly, this diagram is not allowed since the vertex used at $\zeta_5$ is not allowed by our rules. Thus, we have seen that the only consistent assignments at $\zeta_1$ are
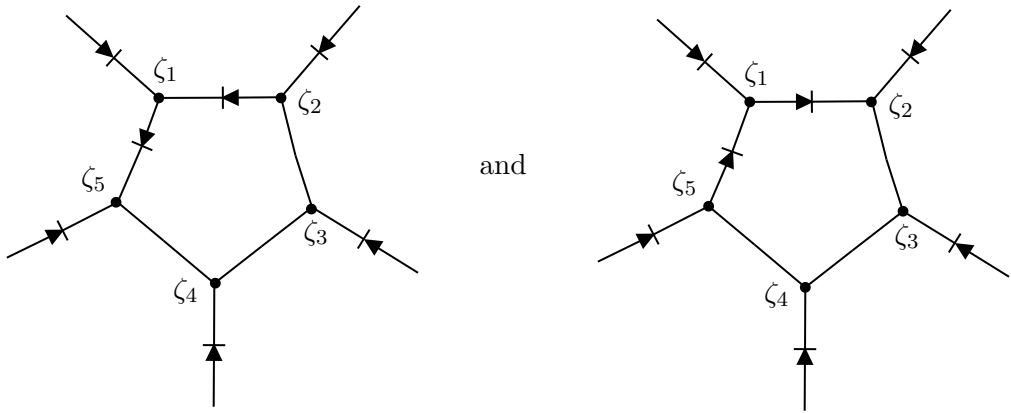
\begin{equation}
\tikzset{every picture/.style={line width=0.75pt}} 
\begin{tikzpicture}[x=0.75pt,y=0.75pt,yscale=-1,xscale=1]

\draw    (306.79,153.98) -- (318.96,125.63) ;
\draw [shift={(318.96,125.63)}, rotate = 113.23] [color={rgb, 255:red, 0; green, 0; blue, 0 }  ][line width=0.75]    (0,5.59) -- (0,-5.59)   ;
\draw    (319.6,123.73) -- (327.54,101.82) ;
\draw [shift={(318.58,126.55)}, rotate = 289.92] [fill={rgb, 255:red, 0; green, 0; blue, 0 }  ][line width=0.08]  [draw opacity=0] (8.93,-4.29) -- (0,0) -- (8.93,4.29) -- cycle    ;
\draw    (327.33,101.51) -- (359.31,101.39) ;
\draw [shift={(359.31,101.39)}, rotate = 179.78] [color={rgb, 255:red, 0; green, 0; blue, 0 }  ][line width=0.75]    (0,5.59) -- (0,-5.59)   ;
\draw    (361.31,101.38) -- (388.33,101.16) ;
\draw [shift={(358.31,101.4)}, rotate = 359.53] [fill={rgb, 255:red, 0; green, 0; blue, 0 }  ][line width=0.08]  [draw opacity=0] (8.93,-4.29) -- (0,0) -- (8.93,4.29) -- cycle    ;
\draw    (388.33,101.16) -- (395.57,130.51) ;
\draw    (395.57,130.51) -- (403.83,156.51) ;
\draw    (403.83,156.51) -- (378.64,176.19) ;
\draw    (378.64,176.19) -- (355.05,194.79) ;
\draw    (355.05,194.79) -- (329.75,173.33) ;
\draw    (329.75,173.33) -- (306.79,153.98) ;
\draw    (456.73,189.45) -- (432.16,174.09) ;
\draw [shift={(429.62,172.5)}, rotate = 32.01] [fill={rgb, 255:red, 0; green, 0; blue, 0 }  ][line width=0.08]  [draw opacity=0] (8.93,-4.29) -- (0,0) -- (8.93,4.29) -- cycle    ;
\draw    (429.62,172.5) -- (403.83,156.51) ;
\draw [shift={(429.62,172.5)}, rotate = 31.81] [color={rgb, 255:red, 0; green, 0; blue, 0 }  ][line width=0.75]    (0,5.59) -- (0,-5.59)   ;
\draw    (354.73,256.83) -- (354.88,227.86) ;
\draw [shift={(354.9,224.86)}, rotate = 90.3] [fill={rgb, 255:red, 0; green, 0; blue, 0 }  ][line width=0.08]  [draw opacity=0] (8.93,-4.29) -- (0,0) -- (8.93,4.29) -- cycle    ;
\draw    (354.91,225.86) -- (355.05,194.79) ;
\draw [shift={(354.91,225.86)}, rotate = 90.27] [color={rgb, 255:red, 0; green, 0; blue, 0 }  ][line width=0.75]    (0,5.59) -- (0,-5.59)   ;
\draw    (251.5,181.87) -- (277.49,169.06) ;
\draw [shift={(280.18,167.74)}, rotate = 153.76] [fill={rgb, 255:red, 0; green, 0; blue, 0 }  ][line width=0.08]  [draw opacity=0] (8.93,-4.29) -- (0,0) -- (8.93,4.29) -- cycle    ;
\draw    (279.29,168.19) -- (306.79,153.98) ;
\draw [shift={(279.29,168.19)}, rotate = 152.67] [color={rgb, 255:red, 0; green, 0; blue, 0 }  ][line width=0.75]    (0,5.59) -- (0,-5.59)   ;
\draw    (427.65,55.15) -- (408.78,77.14) ;
\draw [shift={(406.83,79.41)}, rotate = 310.63] [fill={rgb, 255:red, 0; green, 0; blue, 0 }  ][line width=0.08]  [draw opacity=0] (8.93,-4.29) -- (0,0) -- (8.93,4.29) -- cycle    ;
\draw    (407.47,78.64) -- (388.02,101.52) ;
\draw [shift={(407.47,78.64)}, rotate = 310.38] [color={rgb, 255:red, 0; green, 0; blue, 0 }  ][line width=0.75]    (0,5.59) -- (0,-5.59)   ;

\draw    (281.07,60.04) -- (302.66,79.36) ;
\draw [shift={(304.89,81.36)}, rotate = 221.83] [fill={rgb, 255:red, 0; green, 0; blue, 0 }  ][line width=0.08]  [draw opacity=0] (8.93,-4.29) -- (0,0) -- (8.93,4.29) -- cycle    ;
\draw    (304.89,81.36) -- (327.33,101.51) ;
\draw [shift={(304.89,81.36)}, rotate = 221.92] [color={rgb, 255:red, 0; green, 0; blue, 0 }  ][line width=0.75]    (0,5.59) -- (0,-5.59)   ;
\draw    (601.79,155.98) -- (612.78,130.38) ;
\draw [shift={(613.96,127.63)}, rotate = 113.23] [fill={rgb, 255:red, 0; green, 0; blue, 0 }  ][line width=0.08]  [draw opacity=0] (8.93,-4.29) -- (0,0) -- (8.93,4.29) -- cycle    ;
\draw    (613.58,128.55) -- (622.54,103.82) ;
\draw [shift={(613.58,128.55)}, rotate = 109.92] [color={rgb, 255:red, 0; green, 0; blue, 0 }  ][line width=0.75]    (0,5.59) -- (0,-5.59)   ;
\draw    (622.33,103.51) -- (651.31,103.4) ;
\draw [shift={(654.31,103.39)}, rotate = 179.78] [fill={rgb, 255:red, 0; green, 0; blue, 0 }  ][line width=0.08]  [draw opacity=0] (8.93,-4.29) -- (0,0) -- (8.93,4.29) -- cycle    ;
\draw    (653.31,103.4) -- (683.33,103.16) ;
\draw [shift={(653.31,103.4)}, rotate = 179.53] [color={rgb, 255:red, 0; green, 0; blue, 0 }  ][line width=0.75]    (0,5.59) -- (0,-5.59)   ;

\draw    (683.33,103.16) -- (690.57,132.51) ;
\draw    (690.57,132.51) -- (698.83,158.51) ;
\draw    (698.83,158.51) -- (673.64,178.19) ;
\draw    (673.64,178.19) -- (650.05,196.79) ;
\draw    (650.05,196.79) -- (624.75,175.33) ;
\draw    (624.75,175.33) -- (601.79,155.98) ;
\draw    (751.73,191.45) -- (727.16,176.09) ;
\draw [shift={(724.62,174.5)}, rotate = 32.01] [fill={rgb, 255:red, 0; green, 0; blue, 0 }  ][line width=0.08]  [draw opacity=0] (8.93,-4.29) -- (0,0) -- (8.93,4.29) -- cycle    ;
\draw    (724.62,174.5) -- (698.83,158.51) ;
\draw [shift={(724.62,174.5)}, rotate = 31.81] [color={rgb, 255:red, 0; green, 0; blue, 0 }  ][line width=0.75]    (0,5.59) -- (0,-5.59)   ;
\draw    (649.73,258.83) -- (649.88,229.86) ;
\draw [shift={(649.9,226.86)}, rotate = 90.3] [fill={rgb, 255:red, 0; green, 0; blue, 0 }  ][line width=0.08]  [draw opacity=0] (8.93,-4.29) -- (0,0) -- (8.93,4.29) -- cycle    ;
\draw    (649.91,227.86) -- (650.05,196.79) ;
\draw [shift={(649.91,227.86)}, rotate = 90.27] [color={rgb, 255:red, 0; green, 0; blue, 0 }  ][line width=0.75]    (0,5.59) -- (0,-5.59)   ;
\draw    (546.5,183.87) -- (572.49,171.06) ;
\draw [shift={(575.18,169.74)}, rotate = 153.76] [fill={rgb, 255:red, 0; green, 0; blue, 0 }  ][line width=0.08]  [draw opacity=0] (8.93,-4.29) -- (0,0) -- (8.93,4.29) -- cycle    ;
\draw    (574.29,170.19) -- (601.79,155.98) ;
\draw [shift={(574.29,170.19)}, rotate = 152.67] [color={rgb, 255:red, 0; green, 0; blue, 0 }  ][line width=0.75]    (0,5.59) -- (0,-5.59)   ;
\draw    (722.65,57.15) -- (703.78,79.14) ;
\draw [shift={(701.83,81.41)}, rotate = 310.63] [fill={rgb, 255:red, 0; green, 0; blue, 0 }  ][line width=0.08]  [draw opacity=0] (8.93,-4.29) -- (0,0) -- (8.93,4.29) -- cycle    ;
\draw    (702.47,80.64) -- (683.02,103.52) ;
\draw [shift={(702.47,80.64)}, rotate = 310.38] [color={rgb, 255:red, 0; green, 0; blue, 0 }  ][line width=0.75]    (0,5.59) -- (0,-5.59)   ;

\draw    (576.07,62.04) -- (597.66,81.36) ;
\draw [shift={(599.89,83.36)}, rotate = 221.83] [fill={rgb, 255:red, 0; green, 0; blue, 0 }  ][line width=0.08]  [draw opacity=0] (8.93,-4.29) -- (0,0) -- (8.93,4.29) -- cycle    ;
\draw    (599.89,83.36) -- (622.33,103.51) ;
\draw [shift={(599.89,83.36)}, rotate = 221.92] [color={rgb, 255:red, 0; green, 0; blue, 0 }  ][line width=0.75]    (0,5.59) -- (0,-5.59)   ;

\draw (323,79.4) node [anchor=north west][inner sep=0.75pt]    {$\zeta _{1}$};
\draw (395,95.4) node [anchor=north west][inner sep=0.75pt]    {$\zeta _{2}$};
\draw (398,161.4) node [anchor=north west][inner sep=0.75pt]    {$\zeta _{3}$};
\draw (287,134.4) node [anchor=north west][inner sep=0.75pt]    {$\zeta _{5}$};
\draw (333,190.4) node [anchor=north west][inner sep=0.75pt]    {$\zeta _{4}$};
\draw (618,81.4) node [anchor=north west][inner sep=0.75pt]    {$\zeta _{1}$};
\draw (690,97.4) node [anchor=north west][inner sep=0.75pt]    {$\zeta _{2}$};
\draw (693,163.4) node [anchor=north west][inner sep=0.75pt]    {$\zeta _{3}$};
\draw (582,136.4) node [anchor=north west][inner sep=0.75pt]    {$\zeta _{5}$};
\draw (628,192.4) node [anchor=north west][inner sep=0.75pt]    {$\zeta _{4}$};
\draw (486,128) node [anchor=north west][inner sep=0.75pt]   [align=left] {and};

\draw [shift={(683.02,103.52)}, rotate = 210.82] [color={rgb, 255:red, 0; green, 0; blue, 0 }  ][fill={rgb, 255:red, 0; green, 0; blue, 0 }  ][line width=0.75]      (0, 0) circle [x radius= 2.2, y radius= 2.2]   ;

\draw [shift={(622.33,103.51)}, rotate = 210.82] [color={rgb, 255:red, 0; green, 0; blue, 0 }  ][fill={rgb, 255:red, 0; green, 0; blue, 0 }  ][line width=0.75]      (0, 0) circle [x radius= 2.2, y radius= 2.2]   ;

\draw [shift={(601.79,155.98)}, rotate = 210.82] [color={rgb, 255:red, 0; green, 0; blue, 0 }  ][fill={rgb, 255:red, 0; green, 0; blue, 0 }  ][line width=0.75]      (0, 0) circle [x radius= 2.2, y radius= 2.2]   ;

\draw [shift={(650.05,196.79)}, rotate = 210.82] [color={rgb, 255:red, 0; green, 0; blue, 0 }  ][fill={rgb, 255:red, 0; green, 0; blue, 0 }  ][line width=0.75]      (0, 0) circle [x radius= 2.2, y radius= 2.2]   ;

\draw [shift={(698.83,158.51)}, rotate = 210.82] [color={rgb, 255:red, 0; green, 0; blue, 0 }  ][fill={rgb, 255:red, 0; green, 0; blue, 0 }  ][line width=0.75]      (0, 0) circle [x radius= 2.2, y radius= 2.2]   ;

\draw [shift={(388.33,101.51)}, rotate = 210.82] [color={rgb, 255:red, 0; green, 0; blue, 0 }  ][fill={rgb, 255:red, 0; green, 0; blue, 0 }  ][line width=0.75]      (0, 0) circle [x radius= 2.2, y radius= 2.2]   ;

\draw [shift={(327.33,101.51)}, rotate = 210.82] [color={rgb, 255:red, 0; green, 0; blue, 0 }  ][fill={rgb, 255:red, 0; green, 0; blue, 0 }  ][line width=0.75]      (0, 0) circle [x radius= 2.2, y radius= 2.2]   ;

\draw [shift={(305.99,153.98)}, rotate = 210.82] [color={rgb, 255:red, 0; green, 0; blue, 0 }  ][fill={rgb, 255:red, 0; green, 0; blue, 0 }  ][line width=0.75]      (0, 0) circle [x radius= 2.2, y radius= 2.2]   ;

\draw [shift={(355.55,194.69)}, rotate = 210.82] [color={rgb, 255:red, 0; green, 0; blue, 0 }  ][fill={rgb, 255:red, 0; green, 0; blue, 0 }  ][line width=0.75]      (0, 0) circle [x radius= 2.2, y radius= 2.2]   ;

\draw [shift={(403.5,157.4)}, rotate = 210.82] [color={rgb, 255:red, 0; green, 0; blue, 0 }  ][fill={rgb, 255:red, 0; green, 0; blue, 0 }  ][line width=0.75]      (0, 0) circle [x radius= 2.2, y radius= 2.2]   ;
\end{tikzpicture}
\end{equation}
By the arguments we used above, this choice at $\zeta_1$ fixes the choices at all other $\zeta_i$, and thus results in a diagram with a loop of the arrows. This argument generalises to any polygonal 1-loop diagram. The generalisation from $\phi^3$ to $\phi^n$ is also straightforward since we can think of many incoming external lines at every vertex in the above diagrams, and all the arguments above go through. 

In fact, it is easy to see that this works for higher loops as well. For any given number of loops, if all the external lines are ingoing, then there has to be at least one directed \emph{causal cycle}\footnote{Note that the terms causal cycle and causal loop refer to the same thing and are used interchangeably throughout this note. } (closed loop of causal arrows flowing in the same direction) in the graph. Begin by noting that all the allowed vertices have at least one ingoing line and at least one outgoing line. Therefore, at each vertex, there should be at least one outgoing line. Moreover, since we have taken all the external lines to be ingoing, there should at least be one internal line that is outgoing. Thus, starting from a vertex that has such an outgoing causal line, we can follow it to the next vertex, where it becomes an ingoing line. By the same reasoning, this next vertex must also have at least one outgoing line, and the process can be repeated. Since the graph is finite, continuing this process should eventually end in a causal cycle. 

Since we have already assumed that any diagram containing a causal cycle vanishes, it follows that diagrams with all external legs ingoing vanish at any loop order. By the same argument, diagrams with all external legs outgoing also vanish. Hence, these diagrams satisfy both the SK collapse and KMS conditions to all loop orders, provided the causal cycle argument works. 

The same analysis goes through for any $\phi^n$ vertex, as long as we do not include vertices with all arrows flowing in or all arrows flowing out. Thus, this simple graph-theoretic argument proves that, given the Feynman rules we have, we can never (up to the caveat discussed earlier) have a non-zero correlator for all past or all future operators.

The reader familiar with the Schwinger-Keldysh formalism will recognise this as the statement of the \emph{largest time equation} \cite{Gelis:2019yfm}. The largest time equation states that the sum over all possible cut diagrams of a Feynman amplitude in which the largest time is assigned to the cut (physical states) vanishes \cite{Nastase:1970yyp,Salvio:2024upo}. The important point to note is that all the arrows in our diagrams are to be seen as causal arrows, since we only have causal propagators (ingoing and outgoing boundary-to-bulk propagators, and retarded bulk-to-bulk propagators) in our diagrammatics. From this point of view, any diagram in which all the times are flowing in from the boundary towards the bulk cannot have a consistent time flow and thus has to vanish. In other words, the only diagrams we get for all ingoing (outgoing) boundary lines are diagrams with at least one directed cycle, and one cannot consistently assign a largest time for vertices in this cycle. The argument outlined above is a Witten diagram version of the largest time equation, in the mixed Fourier representation.

The question now is the validity of the causal cycle assumption. While this indeed sounds reasonable, it would be good to explicitly evaluate integrals of such causal cycles (say, in a boundary derivative expansion or with exact answers in the BTZ case) and check.

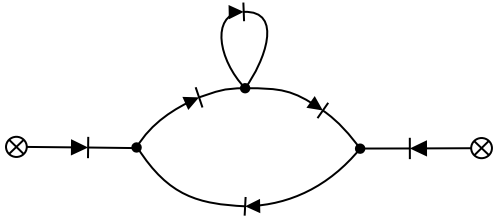
\begin{figure}[H]
    \centering
\tikzset{every picture/.style={line width=0.75pt}} 

\tikzset{every picture/.style={line width=0.75pt}} 

\tikzset{every picture/.style={line width=0.75pt}} 

\begin{tikzpicture}[x=0.75pt,y=0.75pt,yscale=-1,xscale=1,scale=0.6]

\draw [line width=0.75]    (282.33,111.46) .. controls (259.45,87.17) and (254.38,50.63) .. (278.42,47.65) ;
\draw [shift={(281.15,47.46)}, rotate = 178.94] [fill={rgb, 255:red, 0; green, 0; blue, 0 }  ][line width=0.08]  [draw opacity=0] (12.5,-6.01) -- (0,0) -- (12.5,6.01) -- cycle    ;
\draw [shift={(282.33,111.46)}, rotate = 226.71] [color={rgb, 255:red, 0; green, 0; blue, 0 }  ][fill={rgb, 255:red, 0; green, 0; blue, 0 }  ][line width=0.75]      (0, 0) circle [x radius= 3.5, y radius= 3.5]   ;
\draw [line width=0.75]    (282.33,111.46) .. controls (298.55,89.15) and (314.54,46.21) .. (281.15,47.46) ;
\draw [shift={(281.15,47.46)}, rotate = 357.84] [color={rgb, 255:red, 0; green, 0; blue, 0 }  ][line width=0.75]    (0,7.83) -- (0,-7.83)   ;
\draw [line width=0.75]    (192,161.13) .. controls (204.35,140.23) and (223.92,128.35) .. (241.28,120.36) ;
\draw [shift={(244,119.13)}, rotate = 156.04] [fill={rgb, 255:red, 0; green, 0; blue, 0 }  ][line width=0.08]  [draw opacity=0] (12.5,-6.01) -- (0,0) -- (12.5,6.01) -- cycle    ;
\draw [shift={(192,161.13)}, rotate = 300.58] [color={rgb, 255:red, 0; green, 0; blue, 0 }  ][fill={rgb, 255:red, 0; green, 0; blue, 0 }  ][line width=0.75]      (0, 0) circle [x radius= 3.5, y radius= 3.5]   ;
\draw [line width=0.75]    (282.33,111.46) .. controls (263.67,111.46) and (259,114.13) .. (244,119.13) ;
\draw [shift={(244,119.13)}, rotate = 341.57] [color={rgb, 255:red, 0; green, 0; blue, 0 }  ][line width=0.75]    (0,8.94) -- (0,-8.94)   ;
\draw [line width=0.75]    (378,162.13) .. controls (370,151.13) and (360,139.13) .. (347,130.13) ;
\draw [shift={(347,130.13)}, rotate = 34.7] [color={rgb, 255:red, 0; green, 0; blue, 0 }  ][line width=0.75]    (0,7.83) -- (0,-7.83)   ;
\draw [shift={(378,162.13)}, rotate = 233.97] [color={rgb, 255:red, 0; green, 0; blue, 0 }  ][fill={rgb, 255:red, 0; green, 0; blue, 0 }  ][line width=0.75]      (0, 0) circle [x radius= 3.5, y radius= 3.5]   ;
\draw [line width=0.75]    (282.33,111.46) .. controls (312.08,111.46) and (321.75,112.71) .. (344.82,128.61) ;
\draw [shift={(347,130.13)}, rotate = 215.1] [fill={rgb, 255:red, 0; green, 0; blue, 0 }  ][line width=0.08]  [draw opacity=0] (12.5,-6.01) -- (0,0) -- (12.5,6.01) -- cycle    ;
\draw [line width=0.75]    (193,161.67) .. controls (219,202.46) and (243.67,208.46) .. (282.33,210.46) ;
\draw [shift={(282.33,210.46)}, rotate = 182.96] [color={rgb, 255:red, 0; green, 0; blue, 0 }  ][line width=0.75]    (0,7.83) -- (0,-7.83)   ;
\draw [line width=0.75]    (378,162.13) .. controls (351.67,194.3) and (319.33,209.05) .. (284.98,210.38) ;
\draw [shift={(282.33,210.46)}, rotate = 358.92] [fill={rgb, 255:red, 0; green, 0; blue, 0 }  ][line width=0.08]  [draw opacity=0] (12.5,-6.01) -- (0,0) -- (12.5,6.01) -- cycle    ;
\draw [line width=0.75]    (100.68,160.67) -- (148.71,161.14) ;
\draw [shift={(151.71,161.17)}, rotate = 180.56] [fill={rgb, 255:red, 0; green, 0; blue, 0 }  ][line width=0.08]  [draw opacity=0] (14.29,-6.86) -- (0,0) -- (14.29,6.86) -- cycle    ;
\draw [line width=0.75]    (193,161.67) -- (151.71,161.17) ;
\draw [shift={(151.71,161.17)}, rotate = 0.69] [color={rgb, 255:red, 0; green, 0; blue, 0 }  ][line width=0.75]    (0,8.94) -- (0,-8.94)   ;
\draw  [line width=0.75]  (83.4,160.61) .. controls (83.41,155.94) and (87.29,152.17) .. (92.07,152.18) .. controls (96.84,152.2) and (100.69,156) .. (100.68,160.67) .. controls (100.66,165.34) and (96.78,169.11) .. (92.01,169.09) .. controls (87.23,169.07) and (83.38,165.28) .. (83.4,160.61) -- cycle ; \draw  [line width=0.75]  (85.95,154.64) -- (98.12,166.64) ; \draw  [line width=0.75]  (98.17,154.68) -- (85.91,166.59) ;
\draw [line width=0.75]    (470.38,161.76) -- (422.35,161.95) ;
\draw [shift={(419.35,161.96)}, rotate = 359.77] [fill={rgb, 255:red, 0; green, 0; blue, 0 }  ][line width=0.08]  [draw opacity=0] (14.29,-6.86) -- (0,0) -- (14.29,6.86) -- cycle    ;
\draw [line width=0.75]    (378.05,162.03) -- (419.35,161.96) ;
\draw [shift={(419.35,161.96)}, rotate = 179.9] [color={rgb, 255:red, 0; green, 0; blue, 0 }  ][line width=0.75]    (0,8.94) -- (0,-8.94)   ;
\draw  [line width=0.75]  (487.66,161.58) .. controls (487.71,166.25) and (483.88,170.07) .. (479.11,170.12) .. controls (474.34,170.17) and (470.43,166.42) .. (470.38,161.76) .. controls (470.33,157.09) and (474.16,153.26) .. (478.93,153.21) .. controls (483.71,153.16) and (487.61,156.91) .. (487.66,161.58) -- cycle ; \draw  [line width=0.75]  (485.19,167.58) -- (472.85,155.75) ; \draw  [line width=0.75]  (472.97,167.71) -- (485.07,155.63) ;

\end{tikzpicture}
    \caption{Example of a diagram with a causal cycle  of only one retarded propagator.}
    \label{fig:tadpole}
\end{figure}

We will postpone such explicit computations for future work and merely state the result. Our preliminary computations suggest that these arguments work as advertised for any causal cycle with more than one retarded propagator. The above argument however seems to fail when the causal cycle has only one retarded propagator e.g., the tadpole diagrams or the diagrams like the one in Figure~\ref{fig:tadpole}. This issue evidently needs a more detailed analysis.

\section{Discussion}

\subsection*{Summary}

In this work, we have investigated the real-time quantum dynamics of a self-interacting scalar field in the background of an AdS black brane. This is done using the gravitational Schwinger-Keldysh (grSK) formalism in the bulk, which provides a framework for computing real-time/Schwinger–Keldysh boundary correlators. In practice, this amounts to evaluating such correlators in the large but finite-$N$ limit, or equivalently, including bulk loop corrections.

Earlier works \cite{Loganayagam:2024mnj, Martin:2024mdm} proposed a conjecture for carrying out these computations at the tree level. Their conjecture prescribes how to handle the multi-discontinuity integrals that naturally arise in the grSK geometry, reducing them to radial integrals restricted to the black hole exterior. The outcome is an effective classical field theory localised outside the black brane, whose interaction vertices closely resemble those of a thermal field theory.

Here, we have extended this program to the quantum level by proposing a general conjecture for evaluating multiple discontinuity integrals arising from bulk loop diagrams in the grSK geometry. This generalises the tree-level results of \cite{Loganayagam:2024mnj} to arbitrary bulk loops with scalar non-derivative interactions.\footnote{A further caveat remains: the conjecture does not seem to apply to diagrams containing tadpoles, as seen by explicit computation.} Apart from this limitation, the conjecture yields a quantum field theory defined in the exterior of the black brane, with Feynman rules that naturally extend the tree-level picture to include loops.

We tested this conjecture explicitly by computing all one-loop, as well as several two- and three-loop, contributions to two-, three-, and four-point functions. Furthermore, we showed that the resulting diagrammatic rules are consistent with both microscopic unitarity and thermality at arbitrary loop order.

Finally, we emphasise that our construction of the exterior quantum field theory uses a complementary approach based on bulk Schwinger-Dyson equations (SDEs). This provides an alternative route to real-time finite-temperature holography and makes manifest the consistency of our framework. To avoid cluttering the main text, we relegated this discussion to the appendices, where we derive SDEs from the path integral perspective in $\phi^3$ theory and provide explicit evaluations of the monodromy integrals that play a central role throughout the analysis.

\subsection*{Comments on the exterior QFT and its canonical perspective}


We pause here to make a few more comments on the additional structure of the exterior QFT. Broadly speaking, we would like to address three points: (i) what has been achieved so far, (ii) why it is desirable to seek a canonical formulation, and (iii) what such a formulation must achieve.


In this work, we have successfully derived an exterior QFT to arbitrary loop order starting from the grSK geometry. This provides a complete and consistent perturbative framework for describing dynamics in the exterior region. As we have seen, this exterior QFT satisfies microscopic unitarity, causality, and thermality via the KMS conditions. Importantly, we have seen that for non-derivative interactions, there are no terms in the exterior QFT that are localised purely at the horizon. Furthermore, the vertices of this QFT are purely those of a unitary theory, i.e., the vertices are the same as those obtained from a unitary theory. In other words, the vertices do not connect the two legs of the SK contour (there are no Feynman-Vernon type vertices).

All these facts suggest that there is a standard  unitary quantum field theory which can be defined purely in the exterior, and with appropriate boundary conditions at the horizon. We will now argue that it would be good to have a canonical description of such a QFT.\footnote{We thank the anonymous referee for this suggestion that inspired the discussion that follows.}


Despite the success of grSK, it is important to develop an independent derivation of exterior QFT that does not rely on the grSK prescription. The motivation is straightforward: obtaining the same result from a different framework may illuminate why the grSK prescription works in the first place ( and may even point toward a more fundamental understanding of its origin). By an ``independent derivation'', we mean obtaining the same exterior QFT through a \emph{genuinely different} route. Possible alternatives include path-integral approaches without grSK contour, a fully canonical (operator-based) formulation, a euclidean framework based on cigar geometry, or perhaps some other framework that departs from these.

There are additional reasons to pursue such a direction. For instance, a canonical quantisation framework offers some advantages over grSK approach. For example, canonical framework may provide a natural setting for computing higher-OTOCs (useful for quantum chaos).\footnote{Once an operator description is available, generalisations from the standard SK contour to more complicated many-fold contours become conceptually straightforward. Alternately, one might hope that OTOCs can be addressed by directly generalising grSK framework:  after this work came out, authors of \cite{Ammon:2025vod} have suggested precisely such a framework. Whether we still have an exterior QFT in this case remains an open question. } Moreover, many key notions like quantum entanglement and quantum information may be more accessible in such a description. Thus, a canonical formulation could connect the exterior QFT to these subjects.

Finally, there is a conceptual connection to black hole complementarity and the central dogma (i.e., the exterior encodes all information about the interior). While the grSK prescription appears to incorporate this feature implicitly, the underlying mechanism remains unclear. A canonical formulation of exterior QFT could help make this connection more explicit and clear.


By a canonical formulation, we mean a structure of the following type: an exterior mode expansion of the quantum field with some yet to be determined boundary condition at the horizon. Such a mode expansion should satisfy the following criteria :
\begin{itemize}
\item The corresponding creation/annihilation operators should satisfy canonical commutation relations.
\item Evaluated on the thermal state, we should reproduce both the boundary-to-bulk as well as bulk-to-bulk propagators of the exterior QFT quoted in the main text.
\item Dynamics of these operator fields should generate the same exterior QFT Feynman rules that were obtained using the grSK prescription.
\end{itemize}
Such a canonical framework would not only rederive the exterior QFT but also serve as a potential stage for further exploration.

\newpage

\subsection*{Future Directions}

There are several other natural and exciting directions to extend the present work. 
\medskip

\begin{enumerate}
\item 
A natural next step is to understand how \emph{thermal effects from bulk loops} (such as Hawking radiation) renormalise the effective field theory outside the horizon.\footnote{See \cite{Aharony:2016dwx,Giombi:2017hpr, Bertan:2018khc, Bertan:2018afl, Carmi:2018qzm,Carmi:2019ocp, Herderschee:2021jbi,  Banados:2022nhj,MunozSandoval:2023hix} for bulk loops and bulk renormalisation in pure AdS.} Such corrections may shift self-energies, generate thermal masses, and even modify couplings, thereby providing a concrete picture of how black holes ``dress'' nearby quantum fields \cite{Costantino:2020vdu}. For instance, this could reveal how QED parameters---like the running coupling or higher-order interactions---are altered in a black hole background.

\item 
A related question is the possibility of \emph{phase transitions near black holes}, driven by bulk loop effects or boundary finite-$N$ corrections \cite{Dvali:2012rt,Dvali:2012en,Liu:2025iei}. Such corrections could trigger instabilities or qualitatively change the late-time dynamics of fields in the black hole exterior. Connected to it, the Coleman-Mermin-Wagner-Hohenberg (CMWH) mechanism, especially in the presence of derivative interactions, provides a natural framework for symmetry-breaking and its re-emergence at finite $N$ in a thermal setting \cite{Anninos:2010sq}. Clarifying how these effects manifest in the exterior field theory may offer new insights into quantum corrections, horizon-localised contributions, and real-time dynamics.

\item In the fermionic sector, this exploration becomes quite rich. Several \emph{bulk loop-induced effects} are important, a few of which we touch upon now.
\begin{itemize}
    
\item \emph{Quantum oscillations}:
Phenomena such as the Shubnikov-de Haas and de Haas-van Alphen effects \cite{10.1093/oso/9780198505921.001.0001} require loop contributions in the bulk, because tree-level calculations exhibit no oscillatory behaviour \cite{Denef:2009kn, Denef:2009yy, Hartnoll:2009kk}.

\item \emph{Cooper pairing and associated $U(1)$ spontaneous symmetry breaking}:
An effective attractive force between the fermions can be observed due to the underlying condensate formation and spontaneous breaking of $U(1)$ symmetry. However, in the large-$N$ limit, most observables remain blind to this effect, making bulk loop corrections essential in such scenarios \cite{Hartman:2010fk}.

\item \emph{Transport corrections}: Understanding the charge transport in holographic systems with Fermi surfaces requires the loop correction in the bulk, as seen in \cite{Faulkner:2013bna}.

\end{itemize}
Many of the computations described above were performed at one loop level by clever analytic continuations of Euclidean or retarded correlators (see \cite{Hartnoll:2016apf} for a detailed discussion on this subject). We believe the formalism described in this work gives a more straightforward route to rederive the above results, as well as to compute higher loop corrections to these results. 

\item Many recent works on classical black hole scattering/radiation reaction use diagrammatics in the `in-in' or Schwinger-Keldysh formalism.\footnote{A far from exhaustive list of references in this fast-moving subject is \cite{Galley:2009px,Kosower:2018adc,Jakobsen:2022psy,Kalin:2022hph,Porto:2024cwd}. } In order to recover the correct classical physics for particles, we should include arbitrary higher loop contributions in these computations (this is already true for physics in the conservative sector \cite{Goldberger:2004jt,Porto:2016pyg,Levi:2018nxp}). A far less explored direction is to study the dS/AdS analogues of these computations to better understand the effects of the cosmological constant.

\item Finally, our discussion of unitarity and thermality---through Cutkosky rules \cite{Veltman:1994wz, Meltzer:2020qbr} and KMS conditions---suggests a promising direction in the study of cosmological correlators \cite{Ghosh:2024aqd, Goodhew:2020hob}. In this context, the black hole horizon is replaced by the cosmological horizon, and recent work has shown renewed interest in making unitarity manifest in de Sitter computations \cite{Ema:2024hkj, Werth:2024mjg}. The past-future basis we have employed has the additional benefit of simultaneously highlighting unitarity and thermality. 

\end{enumerate}







\section*{Acknowledgements}

We would like to thank Sujay Ashok, Joydeep Chakravarty, Alok Laddha, Hong Liu, Raghu Mahajan, Gautam Mandal, Suvrat Raju, Mukund Rangamani, Ashoke Sen, Omkar Shetye, and Ronak Soni for many insightful discussions on these topics. GM presented this work in the 8th Da Nang Holography and String Theory conference. We acknowledge the support of the Department of Atomic Energy, Government of India, under project no. RTI4001, and the unwavering and generous support provided by the people of India towards research in the fundamental sciences.

\appendix

\section{Deriving Schwinger-Dyson equations from the path integral}\label{app:SDEsfromPI}

In this section, we will derive the Schwinger-Dyson equations for the quantum field theory of one scalar with a $\phi^3$ interaction.We will begin by deriving the Schwinger-Dyson equations for $\phi^3$ theory in flat spacetime. We will then show that a similar structure is inherited by an open quantum field theory. We will explicitly write down the Schwinger-Dyson equations for an open quantum field theory with the sources in the causal (past-future) basis.

The notation we will use in this section will be the one followed in \cite{Cvitanovic:1983eb} and \cite{Rammer:2007zz}. 

\subsection{\texorpdfstring{$\phi^3$}{p3} theory}

Let us begin with $\phi^3$ theory on flat spacetime. The generating function of correlators of this theory is the path integral
\begin{equation}
    Z[J] = \int [D \phi] \ \exp \left\{i \int \diff^{d+1}x \ \left[-\frac{1}{2} \partial_\mu \phi \partial^\mu \phi -\frac{m^2}{2} \phi^2 -\frac{\lambda}{3!} \phi^3 + J \phi\right]\right\} \ .
\end{equation}
The path integral on the RHS is invariant under redefinition of the field $\phi(x)$. In particular, the shift $\phi(x) \mapsto \phi(x) + \epsilon(x)$ should keep the path integral invariant. Performing this shift in the path integration variable on the RHS, we obtain
\begin{equation}
    \begin{split}
        Z[J] &= \int [D\phi] \ \exp \left\{i \int \diff^{d+1}x \ \left[-\frac{1}{2} \partial_\mu \phi \partial^\mu \phi -\frac{m^2}{2} \phi^2 -\frac{\lambda}{3!} \phi^3 + J \phi\right]\right\}\\
        &\qquad \times \left[ 1 + i \int \diff^{d+1}x\ \epsilon(x)\left((\Box - m^2) \phi - \frac{\lambda}{2!} \phi^2 + J \right) \right] \ ,
    \end{split}
\end{equation}
where we have expanded in the shift $\epsilon(x)$, and kept only the linear term. Note that the shift transformation keeps the path integral measure invariant.

The first term on the RHS of the above equation is simply $Z[J]$. Cancelling $Z[J]$ on either side of the equation, we obtain
\begin{equation}
    \begin{split}
        &\int [D\phi] \ \exp \left\{i \int \diff^{d+1}x \ \left[-\frac{1}{2} \partial_\mu \phi \partial^\mu \phi -\frac{m^2}{2} \phi^2 -\frac{\lambda}{3!} \phi^3 + J \phi\right]\right\}\\
        &\times \int \diff^{d+1}x\ \epsilon(x)\left((\Box - m^2) \phi - \frac{\lambda}{2!} \phi^2 + J \right)  = 0 \ .
    \end{split}
\label{eq:SDEoneptstep}
\end{equation}
The above equation can be recast into a functional differential equation for the generating function $Z[J]$:
\begin{equation}
   \left[(\Box - m^2) \frac{1}{i} \frac{\delta}{\delta J(x)} -\frac{\lambda}{2!} \left(\frac{1}{i} \frac{\delta}{\delta J(x)}\right)^2 +J(x) \right] Z[J] = 0 \ .
\end{equation}
This is the one-point Schwinger-Dyson equation for $\phi^3$ theory. Note that we have omitted the integral over spacetime, i.e., set the integrand to zero. We are allowed to do this since the shift function $\epsilon(x)$ is arbitrary.

The one-point Schwinger-Dyson equation is usually written with the spacetime differential operator inverted, i.e., in terms of the propagator as follows:
\begin{equation}
    \frac{1}{i}\frac{\delta}{\delta J(x)} Z[J] = \frac{-i}{(-\Box + m^2)} \left[\frac{-i\lambda}{2!} \left(\frac{1}{i} \frac{\delta}{\delta J(x)}\right)^2 +i J(x) \right] Z[J] \ .
\label{eq:SDE1pt}
\end{equation}
This form of the equations admits a simple diagrammatic interpretation:
\begin{equation}
    \tikzset{every picture/.style={line width=0.75pt}} 
    \begin{tikzpicture}[x=0.75pt,y=0.75pt,yscale=-1,xscale=1]
    \centering
    
    \draw    (111,127) -- (171.33,127) ;
    \draw  [fill={rgb, 255:red, 155; green, 155; blue, 155 }  ,fill opacity=1 ][line width=0.75]  (171.33,127) .. controls (171.33,119.26) and (177.59,112.99) .. (185.33,112.98) .. controls (193.06,112.97) and (199.34,119.23) .. (199.35,126.97) .. controls (199.36,134.71) and (193.09,140.98) .. (185.36,140.99) .. controls (177.62,141) and (171.34,134.74) .. (171.33,127) -- cycle ;
    \draw  [fill={rgb, 255:red, 155; green, 155; blue, 155 }  ,fill opacity=1 ] (339.33,127) .. controls (339.33,119.26) and (345.59,112.99) .. (353.33,112.98) .. controls (361.06,112.97) and (367.34,119.23) .. (367.35,126.97) .. controls (367.36,134.71) and (361.09,140.98) .. (353.36,140.99) .. controls (345.62,141) and (339.34,134.74) .. (339.33,127) -- cycle ;
    \draw    (270.33,128) -- (316.33,128) ;
    \draw    (344.33,116) -- (316.33,128) ;
    \draw    (316.33,128) -- (343.33,137) ;
    \draw    (408.33,128) .. controls (427.16,127.59) and (435.1,124) .. (441.54,119.9) .. controls (450.84,113.98) and (457.02,107) .. (488.33,107) ;
    \draw  [fill={rgb, 255:red, 155; green, 155; blue, 155 }  ,fill opacity=1 ] (463.33,126) .. controls (463.33,118.26) and (469.59,111.99) .. (477.33,111.98) .. controls (485.06,111.97) and (491.34,118.23) .. (491.35,125.97) .. controls (491.36,133.71) and (485.09,139.98) .. (477.36,139.99) .. controls (469.62,140) and (463.34,133.74) .. (463.33,126) -- cycle ;
    
    \draw (210,119.4) node [anchor=north west][inner sep=0.75pt]    {$=$};
    \draw (106,126.4) node [anchor=north west][inner sep=0.75pt]  [font=\scriptsize]  {$\mathnormal{x}$};
    \draw (245,116.4) node [anchor=north west][inner sep=0.75pt]  [font=\scriptsize]  {$\frac{1}{2!}$};
    \draw (266,127.4) node [anchor=north west][inner sep=0.75pt]  [font=\scriptsize]  {$\mathnormal{x}$};
    \draw (379,117.4) node [anchor=north west][inner sep=0.75pt]    {$+$};
    \draw (482,101.4) node [anchor=north west][inner sep=0.75pt]  [font=\scriptsize]  {$\times $};
    \draw (403,126.4) node [anchor=north west][inner sep=0.75pt]  [font=\scriptsize]  {$\mathnormal{x}$};

    \end{tikzpicture} \ .
\end{equation}

Just as we wrote down a functional differential equation for $Z[J]$ above, we can equally well write down a functional differential equation for multiple derivatives of $Z[J]$, i.e., 
\begin{equation}
    \prod_{i=1}^n \frac{1}{i} \frac{\delta}{\delta J(x_i)} Z[J] = \int [D \phi] \ \exp \left\{i \int \diff^{d+1}x  \left[-\frac{1}{2} \partial_\mu \phi \partial^\mu \phi -\frac{m^2}{2} \phi^2 -\frac{\lambda}{3!} \phi^3 + J \phi\right]\right\} \prod_{i=1}^{n} \phi(x_i) \ .
\end{equation}
The analogue of Eq.\eqref{eq:SDEoneptstep} takes the form
\begin{equation}
    \begin{split}
        &\int [D \phi] \ \exp \left\{i \int \diff^{d+1}x  \left[-\frac{1}{2} \partial_\mu \phi \partial^\mu \phi -\frac{m^2}{2} \phi^2 -\frac{\lambda}{3!} \phi^3 + J \phi\right]\right\} \prod_{i=1}^{n} \phi(x_i)\\
        &\qquad \times \int \diff^{d+1}x\ \epsilon(x)\left((\Box - m^2) \phi - \frac{\lambda}{2!} \phi^2 + J + \sum_{i=1}^n \delta^{d+1} (x-x_i) \prod_{j = 1 \atop j \neq i}^{n} \phi(x_j)\right)  = 0 \ .
    \end{split}
\end{equation}
As before, we can express the above equation as a functional differential equation of $Z[J]$ as follows:
\begin{equation}
    \begin{split}
        \frac{1}{i}\frac{\delta}{\delta J(x)}\prod_{i=1}^n \frac{1}{i} \frac{\delta}{\delta J(x_i)} Z[J] &= \frac{-i}{(-\Box + m^2)} \Bigg[\frac{-i\lambda}{2!} \left(\frac{1}{i} \frac{\delta}{\delta J(x)}\right)^2 \prod_{i=1}^n \frac{1}{i} \frac{\delta}{\delta J(x_i)}\\
        &\hspace{1.4cm} +i J(x) \prod_{i=1}^n \frac{1}{i} \frac{\delta}{\delta J(x_i)}+ \sum_{i=1}^n \delta^{d+1} (x-x_i) \prod_{j = 1 \atop j \neq i}^{n} \frac{1}{i} \frac{\delta}{\delta J(x_j)}\Bigg] Z[J] \ .
    \end{split}
\label{eq:SDEn+1pt}
\end{equation}
This is the $(n+1)$-point Schwinger-Dyson equation for $\phi^3$ theory. It is easy to see that this equation reduces to the one-point Schwinger-Dyson equation we had in Eq.\eqref{eq:SDE1pt} for $n=0$. Once again, there is a neat diagrammatic interpretation: 
\begin{equation}
    \tikzset{every picture/.style={line width=0.75pt}} 
    \begin{tikzpicture}[x=0.75pt,y=0.75pt,yscale=-1,xscale=1]
    \centering
    
    \draw    (65.33,64) -- (111.33,64) ;
    \draw  [fill={rgb, 255:red, 155; green, 155; blue, 155 }  ,fill opacity=1 ][line width=0.75]  (111.33,64) .. controls (111.33,56.26) and (117.59,49.99) .. (125.33,49.98) .. controls (133.06,49.97) and (139.34,56.23) .. (139.35,63.97) .. controls (139.36,71.71) and (133.09,77.98) .. (125.36,77.99) .. controls (117.62,78) and (111.34,71.74) .. (111.33,64) -- cycle ;
    \draw    (251.33,66) -- (297.33,66) ;
    \draw    (325.33,54) -- (297.33,66) ;
    \draw    (297.33,66) -- (324.33,75) ;
    \draw    (429.33,65) .. controls (448.16,64.59) and (456.1,61) .. (462.54,56.9) .. controls (468.98,52.8) and (471.02,31) .. (502.33,31) ;
    \draw    (132.33,51) -- (170.33,28) ;
    \draw    (131.33,77) -- (171.33,101) ;
    \draw  [fill={rgb, 255:red, 155; green, 155; blue, 155 }  ,fill opacity=1 ][line width=0.75]  (320.33,65) .. controls (320.33,57.26) and (326.59,50.99) .. (334.33,50.98) .. controls (342.06,50.97) and (348.34,57.23) .. (348.35,64.97) .. controls (348.36,72.71) and (342.09,78.98) .. (334.36,78.99) .. controls (326.62,79) and (320.34,72.74) .. (320.33,65) -- cycle ;
    \draw    (341.33,52) -- (366.14,36.99) -- (379.33,29) ;
    \draw    (340.33,78) -- (380.33,102) ;
    \draw  [fill={rgb, 255:red, 155; green, 155; blue, 155 }  ,fill opacity=1 ][line width=0.75]  (479.33,65) .. controls (479.33,57.26) and (485.59,50.99) .. (493.33,50.98) .. controls (501.06,50.97) and (507.34,57.23) .. (507.35,64.97) .. controls (507.36,72.71) and (501.09,78.98) .. (493.36,78.99) .. controls (485.62,79) and (479.34,72.74) .. (479.33,65) -- cycle ;
    \draw    (500.33,52) -- (538.33,29) ;
    \draw    (499.33,78) -- (539.33,102) ;
    \draw    (253.33,158) .. controls (258.48,157.89) and (262.87,157.32) .. (266.66,156.48) .. controls (270.46,155.63) and (282.65,149.98) .. (287.33,147) .. controls (292.02,144.02) and (331.02,121) .. (362.33,121) ;
    \draw  [fill={rgb, 255:red, 155; green, 155; blue, 155 }  ,fill opacity=1 ][line width=0.75]  (299.33,159) .. controls (299.33,151.26) and (305.59,144.99) .. (313.33,144.98) .. controls (321.06,144.97) and (327.34,151.23) .. (327.35,158.97) .. controls (327.36,166.71) and (321.09,172.98) .. (313.36,172.99) .. controls (305.62,173) and (299.34,166.74) .. (299.33,159) -- cycle ;
    \draw    (318.33,172) -- (358.33,196) ;
    \draw    (320.33,147) -- (360.33,137) ;
    \draw  [fill={rgb, 255:red, 155; green, 155; blue, 155 }  ,fill opacity=1 ][line width=0.75]  (475.33,160) .. controls (475.33,152.26) and (481.59,145.99) .. (489.33,145.98) .. controls (497.06,145.97) and (503.34,152.23) .. (503.35,159.97) .. controls (503.36,167.71) and (497.09,173.98) .. (489.36,173.99) .. controls (481.62,174) and (475.34,167.74) .. (475.33,160) -- cycle ;
    \draw    (494.33,173) -- (534.33,197) ;
    \draw    (496.33,148) -- (536.33,126) ;
    \draw    (431.33,167) .. controls (436.48,166.89) and (440.87,166.32) .. (444.66,165.48) .. controls (448.46,164.63) and (454.33,157) .. (460.33,151) .. controls (466.33,145) and (479.02,135) .. (510.33,135) ;
    \draw    (524.33,137) -- (537.33,138) ;
    \draw  [fill={rgb, 255:red, 155; green, 155; blue, 155 }  ,fill opacity=1 ][line width=0.75]  (389.33,242) .. controls (389.33,234.26) and (395.59,227.99) .. (403.33,227.98) .. controls (411.06,227.97) and (417.34,234.23) .. (417.35,241.97) .. controls (417.36,249.71) and (411.09,255.98) .. (403.36,255.99) .. controls (395.62,256) and (389.34,249.74) .. (389.33,242) -- cycle ;
    \draw    (410.33,229) -- (448.33,206) ;
    \draw    (409.33,255) -- (456.33,265) ;
    \draw    (343.33,248) .. controls (362.16,247.59) and (373.33,249) .. (389.33,260) .. controls (405.33,271) and (400.33,276) .. (456.33,282) ;
    
    \draw (191,57.4) node [anchor=north west][inner sep=0.75pt]    {$=$};
    \draw (58,63.4) node [anchor=north west][inner sep=0.75pt]  [font=\scriptsize]  {$\mathnormal{x}$};
    \draw (226,54.4) node [anchor=north west][inner sep=0.75pt]  [font=\scriptsize]  {$\frac{1}{2!}$};
    \draw (247,65.4) node [anchor=north west][inner sep=0.75pt]  [font=\scriptsize]  {$\mathnormal{x}$};
    \draw (394,54.4) node [anchor=north west][inner sep=0.75pt]    {$+$};
    \draw (497,25.4) node [anchor=north west][inner sep=0.75pt]  [font=\scriptsize]  {$\times $};
    \draw (424,63.4) node [anchor=north west][inner sep=0.75pt]  [font=\scriptsize]  {$\mathnormal{x}$};
    \draw (153,54.4) node [anchor=north west][inner sep=0.75pt]    {$\vdots $};
    \draw (362,55.4) node [anchor=north west][inner sep=0.75pt]    {$\vdots $};
    \draw (521,55.4) node [anchor=north west][inner sep=0.75pt]    {$\vdots $};
    \draw (173,18.4) node [anchor=north west][inner sep=0.75pt]  [font=\scriptsize]  {$x_{1}$};
    \draw (174,98.4) node [anchor=north west][inner sep=0.75pt]  [font=\scriptsize]  {$x_{n}$};
    \draw (380,21.4) node [anchor=north west][inner sep=0.75pt]  [font=\scriptsize]  {$x_{1}$};
    \draw (540,22.4) node [anchor=north west][inner sep=0.75pt]  [font=\scriptsize]  {$x_{1}$};
    \draw (383,97.4) node [anchor=north west][inner sep=0.75pt]  [font=\scriptsize]  {$x_{n}$};
    \draw (542,97.4) node [anchor=north west][inner sep=0.75pt]  [font=\scriptsize]  {$x_{n}$};
    \draw (226,143.4) node [anchor=north west][inner sep=0.75pt]    {$+$};
    \draw (337,156.4) node [anchor=north west][inner sep=0.75pt]    {$\vdots $};
    \draw (360,192.4) node [anchor=north west][inner sep=0.75pt]  [font=\scriptsize]  {$x_{n}$};
    \draw (363,131.4) node [anchor=north west][inner sep=0.75pt]  [font=\scriptsize]  {$x_{2}$};
    \draw (364,116.4) node [anchor=north west][inner sep=0.75pt]  [font=\scriptsize]  {$x_{1}$};
    \draw (246,158.4) node [anchor=north west][inner sep=0.75pt]  [font=\scriptsize]  {$\mathnormal{x}$};
    \draw (394,143.4) node [anchor=north west][inner sep=0.75pt]    {$+$};
    \draw (513,157.4) node [anchor=north west][inner sep=0.75pt]    {$\vdots $};
    \draw (536,193.4) node [anchor=north west][inner sep=0.75pt]  [font=\scriptsize]  {$x_{n}$};
    \draw (539,132.4) node [anchor=north west][inner sep=0.75pt]  [font=\scriptsize]  {$x_{2}$};
    \draw (540,119.4) node [anchor=north west][inner sep=0.75pt]  [font=\scriptsize]  {$x_{1}$};
    \draw (424,166.4) node [anchor=north west][inner sep=0.75pt]  [font=\scriptsize]  {$\mathnormal{x}$};
    \draw (228,234.4) node [anchor=north west][inner sep=0.75pt]    {$+$};
    \draw (336,241.4) node [anchor=north west][inner sep=0.75pt]  [font=\scriptsize]  {$\mathnormal{x}$};
    \draw (431,232.4) node [anchor=north west][inner sep=0.75pt]    {$\vdots $};
    \draw (451,196.4) node [anchor=north west][inner sep=0.75pt]  [font=\scriptsize]  {$x_{1}$};
    \draw (457,263.4) node [anchor=north west][inner sep=0.75pt]  [font=\scriptsize]  {$x_{n-1}$};
    \draw (311,233.4) node [anchor=north west][inner sep=0.75pt]    {$+$};
    \draw (264,231.4) node [anchor=north west][inner sep=0.75pt]    {$\dotsc $};
    \draw (458,277.4) node [anchor=north west][inner sep=0.75pt]  [font=\scriptsize]  {$x_{n}$};
    \end{tikzpicture} \ .
\end{equation}

The reader might have noticed that the generalisation to any scalar theory is quite straightforward. For example, $\phi^k$ theory will introduce a $k$-point vertex in the above equation.

The above Schwinger-Dyson equations for the generating function $Z[J]$ can be recast into similar equations for the generating function of connected correlators $W[J]$ defined as
\begin{equation}
    Z[J] \equiv e^{i W[J]} \ .
\end{equation}
All we need to do to derive these equations is to substitute the above definition into the $(n+1)$-point Schwinger-Dyson equation in Eq.\eqref{eq:SDEn+1pt}. Doing this for $n=0,1,2$, we obtain:
\begin{equation}
    \begin{split}
         \frac{1}{i}\frac{\delta}{\delta J(x)} iW[J] = \frac{-i}{(-\Box + m^2)} \left[\frac{-i\lambda}{2!} \left(\frac{1}{i} \frac{\delta}{\delta J(x)}\right)^2 iW[J] -\frac{i \lambda}{2!} \left(\frac{1}{i} \frac{\delta}{\delta J(x)} iW[J] \right)^2 +i J(x) \right]\ ,
    \end{split}
\end{equation}
\begin{equation}
    \begin{split}
         \frac{1}{i}\frac{\delta}{\delta J(x)} \frac{1}{i}\frac{\delta}{\delta J(x_1)} iW[J] &= \frac{-i}{(-\Box + m^2)} \Bigg[\frac{-i\lambda}{2!} \left(\frac{1}{i} \frac{\delta}{\delta J(x)}\right)^2 \frac{1}{i}\frac{\delta}{\delta J(x_1)} iW[J]\\
         &\hspace{1.5cm} - i \lambda \frac{1}{i} \frac{\delta}{\delta J(x)} iW[J] \frac{1}{i} \frac{\delta}{\delta J(x)}  \frac{1}{i}\frac{\delta}{\delta J(x_1)} iW[J] + \delta^4(x-x_1)\Bigg]\ ,
    \end{split}
\end{equation}
\begin{equation}
    \begin{split}
        \frac{1}{i}\frac{\delta}{\delta J(x)} \frac{1}{i}\frac{\delta}{\delta J(x_1)} \frac{1}{i}\frac{\delta}{\delta J(x_2)} iW[J] &= \frac{-i}{(-\Box + m^2)} \Bigg[ \frac{-i\lambda}{2!} \left(\frac{1}{i} \frac{\delta}{\delta J(x)}\right)^2 \frac{1}{i}\frac{\delta}{\delta J(x_1)} \frac{1}{i}\frac{\delta}{\delta J(x_2)} iW[J]\\
        &\qquad-i\lambda \frac{1}{i} \frac{\delta}{\delta J(x)}  \frac{1}{i}\frac{\delta}{\delta J(x_1)} iW[J] \frac{1}{i} \frac{\delta}{\delta J(x)}  \frac{1}{i}\frac{\delta}{\delta J(x_2)} iW[J]\\
        &\qquad- i \lambda \frac{1}{i} \frac{\delta}{\delta J(x)} iW[J]  \frac{1}{i} \frac{\delta}{\delta J(x)}  \frac{1}{i}\frac{\delta}{\delta J(x_1)} \frac{1}{i}\frac{\delta}{\delta J(x_2)} iW[J] \Bigg] \ .
    \end{split}
\end{equation}

\begin{equation}
    \begin{split}
        \frac{1}{i}\frac{\delta}{\delta J(x)} \prod_{i=1}^{3}\frac{1}{i}\frac{\delta}{\delta J(x_i)}  iW[J] &= \frac{-i}{(-\Box + m^2)} \Bigg[ \frac{-i\lambda}{2!} \left(\frac{1}{i} \frac{\delta}{\delta J(x)}\right)^2  \prod_{i=1}^{3}\frac{1}{i}\frac{\delta}{\delta J(x_i)}  iW[J]\\
        &\qquad-i\lambda \frac{1}{i} \frac{\delta}{\delta J(x)}  \frac{1}{i}\frac{\delta}{\delta J(x_1)} \frac{1}{i}\frac{\delta}{\delta J(x_3)}  iW[J] \frac{1}{i} \frac{\delta}{\delta J(x)}  \frac{1}{i}\frac{\delta}{\delta J(x_2)} iW[J]\\
        &\qquad-i\lambda \frac{1}{i} \frac{\delta}{\delta J(x)}  \frac{1}{i}\frac{\delta}{\delta J(x_1)} iW[J] \frac{1}{i} \frac{\delta}{\delta J(x)}  \frac{1}{i}\frac{\delta}{\delta J(x_2)} \frac{1}{i}\frac{\delta}{\delta J(x_3)}  iW[J]\\
        &\qquad- i \lambda \frac{1}{i} \frac{\delta}{\delta J(x)} \frac{1}{i}\frac{\delta}{\delta J(x_3)}  iW[J]  \frac{1}{i} \frac{\delta}{\delta J(x)}  \frac{1}{i}\frac{\delta}{\delta J(x_1)} \frac{1}{i}\frac{\delta}{\delta J(x_2)} iW[J] \\
        &\qquad- i \lambda \frac{1}{i} \frac{\delta}{\delta J(x)}   iW[J]   \prod_{i=1}^{3}\frac{1}{i}\frac{\delta}{\delta J(x_i)} iW[J]\Bigg] \ .
    \end{split}
\end{equation}
Here, too, there is a clear diagrammatic interpretation. Since $W[J]$ is the generator of connected correlators, only connected diagrams enter into the equation. Another major difference is that the equation is not linear in $W[J]$, and thus the diagrams can have multiple $W[J]$ blobs.
\begin{equation}
    \tikzset{every picture/.style={line width=0.75pt}} 
 \ .
\end{equation}
It is clear from the above diagrammatic equation how the Schwinger-Dyson equation generalises to higher $n$, as well as for $\phi^k$ theories.

\subsection{Open \texorpdfstring{$\phi^3$}{p3} theory}

Now that we understand how the Schwinger-Dyson equations are derived in the simple case of closed scalar quantum field theories, we would like to move to the case of open quantum field theories. Open quantum field theories can be efficiently and neatly understood using the Schwinger-Keldysh formalism, which we will employ here. Therefore, the reader will notice below that there are two-copies of the fields as well as the sources. 

Since there is more than one field, we can choose to work in many different bases of the field and the sources. With a view towards our eventual goal, here we will work with the causal (past-future) basis, which is defined by 
\begin{alignat*}{4}
        &\phi_{\Psmall}(k) &&= -[\phi_{\sR}(k)-\phi_{\sL}(k)] \ , \qquad &&\phi_{\Fsmall}(k)&&= \phi_{\sR}(k) - \frac{\nbe_k}{1+\nbe_k} \phi_{\sL}(k) \ ,\\
        &J_{\Pb}(k) &&= -\nbe_k [ J_{\sR}(k) -J_{\sL}(k)]\ , \qquad &&J_{\Fb}(k) &&= -(1+\nbe_k) J_{\sR}(k) + \nbe_k J_{\sL}(k) \ .
\end{alignat*}
for fields $\phi_{\sR}$ and $\phi_{\sL}$ and sources $J_{\sR}$ and $J_{\sL}$. Here $k$ is the Fourier conjugate to the spacetime position $x$, and $\nbe_k$ denotes the Bose-Einstein defined by
\begin{equation}
    \nbe_k = \frac{1}{e^{\beta k^0}-1} \ .
\end{equation}

Once again, we will begin our analysis with the simplest model: open $\phi^3$ theory. The Schwinger-Keldysh path integral for this theory (written in the causal basis of fields and sources) takes the form
\begin{equation}
    \begin{split}
        Z_{\rm SK}[J_{\Pb},J_{\Fb}] &= \int [D \phi_{\Psmall}] [D \phi_{\Fsmall}] \ \exp \Bigg\{ i  S_{\rm SK}[\phi_{\Psmall}, \phi_{\Fsmall}, J_{\Pb}, J_{\Fb}]  \Bigg\} \ ,
    \end{split}
\end{equation}
where
\begin{equation}
    \begin{split}
        S_{\rm SK}[\phi_{\Psmall}, \phi_{\Fsmall}, J_{\Pb}, J_{\Fb}] &\equiv - \int_p \phi_{\Psmall}(-p) G^{-1}(p) \phi_{\Fsmall}(p) + \int_p [J_{\Pb}(-p) \phi_{\Fsmall}(p)+ J_{\Fb}(-p) \phi_{\Psmall}(p)]\\
        &-\frac{1}{2!} \int_{p_{1,2,3}}{(2\pi)^{d+1}} \delta^{d+1} (p_1+p_2+p_3)\\
        &\times \left[\lambda_{\Fsmall \Fsmall \Psmall}(p_1,p_2,p_3) \phi_{\Fsmall}(p_1) \phi_{\Fsmall}(p_2)  + \lambda_{\Fsmall \Psmall \Psmall}(p_1,p_2,p_3) \phi_{\Fsmall}(p_1) \phi_{\Psmall}(p_2)  \right]\phi_{\Psmall}(p_3) \ .
    \end{split}
\label{eq:ActionSKwithsources}
\end{equation}
Here we have introduced the useful notation for momentum integrals:
\begin{equation}
     \int_{p_{1, \ldots, n}} \equiv \int \frac{\diff^{d+1} p_1}{(2\pi)^{d+1}} \ldots \frac{\diff^{d+1} p_n}{(2\pi)^{d+1}} \ .
\end{equation}

The Schwinger-Dyson equations can now be derived as we did before by varying the path integration variables. Already, we see that we now have two sets of Schwinger-Dyson equations corresponding to differentiation with respect to the past source and the future source. We will begin by deriving the one-point Schwinger-Dyson equation and then moving on to the $(n+1)$-point ones. 

Performing the dummy variable shifts $\phi_{\Psmall}(p) \mapsto \phi_{\Psmall}(p) + \epsilon_{\Psmall}(p)$ and $\phi_{\Fsmall}(p) \mapsto \phi_{\Fsmall}(p) + \epsilon_{\Fsmall}(p)$ on the RHS of the path integral above, we obtain at first-order in the shifts
\begin{equation}
    \begin{split}
        &\int [D \phi_{\Psmall}] [D \phi_{\Fsmall}] \ \exp \Bigg\{ iS_{\rm SK}[\phi_{\Psmall}, \phi_{\Fsmall}, J_{\Pb}, J_{\Fb}] \Bigg\} \int_{p_1} \Bigg\{\epsilon_{\Psmall}(p_1) \Bigg[- G^{-1}(-p_1) \phi_{\Fsmall}(-p_1) + J_{\Fb} (-p_1)\\
        &-\int_{p_{2,3}}(2\pi)^{d+1} \delta^{d+1}(p_1+p_2+p_3)\\
        &\qquad\times \Bigg(\frac{1}{2!} \lambda_{\Fsmall \Fsmall \Psmall}(p_2,p_3,p_1) \phi_{\Fsmall}(p_2) \phi_{\Fsmall}(p_3) +  \lambda_{\Fsmall \Psmall \Psmall}(p_2,p_1,p_3) \phi_{\Fsmall}(p_2) \phi_{\Psmall}(p_3)\Bigg)\Bigg]  \Bigg\} = 0 \ ,
    \end{split}
\end{equation}
\begin{equation}
    \begin{split}
        &\int [D \phi_{\Psmall}] [D \phi_{\Fsmall}] \ \exp \Bigg\{ iS_{\rm SK}[\phi_{\Psmall}, \phi_{\Fsmall}, J_{\Pb}, J_{\Fb}] \Bigg\} \int_{p_1} \Bigg\{\epsilon_{\Fsmall}(p_1) \Bigg[ -G^{-1}(p_1) \phi_{\Psmall}(-p_1) + J_{\Pb} (-p_1)\\
        &-\int_{p_{2,3}}(2\pi)^{d+1} \delta^{d+1}(p_1+p_2+p_3)\\
        &\qquad\times \Bigg(\lambda_{\Fsmall \Fsmall \Psmall}(p_1,p_2,p_3) \phi_{\Fsmall}(p_2) \phi_{\Psmall}(p_3) +  \frac{1}{2!} \lambda_{\Fsmall \Psmall \Psmall}(p_1,p_2,p_3) \phi_{\Psmall}(p_2) \phi_{\Psmall}(p_3)\Bigg)  \Bigg]\Bigg\} = 0 \ .
    \end{split}
\end{equation}
Since these equations hold for arbitrary shifts $\epsilon_{\Psmall}$ and $\epsilon_{\Fsmall}$, we have the following Schwinger-Dyson equations that the Schwinger-Keldysh generating function $Z_{\rm SK}[J_{\Pb}, J_{\Fb}]$ satisfies:
\begin{equation}
    \begin{split}
        &\frac{1}{i}\frac{\delta}{\delta J_{\Pb}(p_1)}Z_{\rm SK} = -i G(-p_1) \Bigg[i J_{\Fb} (-p_1)-i \int_{p_{2,3}}(2\pi)^{d+1} \delta^{d+1}(p_1+p_2+p_3)\\
        &\hspace{1cm}\times \Bigg(\frac{1}{2!} \lambda_{\Fsmall \Fsmall \Psmall}(p_2,p_3,p_1) \frac{1}{i}\frac{\delta}{\delta J_{\Pb}(-p_3)} +  \lambda_{\Fsmall \Psmall \Psmall}(p_2,p_1,p_3) \frac{1}{i}\frac{\delta}{\delta J_{\Fb}(-p_3)}\Bigg)\frac{1}{i}\frac{\delta}{\delta J_{\Pb}(-p_2)}\Bigg] Z_{\rm SK}\ ,
    \end{split}
\end{equation}
\begin{equation}
    \begin{split}
        &\frac{1}{i}\frac{\delta}{\delta J_{\Fb}(p_1)}Z_{\rm SK} = - i G(p_1) \Bigg[i J_{\Pb} (-p_1)-i \int_{p_{2,3}}(2\pi)^{d+1} \delta^{d+1}(p_1+p_2+p_3)\\
        &\hspace{1cm}\times \Bigg(\lambda_{\Fsmall \Fsmall \Psmall}(p_1,p_2,p_3) \frac{1}{i}\frac{\delta}{\delta J_{\Pb}(-p_2)} +  \frac{1}{2!} \lambda_{\Fsmall \Psmall \Psmall}(p_1,p_2,p_3) \frac{1}{i}\frac{\delta}{\delta J_{\Fb}(-p_2)}\Bigg)\frac{1}{i}\frac{\delta}{\delta J_{\Fb}(-p_3)} \Bigg] Z_{\rm SK} \ .
    \end{split}
\end{equation}
As is very familiar by now, these equations also have a neat diagrammatic interpretation:
\begin{equation}
    \tikzset{every picture/.style={line width=0.75pt}} 
    \begin{tikzpicture}[x=0.75pt,y=0.75pt,yscale=-1,xscale=1]
    \centering
    
    \draw    (69,127) -- (126.33,127) ;
    \draw [shift={(129.33,127)}, rotate = 180] [fill={rgb, 255:red, 0; green, 0; blue, 0 }  ][line width=0.08]  [draw opacity=0] (8.93,-4.29) -- (0,0) -- (8.93,4.29) -- cycle    ;
    \draw  [fill={rgb, 255:red, 155; green, 155; blue, 155 }  ,fill opacity=1 ][line width=0.75]  (129.33,127) .. controls (129.33,119.26) and (135.59,112.99) .. (143.33,112.98) .. controls (151.06,112.97) and (157.34,119.23) .. (157.35,126.97) .. controls (157.36,134.71) and (151.09,140.98) .. (143.36,140.99) .. controls (135.62,141) and (129.34,134.74) .. (129.33,127) -- cycle ;
    \draw  [fill={rgb, 255:red, 155; green, 155; blue, 155 }  ,fill opacity=1 ] (297.33,127) .. controls (297.33,119.26) and (303.59,112.99) .. (311.33,112.98) .. controls (319.06,112.97) and (325.34,119.23) .. (325.35,126.97) .. controls (325.36,134.71) and (319.09,140.98) .. (311.36,140.99) .. controls (303.62,141) and (297.34,134.74) .. (297.33,127) -- cycle ;
    \draw    (228.33,128) -- (247.33,128.29) ;
    \draw [shift={(250.33,128.33)}, rotate = 180.87] [fill={rgb, 255:red, 0; green, 0; blue, 0 }  ][line width=0.08]  [draw opacity=0] (8.93,-4.29) -- (0,0) -- (8.93,4.29) -- cycle    ;
    \draw    (515.33,131) .. controls (531.81,130.64) and (539.94,127.85) .. (546.04,124.41) ;
    \draw [shift={(548.54,122.9)}, rotate = 147.52] [fill={rgb, 255:red, 0; green, 0; blue, 0 }  ][line width=0.08]  [draw opacity=0] (8.93,-4.29) -- (0,0) -- (8.93,4.29) -- cycle    ;
    \draw  [fill={rgb, 255:red, 155; green, 155; blue, 155 }  ,fill opacity=1 ] (570.33,129) .. controls (570.33,121.26) and (576.59,114.99) .. (584.33,114.98) .. controls (592.06,114.97) and (598.34,121.23) .. (598.35,128.97) .. controls (598.36,136.71) and (592.09,142.98) .. (584.36,142.99) .. controls (576.62,143) and (570.34,136.74) .. (570.33,129) -- cycle ;
    \draw    (250.33,128.33) -- (265.33,127.86) -- (274.33,128) ;
    \draw [shift={(250.33,128.33)}, rotate = 178.21] [color={rgb, 255:red, 0; green, 0; blue, 0 }  ][line width=0.75]    (0,5.59) -- (0,-5.59)   ;
    \draw    (274.33,128) .. controls (277.2,120.68) and (287.36,101.2) .. (302.2,112.52) ;
    \draw [shift={(304.33,114.33)}, rotate = 223.15] [fill={rgb, 255:red, 0; green, 0; blue, 0 }  ][line width=0.08]  [draw opacity=0] (8.93,-4.29) -- (0,0) -- (8.93,4.29) -- cycle    ;
    \draw    (274.33,128) .. controls (284.07,150.42) and (296.94,147.52) .. (303.23,142.39) ;
    \draw [shift={(305.33,140.33)}, rotate = 129.81] [fill={rgb, 255:red, 0; green, 0; blue, 0 }  ][line width=0.08]  [draw opacity=0] (8.93,-4.29) -- (0,0) -- (8.93,4.29) -- cycle    ;
    \draw  [fill={rgb, 255:red, 155; green, 155; blue, 155 }  ,fill opacity=1 ] (439.33,127) .. controls (439.33,119.26) and (445.59,112.99) .. (453.33,112.98) .. controls (461.06,112.97) and (467.34,119.23) .. (467.35,126.97) .. controls (467.36,134.71) and (461.09,140.98) .. (453.36,140.99) .. controls (445.62,141) and (439.34,134.74) .. (439.33,127) -- cycle ;
    \draw    (370.33,128) -- (389.33,128.29) ;
    \draw [shift={(392.33,128.33)}, rotate = 180.87] [fill={rgb, 255:red, 0; green, 0; blue, 0 }  ][line width=0.08]  [draw opacity=0] (8.93,-4.29) -- (0,0) -- (8.93,4.29) -- cycle    ;
    \draw    (392.33,128.33) -- (407.33,127.86) -- (416.33,128) ;
    \draw [shift={(392.33,128.33)}, rotate = 178.21] [color={rgb, 255:red, 0; green, 0; blue, 0 }  ][line width=0.75]    (0,5.59) -- (0,-5.59)   ;
    \draw    (416.33,128) .. controls (419.33,120.33) and (438.33,98.33) .. (446.33,114.33) ;
    \draw [shift={(446.33,114.33)}, rotate = 243.43] [color={rgb, 255:red, 0; green, 0; blue, 0 }  ][line width=0.75]    (0,5.59) -- (0,-5.59)   ;
    \draw    (416.33,128) .. controls (426.07,150.42) and (438.94,147.52) .. (445.23,142.39) ;
    \draw [shift={(447.33,140.33)}, rotate = 129.81] [fill={rgb, 255:red, 0; green, 0; blue, 0 }  ][line width=0.08]  [draw opacity=0] (8.93,-4.29) -- (0,0) -- (8.93,4.29) -- cycle    ;
    \draw    (548.54,122.9) .. controls (556.33,118.33) and (553.33,109.33) .. (596.33,110.33) ;
    \draw [shift={(548.54,122.9)}, rotate = 149.64] [color={rgb, 255:red, 0; green, 0; blue, 0 }  ][line width=0.75]    (0,5.59) -- (0,-5.59)   ;
    
    \draw (168,119.4) node [anchor=north west][inner sep=0.75pt]    {$=$};
    \draw (64,126.4) node [anchor=north west][inner sep=0.75pt]  [font=\scriptsize]  {$\mathnormal{x}$};
    \draw (203,116.4) node [anchor=north west][inner sep=0.75pt]  [font=\scriptsize]  {$\frac{1}{2!}$};
    \draw (224,127.4) node [anchor=north west][inner sep=0.75pt]  [font=\scriptsize]  {$\mathnormal{x}$};
    \draw (337,117.4) node [anchor=north west][inner sep=0.75pt]    {$+$};
    \draw (589,104.4) node [anchor=north west][inner sep=0.75pt]  [font=\scriptsize]  {$\times $};
    \draw (510,129.4) node [anchor=north west][inner sep=0.75pt]  [font=\scriptsize]  {$\mathnormal{x}$};
    \draw (484,117.4) node [anchor=north west][inner sep=0.75pt]    {$+$};
    \draw (366,127.4) node [anchor=north west][inner sep=0.75pt]  [font=\scriptsize]  {$\mathnormal{x}$};

    \end{tikzpicture} \ ,
\end{equation}
\begin{equation}
    \tikzset{every picture/.style={line width=0.75pt}} 
    \begin{tikzpicture}[x=0.75pt,y=0.75pt,yscale=-1,xscale=1]
    \centering
    
    \draw    (67,134.67) -- (127.33,134.67) ;
    \draw [shift={(127.33,134.67)}, rotate = 180] [color={rgb, 255:red, 0; green, 0; blue, 0 }  ][line width=0.75]    (0,5.59) -- (0,-5.59)   ;
    \draw  [fill={rgb, 255:red, 155; green, 155; blue, 155 }  ,fill opacity=1 ][line width=0.75]  (127.33,134.67) .. controls (127.33,126.93) and (133.59,120.65) .. (141.33,120.64) .. controls (149.06,120.64) and (155.34,126.9) .. (155.35,134.64) .. controls (155.36,142.37) and (149.09,148.65) .. (141.36,148.66) .. controls (133.62,148.67) and (127.34,142.4) .. (127.33,134.67) -- cycle ;
    \draw  [fill={rgb, 255:red, 155; green, 155; blue, 155 }  ,fill opacity=1 ] (440.33,133.67) .. controls (440.33,125.93) and (446.59,119.65) .. (454.33,119.64) .. controls (462.06,119.64) and (468.34,125.9) .. (468.35,133.64) .. controls (468.36,141.37) and (462.09,147.65) .. (454.36,147.66) .. controls (446.62,147.67) and (440.34,141.4) .. (440.33,133.67) -- cycle ;
    \draw    (371.33,134.67) -- (393.33,135) ;
    \draw [shift={(393.33,135)}, rotate = 180.87] [color={rgb, 255:red, 0; green, 0; blue, 0 }  ][line width=0.75]    (0,5.59) -- (0,-5.59)   ;
    \draw    (513.33,138.67) .. controls (532.16,138.26) and (540.1,134.67) .. (546.54,130.56) ;
    \draw [shift={(546.54,130.56)}, rotate = 147.52] [color={rgb, 255:red, 0; green, 0; blue, 0 }  ][line width=0.75]    (0,5.59) -- (0,-5.59)   ;
    \draw  [fill={rgb, 255:red, 155; green, 155; blue, 155 }  ,fill opacity=1 ] (568.33,136.67) .. controls (568.33,128.93) and (574.59,122.65) .. (582.33,122.64) .. controls (590.06,122.64) and (596.34,128.9) .. (596.35,136.64) .. controls (596.36,144.37) and (590.09,150.65) .. (582.36,150.66) .. controls (574.62,150.67) and (568.34,144.4) .. (568.33,136.67) -- cycle ;
    \draw    (396.33,134.91) -- (408.33,134.53) -- (417.33,134.67) ;
    \draw [shift={(393.33,135)}, rotate = 358.21] [fill={rgb, 255:red, 0; green, 0; blue, 0 }  ][line width=0.08]  [draw opacity=0] (8.93,-4.29) -- (0,0) -- (8.93,4.29) -- cycle    ;
    \draw    (417.33,134.67) .. controls (420.33,127) and (435.33,103) .. (447.33,121) ;
    \draw [shift={(447.33,121)}, rotate = 236.31] [color={rgb, 255:red, 0; green, 0; blue, 0 }  ][line width=0.75]    (0,5.59) -- (0,-5.59)   ;
    \draw    (417.33,134.67) .. controls (428.33,160) and (444.33,157) .. (448.33,147) ;
    \draw [shift={(448.33,147)}, rotate = 111.8] [color={rgb, 255:red, 0; green, 0; blue, 0 }  ][line width=0.75]    (0,5.59) -- (0,-5.59)   ;
    \draw  [fill={rgb, 255:red, 155; green, 155; blue, 155 }  ,fill opacity=1 ] (278.33,133.67) .. controls (278.33,125.93) and (284.59,119.65) .. (292.33,119.64) .. controls (300.06,119.64) and (306.34,125.9) .. (306.35,133.64) .. controls (306.36,141.37) and (300.09,147.65) .. (292.36,147.66) .. controls (284.62,147.67) and (278.34,141.4) .. (278.33,133.67) -- cycle ;
    \draw    (209.33,134.67) -- (231.33,135) ;
    \draw [shift={(231.33,135)}, rotate = 180.87] [color={rgb, 255:red, 0; green, 0; blue, 0 }  ][line width=0.75]    (0,5.59) -- (0,-5.59)   ;
    \draw    (234.33,134.91) -- (246.33,134.53) -- (255.33,134.67) ;
    \draw [shift={(231.33,135)}, rotate = 358.21] [fill={rgb, 255:red, 0; green, 0; blue, 0 }  ][line width=0.08]  [draw opacity=0] (8.93,-4.29) -- (0,0) -- (8.93,4.29) -- cycle    ;
    \draw    (255.33,134.67) .. controls (258.2,127.34) and (272.01,105.13) .. (283.7,118.82) ;
    \draw [shift={(285.33,121)}, rotate = 236.31] [fill={rgb, 255:red, 0; green, 0; blue, 0 }  ][line width=0.08]  [draw opacity=0] (8.93,-4.29) -- (0,0) -- (8.93,4.29) -- cycle    ;
    \draw    (255.33,134.67) .. controls (266.33,160) and (282.33,157) .. (286.33,147) ;
    \draw [shift={(286.33,147)}, rotate = 111.8] [color={rgb, 255:red, 0; green, 0; blue, 0 }  ][line width=0.75]    (0,5.59) -- (0,-5.59)   ;
    \draw    (549.12,128.8) .. controls (566.73,117.01) and (571.37,120.53) .. (593.33,117.67) ;
    \draw [shift={(546.54,130.56)}, rotate = 324.92] [fill={rgb, 255:red, 0; green, 0; blue, 0 }  ][line width=0.08]  [draw opacity=0] (8.93,-4.29) -- (0,0) -- (8.93,4.29) -- cycle    ;
    
    \draw (166,127.07) node [anchor=north west][inner sep=0.75pt]    {$=$};
    \draw (62,134.07) node [anchor=north west][inner sep=0.75pt]  [font=\scriptsize]  {$\mathnormal{x}$};
    \draw (346,123.07) node [anchor=north west][inner sep=0.75pt]  [font=\scriptsize]  {$\frac{1}{2!}$};
    \draw (367,134.07) node [anchor=north west][inner sep=0.75pt]  [font=\scriptsize]  {$\mathnormal{x}$};
    \draw (335,125.07) node [anchor=north west][inner sep=0.75pt]    {$+$};
    \draw (587,112.07) node [anchor=north west][inner sep=0.75pt]  [font=\scriptsize]  {$\times $};
    \draw (508,137.07) node [anchor=north west][inner sep=0.75pt]  [font=\scriptsize]  {$\mathnormal{x}$};
    \draw (482,125.07) node [anchor=north west][inner sep=0.75pt]    {$+$};
    \draw (205,134.07) node [anchor=north west][inner sep=0.75pt]  [font=\scriptsize]  {$\mathnormal{x}$};

    \end{tikzpicture} \ .
\end{equation}

We will now derive the $(n+1)$-point Schwinger-Dyson equation for open $\phi^3$ theory in the causal basis. To this end, we start by performing the dummy variable shift in the path integral
\begin{equation}
    \begin{split}
        &\prod_{l=1}^p \frac{1}{i} \frac{\delta}{\delta J_{\Fb}(\bar{k}_l)} \prod_{m=p+1}^n \frac{1}{i} \frac{\delta}{\delta J_{\Pb}(\bar{k}_m)} Z_{\rm SK}[J_{\Pb},J_{\Fb}]\\
        &\hspace{2cm} = \int [D \phi_{\Psmall}] [D \phi_{\Fsmall}] \ \exp \Bigg\{ i  S_{\rm SK}[\phi_{\Psmall}, \phi_{\Fsmall}, J_{\Pb}, J_{\Fb}]  \Bigg\} \prod_{l=1}^p \phi_{\Psmall}(k_l) \prod_{m=p+1}^n \phi_{\Fsmall}(k_m) \ .
    \end{split}
\end{equation}
Using similar arguments as in the $n=0$ case above, we obtain the two $(n+1)$-point Schwinger-Dyson equations to be
\begin{equation}
    \begin{split}
        &\frac{1}{i} \frac{\delta}{\delta J_{\Pb}(\bar{k})} \prod_{l=1}^p \frac{1}{i} \frac{\delta}{\delta J_{\Fb}(\bar{k}_l)} \prod_{m=p+1}^n \frac{1}{i} \frac{\delta}{\delta J_{\Pb}(\bar{k}_m)} Z_{\rm SK}\\
        &= -  i G(\bar{k}) \Bigg\{i J_{\Fb} (\bar{k}) \prod_{l=1}^p \frac{1}{i} \frac{\delta}{\delta J_{\Fb}(\bar{k}_l)}+ \sum_{r=1}^{p} (2\pi)^{d+1}\delta^{d+1}(k+k_r) \prod_{l=1 \atop l \neq r}^p \frac{1}{i} \frac{\delta}{\delta J_{\Fb}(\bar{k}_l)}\\
        &-i \int_{p_{2,3}}(2\pi)^{d+1} \delta^{d+1}(k+p_2+p_3) \Bigg[\frac{1}{2!} \lambda_{\Fsmall \Fsmall \Psmall}(p_2,p_3,k) \frac{1}{i}\frac{\delta}{\delta J_{\Pb}(\bar{p}_3)} +  \lambda_{\Fsmall \Psmall \Psmall}(p_2,k,p_3) \frac{1}{i}\frac{\delta}{\delta J_{\Fb}(\bar{p}_3)}\Bigg]\\
        &\hspace{6cm}\times \frac{1}{i}\frac{\delta}{\delta J_{\Pb}(\bar{p}_2)} \prod_{l=1}^p \frac{1}{i} \frac{\delta}{\delta J_{\Fb}(\bar{k}_l)}  \Bigg\}  \prod_{m=p+1}^n \frac{1}{i} \frac{\delta}{\delta J_{\Pb}(\bar{k}_m)} Z_{\rm SK}\ ,
    \end{split}
\end{equation}
\begin{equation}
    \begin{split}
        &\frac{1}{i} \frac{\delta}{\delta J_{\Fb}(\bar{k})} \prod_{l=1}^p \frac{1}{i} \frac{\delta}{\delta J_{\Fb}(\bar{k}_l)} \prod_{m=p+1}^n \frac{1}{i} \frac{\delta}{\delta J_{\Pb}(\bar{k}_m)} Z_{\rm SK}\\
        &= -i G(k) \Bigg\{i J_{\Pb} (\bar{k}) \prod_{m=p+1}^n \frac{1}{i} \frac{\delta}{\delta J_{\Pb}(\bar{k}_m)}+ \sum_{r=p+1}^{n} (2\pi)^{d+1}\delta^{d+1}(k+k_r) \prod_{m=p+1 \atop m \neq r}^n \frac{1}{i} \frac{\delta}{\delta J_{\Pb}(\bar{k}_m)}\\
        &-i \int_{p_{2,3}}(2\pi)^{d+1} \delta^{d+1}(k+p_2+p_3) \Bigg[\lambda_{\Fsmall \Fsmall \Psmall}(p_1,p_2,p_3) \frac{1}{i}\frac{\delta}{\delta J_{\Pb}(\bar{p}_2)} +  \frac{1}{2!} \lambda_{\Fsmall \Psmall \Psmall}(p_1,p_2,p_3) \frac{1}{i}\frac{\delta}{\delta J_{\Fb}(\bar{p}_2)}\Bigg]\\
        &\hspace{6cm}\times \frac{1}{i}\frac{\delta}{\delta J_{\Fb}(\bar{p}_2)}   \prod_{m=p+1}^n \frac{1}{i} \frac{\delta}{\delta J_{\Pb}(\bar{k}_m)}\Bigg\} \prod_{l=1 }^p \frac{1}{i} \frac{\delta}{\delta J_{\Fb}(\bar{k}_l)}  Z_{\rm SK}\ ,
    \end{split}
\end{equation}
where for ease of notation, we have defined $\bar{k} \equiv -k$. The diagrammatic versions of the above equations take the form:
\begin{equation}
    \tikzset{every picture/.style={line width=0.75pt}} 
    \begin{tikzpicture}[x=0.75pt,y=0.75pt,yscale=-0.9,xscale=0.9]
    \centering
    
    \draw    (10.33,58) -- (53.33,58) ;
    \draw [shift={(56.33,58)}, rotate = 180] [fill={rgb, 255:red, 0; green, 0; blue, 0 }  ][line width=0.08]  [draw opacity=0] (8.93,-4.29) -- (0,0) -- (8.93,4.29) -- cycle    ;
    \draw  [fill={rgb, 255:red, 155; green, 155; blue, 155 }  ,fill opacity=1 ][line width=0.75]  (56.33,58) .. controls (56.33,50.26) and (62.59,43.99) .. (70.33,43.98) .. controls (78.06,43.97) and (84.34,50.23) .. (84.35,57.97) .. controls (84.36,65.71) and (78.09,71.98) .. (70.36,71.99) .. controls (62.62,72) and (56.34,65.74) .. (56.33,58) -- cycle ;
    \draw    (174.33,57) .. controls (190.71,56.64) and (194.91,50.17) .. (199.97,45.79) ;
    \draw [shift={(202.33,44)}, rotate = 147.52] [fill={rgb, 255:red, 0; green, 0; blue, 0 }  ][line width=0.08]  [draw opacity=0] (8.93,-4.29) -- (0,0) -- (8.93,4.29) -- cycle    ;
    \draw    (79.17,42.63) -- (101.33,14) ;
    \draw [shift={(77.33,45)}, rotate = 307.75] [fill={rgb, 255:red, 0; green, 0; blue, 0 }  ][line width=0.08]  [draw opacity=0] (8.93,-4.29) -- (0,0) -- (8.93,4.29) -- cycle    ;
    \draw    (76.33,71) -- (98.33,113) ;
    \draw [shift={(76.33,71)}, rotate = 242.35] [color={rgb, 255:red, 0; green, 0; blue, 0 }  ][line width=0.75]    (0,5.59) -- (0,-5.59)   ;
    \draw    (86.32,52.76) -- (121.33,50) ;
    \draw [shift={(83.33,53)}, rotate = 355.49] [fill={rgb, 255:red, 0; green, 0; blue, 0 }  ][line width=0.08]  [draw opacity=0] (8.93,-4.29) -- (0,0) -- (8.93,4.29) -- cycle    ;
    \draw    (83.33,64) -- (119.33,78) ;
    \draw [shift={(83.33,64)}, rotate = 201.25] [color={rgb, 255:red, 0; green, 0; blue, 0 }  ][line width=0.75]    (0,5.59) -- (0,-5.59)   ;
    \draw  [fill={rgb, 255:red, 155; green, 155; blue, 155 }  ,fill opacity=1 ][line width=0.75]  (209.33,58) .. controls (209.33,50.26) and (215.59,43.99) .. (223.33,43.98) .. controls (231.06,43.97) and (237.34,50.23) .. (237.35,57.97) .. controls (237.36,65.71) and (231.09,71.98) .. (223.36,71.99) .. controls (215.62,72) and (209.34,65.74) .. (209.33,58) -- cycle ;
    \draw    (232.57,43) -- (259.33,19) ;
    \draw [shift={(230.33,45)}, rotate = 318.12] [fill={rgb, 255:red, 0; green, 0; blue, 0 }  ][line width=0.08]  [draw opacity=0] (8.93,-4.29) -- (0,0) -- (8.93,4.29) -- cycle    ;
    \draw    (229.33,71) -- (251.33,113) ;
    \draw [shift={(229.33,71)}, rotate = 242.35] [color={rgb, 255:red, 0; green, 0; blue, 0 }  ][line width=0.75]    (0,5.59) -- (0,-5.59)   ;
    \draw    (239.32,52.76) -- (274.33,50) ;
    \draw [shift={(236.33,53)}, rotate = 355.49] [fill={rgb, 255:red, 0; green, 0; blue, 0 }  ][line width=0.08]  [draw opacity=0] (8.93,-4.29) -- (0,0) -- (8.93,4.29) -- cycle    ;
    \draw    (236.33,64) -- (272.33,78) ;
    \draw [shift={(236.33,64)}, rotate = 201.25] [color={rgb, 255:red, 0; green, 0; blue, 0 }  ][line width=0.75]    (0,5.59) -- (0,-5.59)   ;
    \draw  [fill={rgb, 255:red, 155; green, 155; blue, 155 }  ,fill opacity=1 ][line width=0.75]  (340.33,67) .. controls (340.33,59.26) and (346.59,52.99) .. (354.33,52.98) .. controls (362.06,52.97) and (368.34,59.23) .. (368.35,66.97) .. controls (368.36,74.71) and (362.09,80.98) .. (354.36,80.99) .. controls (346.62,81) and (340.34,74.74) .. (340.33,67) -- cycle ;
    \draw    (363.17,51.63) -- (385.33,23) ;
    \draw [shift={(361.33,54)}, rotate = 307.75] [fill={rgb, 255:red, 0; green, 0; blue, 0 }  ][line width=0.08]  [draw opacity=0] (8.93,-4.29) -- (0,0) -- (8.93,4.29) -- cycle    ;
    \draw    (360.33,80) -- (382.33,122) ;
    \draw [shift={(360.33,80)}, rotate = 242.35] [color={rgb, 255:red, 0; green, 0; blue, 0 }  ][line width=0.75]    (0,5.59) -- (0,-5.59)   ;
    \draw    (370.32,61.76) -- (405.33,59) ;
    \draw [shift={(367.33,62)}, rotate = 355.49] [fill={rgb, 255:red, 0; green, 0; blue, 0 }  ][line width=0.08]  [draw opacity=0] (8.93,-4.29) -- (0,0) -- (8.93,4.29) -- cycle    ;
    \draw    (384.33,82) -- (402.33,82) ;
    \draw [shift={(384.33,82)}, rotate = 180] [color={rgb, 255:red, 0; green, 0; blue, 0 }  ][line width=0.75]    (0,5.59) -- (0,-5.59)   ;
    \draw    (366.33,86) -- (381.4,82.65) ;
    \draw [shift={(384.33,82)}, rotate = 167.47] [fill={rgb, 255:red, 0; green, 0; blue, 0 }  ][line width=0.08]  [draw opacity=0] (8.93,-4.29) -- (0,0) -- (8.93,4.29) -- cycle    ;
    \draw    (310,64) .. controls (356.33,61) and (317.33,95) .. (358.33,89) ;
    \draw  [fill={rgb, 255:red, 155; green, 155; blue, 155 }  ,fill opacity=1 ][line width=0.75]  (531.33,67) .. controls (531.33,59.26) and (537.59,52.99) .. (545.33,52.98) .. controls (553.06,52.97) and (559.34,59.23) .. (559.35,66.97) .. controls (559.36,74.71) and (553.09,80.98) .. (545.36,80.99) .. controls (537.62,81) and (531.34,74.74) .. (531.33,67) -- cycle ;
    \draw    (554.17,51.63) -- (576.33,23) ;
    \draw [shift={(552.33,54)}, rotate = 307.75] [fill={rgb, 255:red, 0; green, 0; blue, 0 }  ][line width=0.08]  [draw opacity=0] (8.93,-4.29) -- (0,0) -- (8.93,4.29) -- cycle    ;
    \draw    (554.33,101) -- (573.33,122) ;
    \draw [shift={(554.33,101)}, rotate = 227.86] [color={rgb, 255:red, 0; green, 0; blue, 0 }  ][line width=0.75]    (0,5.59) -- (0,-5.59)   ;
    \draw    (561.32,61.76) -- (596.33,59) ;
    \draw [shift={(558.33,62)}, rotate = 355.49] [fill={rgb, 255:red, 0; green, 0; blue, 0 }  ][line width=0.08]  [draw opacity=0] (8.93,-4.29) -- (0,0) -- (8.93,4.29) -- cycle    ;
    \draw    (497,64) .. controls (529.17,75.58) and (533.09,82.5) .. (552.19,99.15) ;
    \draw [shift={(554.33,101)}, rotate = 220.6] [fill={rgb, 255:red, 0; green, 0; blue, 0 }  ][line width=0.08]  [draw opacity=0] (8.93,-4.29) -- (0,0) -- (8.93,4.29) -- cycle    ;
    \draw    (559.33,72) -- (595.33,86) ;
    \draw [shift={(559.33,72)}, rotate = 201.25] [color={rgb, 255:red, 0; green, 0; blue, 0 }  ][line width=0.75]    (0,5.59) -- (0,-5.59)   ;
    \draw    (179,198.17) -- (195,198.17) ;
    \draw [shift={(198,198.17)}, rotate = 180] [fill={rgb, 255:red, 0; green, 0; blue, 0 }  ][line width=0.08]  [draw opacity=0] (8.93,-4.29) -- (0,0) -- (8.93,4.29) -- cycle    ;
    \draw  [fill={rgb, 255:red, 155; green, 155; blue, 155 }  ,fill opacity=1 ][line width=0.75]  (228.33,198) .. controls (228.33,190.26) and (234.59,183.99) .. (242.33,183.98) .. controls (250.06,183.97) and (256.34,190.23) .. (256.35,197.97) .. controls (256.36,205.71) and (250.09,211.98) .. (242.36,211.99) .. controls (234.62,212) and (228.34,205.74) .. (228.33,198) -- cycle ;
    \draw    (251.17,182.63) -- (273.33,154) ;
    \draw [shift={(249.33,185)}, rotate = 307.75] [fill={rgb, 255:red, 0; green, 0; blue, 0 }  ][line width=0.08]  [draw opacity=0] (8.93,-4.29) -- (0,0) -- (8.93,4.29) -- cycle    ;
    \draw    (248.33,211) -- (270.33,253) ;
    \draw [shift={(248.33,211)}, rotate = 242.35] [color={rgb, 255:red, 0; green, 0; blue, 0 }  ][line width=0.75]    (0,5.59) -- (0,-5.59)   ;
    \draw    (258.32,192.76) -- (293.33,190) ;
    \draw [shift={(255.33,193)}, rotate = 355.49] [fill={rgb, 255:red, 0; green, 0; blue, 0 }  ][line width=0.08]  [draw opacity=0] (8.93,-4.29) -- (0,0) -- (8.93,4.29) -- cycle    ;
    \draw    (255.33,204) -- (291.33,218) ;
    \draw [shift={(255.33,204)}, rotate = 201.25] [color={rgb, 255:red, 0; green, 0; blue, 0 }  ][line width=0.75]    (0,5.59) -- (0,-5.59)   ;
    \draw    (198,198.17) -- (211,198.17) ;
    \draw [shift={(198,198.17)}, rotate = 180] [color={rgb, 255:red, 0; green, 0; blue, 0 }  ][line width=0.75]    (0,5.59) -- (0,-5.59)   ;
    \draw    (211,198.17) .. controls (220.92,183.61) and (221.32,168.28) .. (238.73,182.56) ;
    \draw [shift={(241,184.5)}, rotate = 221.66] [fill={rgb, 255:red, 0; green, 0; blue, 0 }  ][line width=0.08]  [draw opacity=0] (8.93,-4.29) -- (0,0) -- (8.93,4.29) -- cycle    ;
    \draw    (211,198.17) .. controls (221.48,227.61) and (231.68,224.69) .. (236.32,213.79) ;
    \draw [shift={(237.33,211)}, rotate = 107.1] [fill={rgb, 255:red, 0; green, 0; blue, 0 }  ][line width=0.08]  [draw opacity=0] (8.93,-4.29) -- (0,0) -- (8.93,4.29) -- cycle    ;
    \draw    (349,199.17) -- (365,199.17) ;
    \draw [shift={(368,199.17)}, rotate = 180] [fill={rgb, 255:red, 0; green, 0; blue, 0 }  ][line width=0.08]  [draw opacity=0] (8.93,-4.29) -- (0,0) -- (8.93,4.29) -- cycle    ;
    \draw  [fill={rgb, 255:red, 155; green, 155; blue, 155 }  ,fill opacity=1 ][line width=0.75]  (398.33,199) .. controls (398.33,191.26) and (404.59,184.99) .. (412.33,184.98) .. controls (420.06,184.97) and (426.34,191.23) .. (426.35,198.97) .. controls (426.36,206.71) and (420.09,212.98) .. (412.36,212.99) .. controls (404.62,213) and (398.34,206.74) .. (398.33,199) -- cycle ;
    \draw    (421.17,183.63) -- (443.33,155) ;
    \draw [shift={(419.33,186)}, rotate = 307.75] [fill={rgb, 255:red, 0; green, 0; blue, 0 }  ][line width=0.08]  [draw opacity=0] (8.93,-4.29) -- (0,0) -- (8.93,4.29) -- cycle    ;
    \draw    (418.33,212) -- (440.33,254) ;
    \draw [shift={(418.33,212)}, rotate = 242.35] [color={rgb, 255:red, 0; green, 0; blue, 0 }  ][line width=0.75]    (0,5.59) -- (0,-5.59)   ;
    \draw    (428.32,193.76) -- (463.33,191) ;
    \draw [shift={(425.33,194)}, rotate = 355.49] [fill={rgb, 255:red, 0; green, 0; blue, 0 }  ][line width=0.08]  [draw opacity=0] (8.93,-4.29) -- (0,0) -- (8.93,4.29) -- cycle    ;
    \draw    (425.33,205) -- (461.33,219) ;
    \draw [shift={(425.33,205)}, rotate = 201.25] [color={rgb, 255:red, 0; green, 0; blue, 0 }  ][line width=0.75]    (0,5.59) -- (0,-5.59)   ;
    \draw    (368,199.17) -- (381,199.17) ;
    \draw [shift={(368,199.17)}, rotate = 180] [color={rgb, 255:red, 0; green, 0; blue, 0 }  ][line width=0.75]    (0,5.59) -- (0,-5.59)   ;
    \draw    (381,199.17) .. controls (390.92,184.61) and (391.32,169.28) .. (408.73,183.56) ;
    \draw [shift={(411,185.5)}, rotate = 221.66] [fill={rgb, 255:red, 0; green, 0; blue, 0 }  ][line width=0.08]  [draw opacity=0] (8.93,-4.29) -- (0,0) -- (8.93,4.29) -- cycle    ;
    \draw    (381,199.17) .. controls (392.33,231) and (403.33,225) .. (407.33,212) ;
    \draw [shift={(407.33,212)}, rotate = 107.1] [color={rgb, 255:red, 0; green, 0; blue, 0 }  ][line width=0.75]    (0,5.59) -- (0,-5.59)   ;
    \draw    (202.33,44) .. controls (226.33,24.33) and (246.89,8.43) .. (253.33,4.33) ;
    \draw [shift={(202.33,44)}, rotate = 140.67] [color={rgb, 255:red, 0; green, 0; blue, 0 }  ][line width=0.75]    (0,5.59) -- (0,-5.59)   ;
    
    \draw (140,51.4) node [anchor=north west][inner sep=0.75pt]    {$=$};
    \draw (246,-0.6) node [anchor=north west][inner sep=0.75pt]  [font=\scriptsize]  {$\times $};
    \draw (97.38,28.24) node [anchor=north west][inner sep=0.75pt]  [font=\footnotesize,rotate=-338.04]  {$\vdots $};
    \draw (105,2.4) node [anchor=north west][inner sep=0.75pt]  [font=\scriptsize]  {$k_{1}$};
    \draw (100.33,116.4) node [anchor=north west][inner sep=0.75pt]  [font=\scriptsize]  {$k_{n}$};
    \draw (103.37,85.27) node [anchor=north west][inner sep=0.75pt]  [font=\footnotesize,rotate=-32.77,xslant=0]  {$\vdots $};
    \draw (125,43.4) node [anchor=north west][inner sep=0.75pt]  [font=\scriptsize]  {$k_{p}$};
    \draw (120.33,75.4) node [anchor=north west][inner sep=0.75pt]  [font=\scriptsize]  {$k_{p+1}$};
    \draw (5,58.4) node [anchor=north west][inner sep=0.75pt]  [font=\footnotesize]  {$k$};
    \draw (168,55.4) node [anchor=north west][inner sep=0.75pt]  [font=\footnotesize]  {$k$};
    \draw (247.96,34.41) node [anchor=north west][inner sep=0.75pt]  [font=\footnotesize,rotate=-338.04]  {$\vdots $};
    \draw (262.33,10.4) node [anchor=north west][inner sep=0.75pt]  [font=\scriptsize]  {$k_{1}$};
    \draw (253.33,116.4) node [anchor=north west][inner sep=0.75pt]  [font=\scriptsize]  {$k_{n}$};
    \draw (256.37,85.27) node [anchor=north west][inner sep=0.75pt]  [font=\footnotesize,rotate=-32.77,xslant=0]  {$\vdots $};
    \draw (277,45.4) node [anchor=north west][inner sep=0.75pt]  [font=\scriptsize]  {$k_{p}$};
    \draw (274.33,81.4) node [anchor=north west][inner sep=0.75pt]  [font=\scriptsize]  {$k_{p+1}$};
    \draw (291,55.4) node [anchor=north west][inner sep=0.75pt]    {$+$};
    \draw (381.38,37.24) node [anchor=north west][inner sep=0.75pt]  [font=\footnotesize,rotate=-338.04]  {$\vdots $};
    \draw (389,11.4) node [anchor=north west][inner sep=0.75pt]  [font=\scriptsize]  {$k_{1}$};
    \draw (384.33,125.4) node [anchor=north west][inner sep=0.75pt]  [font=\scriptsize]  {$k_{n}$};
    \draw (387.37,94.27) node [anchor=north west][inner sep=0.75pt]  [font=\footnotesize,rotate=-32.77,xslant=0]  {$\vdots $};
    \draw (409,52.4) node [anchor=north west][inner sep=0.75pt]  [font=\scriptsize]  {$k_{p}$};
    \draw (404.33,76.4) node [anchor=north west][inner sep=0.75pt]  [font=\scriptsize]  {$k_{p+1}$};
    \draw (307,67.4) node [anchor=north west][inner sep=0.75pt]  [font=\footnotesize]  {$k$};
    \draw (482,55.4) node [anchor=north west][inner sep=0.75pt]    {$+$};
    \draw (572.38,37.24) node [anchor=north west][inner sep=0.75pt]  [font=\footnotesize,rotate=-338.04]  {$\vdots $};
    \draw (580,11.4) node [anchor=north west][inner sep=0.75pt]  [font=\scriptsize]  {$k_{1}$};
    \draw (575.33,125.4) node [anchor=north west][inner sep=0.75pt]  [font=\scriptsize]  {$k_{n}$};
    \draw (578.37,94.27) node [anchor=north west][inner sep=0.75pt]  [font=\footnotesize,rotate=-32.77,xslant=0]  {$\vdots $};
    \draw (600,52.4) node [anchor=north west][inner sep=0.75pt]  [font=\scriptsize]  {$k_{p}$};
    \draw (598.33,83.4) node [anchor=north west][inner sep=0.75pt]  [font=\scriptsize]  {$k_{p+1}$};
    \draw (498,67.4) node [anchor=north west][inner sep=0.75pt]  [font=\footnotesize]  {$k$};
    \draw (439,55.4) node [anchor=north west][inner sep=0.75pt]    {$+$};
    \draw (454,55.4) node [anchor=north west][inner sep=0.75pt]    {$\dotsc $};
    \draw (161,187.4) node [anchor=north west][inner sep=0.75pt]  [font=\scriptsize]  {$\frac{1}{2!}$};
    \draw (328,188.4) node [anchor=north west][inner sep=0.75pt]    {$+$};
    \draw (269.38,168.24) node [anchor=north west][inner sep=0.75pt]  [font=\footnotesize,rotate=-338.04]  {$\vdots $};
    \draw (277,142.4) node [anchor=north west][inner sep=0.75pt]  [font=\scriptsize]  {$k_{1}$};
    \draw (272.33,256.4) node [anchor=north west][inner sep=0.75pt]  [font=\scriptsize]  {$k_{n}$};
    \draw (275.37,225.27) node [anchor=north west][inner sep=0.75pt]  [font=\footnotesize,rotate=-32.77,xslant=0]  {$\vdots $};
    \draw (297,183.4) node [anchor=north west][inner sep=0.75pt]  [font=\scriptsize]  {$k_{p}$};
    \draw (293.33,215.4) node [anchor=north west][inner sep=0.75pt]  [font=\scriptsize]  {$k_{p+1}$};
    \draw (439.38,169.24) node [anchor=north west][inner sep=0.75pt]  [font=\footnotesize,rotate=-338.04]  {$\vdots $};
    \draw (447,143.4) node [anchor=north west][inner sep=0.75pt]  [font=\scriptsize]  {$k_{1}$};
    \draw (442.33,257.4) node [anchor=north west][inner sep=0.75pt]  [font=\scriptsize]  {$k_{n}$};
    \draw (445.37,226.27) node [anchor=north west][inner sep=0.75pt]  [font=\footnotesize,rotate=-32.77,xslant=0]  {$\vdots $};
    \draw (467,184.4) node [anchor=north west][inner sep=0.75pt]  [font=\scriptsize]  {$k_{p}$};
    \draw (463.33,216.4) node [anchor=north west][inner sep=0.75pt]  [font=\scriptsize]  {$k_{p+1}$};
    \draw (174,199.4) node [anchor=north west][inner sep=0.75pt]  [font=\footnotesize]  {$k$};
    \draw (344,200.4) node [anchor=north west][inner sep=0.75pt]  [font=\footnotesize]  {$k$};
    \draw (145,190.4) node [anchor=north west][inner sep=0.75pt]    {$+$};

    \end{tikzpicture} \ ,
\end{equation}
\begin{equation}
    \tikzset{every picture/.style={line width=0.75pt}} 
    \begin{tikzpicture}[x=0.75pt,y=0.75pt,yscale=-0.9,xscale=0.9]
    \centering
    
    \draw    (24.33,65) -- (70.33,65) ;
    \draw [shift={(70.33,65)}, rotate = 180] [color={rgb, 255:red, 0; green, 0; blue, 0 }  ][line width=0.75]    (0,5.59) -- (0,-5.59)   ;
    \draw  [fill={rgb, 255:red, 155; green, 155; blue, 155 }  ,fill opacity=1 ][line width=0.75]  (70.33,65) .. controls (70.33,57.26) and (76.59,50.99) .. (84.33,50.98) .. controls (92.06,50.97) and (98.34,57.23) .. (98.35,64.97) .. controls (98.36,72.71) and (92.09,78.98) .. (84.36,78.99) .. controls (76.62,79) and (70.34,72.74) .. (70.33,65) -- cycle ;
    \draw    (188.33,64) .. controls (203.33,63.67) and (209.89,55.1) .. (216.33,51) ;
    \draw [shift={(216.33,51)}, rotate = 147.52] [color={rgb, 255:red, 0; green, 0; blue, 0 }  ][line width=0.75]    (0,5.59) -- (0,-5.59)   ;
    \draw    (93.17,49.63) -- (115.33,21) ;
    \draw [shift={(91.33,52)}, rotate = 307.75] [fill={rgb, 255:red, 0; green, 0; blue, 0 }  ][line width=0.08]  [draw opacity=0] (8.93,-4.29) -- (0,0) -- (8.93,4.29) -- cycle    ;
    \draw    (90.33,78) -- (112.33,120) ;
    \draw [shift={(90.33,78)}, rotate = 242.35] [color={rgb, 255:red, 0; green, 0; blue, 0 }  ][line width=0.75]    (0,5.59) -- (0,-5.59)   ;
    \draw    (100.32,59.76) -- (135.33,57) ;
    \draw [shift={(97.33,60)}, rotate = 355.49] [fill={rgb, 255:red, 0; green, 0; blue, 0 }  ][line width=0.08]  [draw opacity=0] (8.93,-4.29) -- (0,0) -- (8.93,4.29) -- cycle    ;
    \draw    (97.33,71) -- (133.33,85) ;
    \draw [shift={(97.33,71)}, rotate = 201.25] [color={rgb, 255:red, 0; green, 0; blue, 0 }  ][line width=0.75]    (0,5.59) -- (0,-5.59)   ;
    \draw  [fill={rgb, 255:red, 155; green, 155; blue, 155 }  ,fill opacity=1 ][line width=0.75]  (223.33,65) .. controls (223.33,57.26) and (229.59,50.99) .. (237.33,50.98) .. controls (245.06,50.97) and (251.34,57.23) .. (251.35,64.97) .. controls (251.36,72.71) and (245.09,78.98) .. (237.36,78.99) .. controls (229.62,79) and (223.34,72.74) .. (223.33,65) -- cycle ;
    \draw    (246.57,50) -- (273.33,26) ;
    \draw [shift={(244.33,52)}, rotate = 318.12] [fill={rgb, 255:red, 0; green, 0; blue, 0 }  ][line width=0.08]  [draw opacity=0] (8.93,-4.29) -- (0,0) -- (8.93,4.29) -- cycle    ;
    \draw    (243.33,78) -- (265.33,120) ;
    \draw [shift={(243.33,78)}, rotate = 242.35] [color={rgb, 255:red, 0; green, 0; blue, 0 }  ][line width=0.75]    (0,5.59) -- (0,-5.59)   ;
    \draw    (253.32,59.76) -- (288.33,57) ;
    \draw [shift={(250.33,60)}, rotate = 355.49] [fill={rgb, 255:red, 0; green, 0; blue, 0 }  ][line width=0.08]  [draw opacity=0] (8.93,-4.29) -- (0,0) -- (8.93,4.29) -- cycle    ;
    \draw    (250.33,71) -- (286.33,85) ;
    \draw [shift={(250.33,71)}, rotate = 201.25] [color={rgb, 255:red, 0; green, 0; blue, 0 }  ][line width=0.75]    (0,5.59) -- (0,-5.59)   ;
    \draw  [fill={rgb, 255:red, 155; green, 155; blue, 155 }  ,fill opacity=1 ][line width=0.75]  (346.33,68) .. controls (346.33,60.26) and (352.59,53.99) .. (360.33,53.98) .. controls (368.06,53.97) and (374.34,60.23) .. (374.35,67.97) .. controls (374.36,75.71) and (368.09,81.98) .. (360.36,81.99) .. controls (352.62,82) and (346.34,75.74) .. (346.33,68) -- cycle ;
    \draw    (370.06,43.42) -- (399.33,30) ;
    \draw [shift={(367.33,44.67)}, rotate = 335.38] [fill={rgb, 255:red, 0; green, 0; blue, 0 }  ][line width=0.08]  [draw opacity=0] (8.93,-4.29) -- (0,0) -- (8.93,4.29) -- cycle    ;
    \draw    (366.33,81) -- (388.33,123) ;
    \draw [shift={(366.33,81)}, rotate = 242.35] [color={rgb, 255:red, 0; green, 0; blue, 0 }  ][line width=0.75]    (0,5.59) -- (0,-5.59)   ;
    \draw    (376.32,62.76) -- (411.33,60) ;
    \draw [shift={(373.33,63)}, rotate = 355.49] [fill={rgb, 255:red, 0; green, 0; blue, 0 }  ][line width=0.08]  [draw opacity=0] (8.93,-4.29) -- (0,0) -- (8.93,4.29) -- cycle    ;
    \draw    (374.33,73.67) -- (405.33,86.67) ;
    \draw [shift={(374.33,73.67)}, rotate = 202.75] [color={rgb, 255:red, 0; green, 0; blue, 0 }  ][line width=0.75]    (0,5.59) -- (0,-5.59)   ;
    \draw    (324,71) .. controls (343.33,54) and (332.33,58) .. (367.33,44.67) ;
    \draw [shift={(367.33,44.67)}, rotate = 159.15] [color={rgb, 255:red, 0; green, 0; blue, 0 }  ][line width=0.75]    (0,5.59) -- (0,-5.59)   ;
    \draw  [fill={rgb, 255:red, 155; green, 155; blue, 155 }  ,fill opacity=1 ][line width=0.75]  (545.33,74) .. controls (545.33,66.26) and (551.59,59.99) .. (559.33,59.98) .. controls (567.06,59.97) and (573.34,66.23) .. (573.35,73.97) .. controls (573.36,81.71) and (567.09,87.98) .. (559.36,87.99) .. controls (551.62,88) and (545.34,81.74) .. (545.33,74) -- cycle ;
    \draw    (568.17,58.63) -- (590.33,30) ;
    \draw [shift={(566.33,61)}, rotate = 307.75] [fill={rgb, 255:red, 0; green, 0; blue, 0 }  ][line width=0.08]  [draw opacity=0] (8.93,-4.29) -- (0,0) -- (8.93,4.29) -- cycle    ;
    \draw    (566.33,87) -- (587.33,129) ;
    \draw [shift={(566.33,87)}, rotate = 243.43] [color={rgb, 255:red, 0; green, 0; blue, 0 }  ][line width=0.75]    (0,5.59) -- (0,-5.59)   ;
    \draw    (599.58,63.71) -- (618.33,72) ;
    \draw [shift={(596.83,62.5)}, rotate = 23.84] [fill={rgb, 255:red, 0; green, 0; blue, 0 }  ][line width=0.08]  [draw opacity=0] (8.93,-4.29) -- (0,0) -- (8.93,4.29) -- cycle    ;
    \draw    (513,71) .. controls (527.18,52.5) and (548.78,32.29) .. (568.56,43.61) .. controls (570.16,44.52) and (571.76,45.65) .. (573.33,47) ;
    \draw    (573.33,79) -- (609.33,93) ;
    \draw [shift={(573.33,79)}, rotate = 201.25] [color={rgb, 255:red, 0; green, 0; blue, 0 }  ][line width=0.75]    (0,5.59) -- (0,-5.59)   ;
    \draw    (379,209.17) -- (398,209.17) ;
    \draw [shift={(398,209.17)}, rotate = 180] [color={rgb, 255:red, 0; green, 0; blue, 0 }  ][line width=0.75]    (0,5.59) -- (0,-5.59)   ;
    \draw  [fill={rgb, 255:red, 155; green, 155; blue, 155 }  ,fill opacity=1 ][line width=0.75]  (428.33,209) .. controls (428.33,201.26) and (434.59,194.99) .. (442.33,194.98) .. controls (450.06,194.97) and (456.34,201.23) .. (456.35,208.97) .. controls (456.36,216.71) and (450.09,222.98) .. (442.36,222.99) .. controls (434.62,223) and (428.34,216.74) .. (428.33,209) -- cycle ;
    \draw    (451.17,193.63) -- (473.33,165) ;
    \draw [shift={(449.33,196)}, rotate = 307.75] [fill={rgb, 255:red, 0; green, 0; blue, 0 }  ][line width=0.08]  [draw opacity=0] (8.93,-4.29) -- (0,0) -- (8.93,4.29) -- cycle    ;
    \draw    (448.33,222) -- (470.33,264) ;
    \draw [shift={(448.33,222)}, rotate = 242.35] [color={rgb, 255:red, 0; green, 0; blue, 0 }  ][line width=0.75]    (0,5.59) -- (0,-5.59)   ;
    \draw    (458.32,203.76) -- (493.33,201) ;
    \draw [shift={(455.33,204)}, rotate = 355.49] [fill={rgb, 255:red, 0; green, 0; blue, 0 }  ][line width=0.08]  [draw opacity=0] (8.93,-4.29) -- (0,0) -- (8.93,4.29) -- cycle    ;
    \draw    (455.33,215) -- (491.33,229) ;
    \draw [shift={(455.33,215)}, rotate = 201.25] [color={rgb, 255:red, 0; green, 0; blue, 0 }  ][line width=0.75]    (0,5.59) -- (0,-5.59)   ;
    \draw    (401,209.17) -- (411,209.17) ;
    \draw [shift={(398,209.17)}, rotate = 0] [fill={rgb, 255:red, 0; green, 0; blue, 0 }  ][line width=0.08]  [draw opacity=0] (8.93,-4.29) -- (0,0) -- (8.93,4.29) -- cycle    ;
    \draw    (411,209.17) .. controls (420.33,186) and (426.33,183) .. (434.33,197) ;
    \draw [shift={(434.33,197)}, rotate = 240.26] [color={rgb, 255:red, 0; green, 0; blue, 0 }  ][line width=0.75]    (0,5.59) -- (0,-5.59)   ;
    \draw    (411,209.17) .. controls (422.33,241) and (433.33,235) .. (437.33,222) ;
    \draw [shift={(437.33,222)}, rotate = 107.1] [color={rgb, 255:red, 0; green, 0; blue, 0 }  ][line width=0.75]    (0,5.59) -- (0,-5.59)   ;
    \draw    (198,209.17) -- (217,209.17) ;
    \draw [shift={(217,209.17)}, rotate = 180] [color={rgb, 255:red, 0; green, 0; blue, 0 }  ][line width=0.75]    (0,5.59) -- (0,-5.59)   ;
    \draw  [fill={rgb, 255:red, 155; green, 155; blue, 155 }  ,fill opacity=1 ][line width=0.75]  (247.33,209) .. controls (247.33,201.26) and (253.59,194.99) .. (261.33,194.98) .. controls (269.06,194.97) and (275.34,201.23) .. (275.35,208.97) .. controls (275.36,216.71) and (269.09,222.98) .. (261.36,222.99) .. controls (253.62,223) and (247.34,216.74) .. (247.33,209) -- cycle ;
    \draw    (270.17,193.63) -- (292.33,165) ;
    \draw [shift={(268.33,196)}, rotate = 307.75] [fill={rgb, 255:red, 0; green, 0; blue, 0 }  ][line width=0.08]  [draw opacity=0] (8.93,-4.29) -- (0,0) -- (8.93,4.29) -- cycle    ;
    \draw    (267.33,222) -- (289.33,264) ;
    \draw [shift={(267.33,222)}, rotate = 242.35] [color={rgb, 255:red, 0; green, 0; blue, 0 }  ][line width=0.75]    (0,5.59) -- (0,-5.59)   ;
    \draw    (277.32,203.76) -- (312.33,201) ;
    \draw [shift={(274.33,204)}, rotate = 355.49] [fill={rgb, 255:red, 0; green, 0; blue, 0 }  ][line width=0.08]  [draw opacity=0] (8.93,-4.29) -- (0,0) -- (8.93,4.29) -- cycle    ;
    \draw    (274.33,215) -- (310.33,229) ;
    \draw [shift={(274.33,215)}, rotate = 201.25] [color={rgb, 255:red, 0; green, 0; blue, 0 }  ][line width=0.75]    (0,5.59) -- (0,-5.59)   ;
    \draw    (220,209.17) -- (230,209.17) ;
    \draw [shift={(217,209.17)}, rotate = 0] [fill={rgb, 255:red, 0; green, 0; blue, 0 }  ][line width=0.08]  [draw opacity=0] (8.93,-4.29) -- (0,0) -- (8.93,4.29) -- cycle    ;
    \draw    (230,209.17) .. controls (239.92,194.61) and (240.32,179.28) .. (257.73,193.56) ;
    \draw [shift={(260,195.5)}, rotate = 221.66] [fill={rgb, 255:red, 0; green, 0; blue, 0 }  ][line width=0.08]  [draw opacity=0] (8.93,-4.29) -- (0,0) -- (8.93,4.29) -- cycle    ;
    \draw    (230,209.17) .. controls (241.33,241) and (252.33,235) .. (256.33,222) ;
    \draw [shift={(256.33,222)}, rotate = 107.1] [color={rgb, 255:red, 0; green, 0; blue, 0 }  ][line width=0.75]    (0,5.59) -- (0,-5.59)   ;
    \draw    (580.33,52) -- (596.83,62.5) ;
    \draw [shift={(596.83,62.5)}, rotate = 212.47] [color={rgb, 255:red, 0; green, 0; blue, 0 }  ][line width=0.75]    (0,5.59) -- (0,-5.59)   ;
    \draw    (218.93,49.08) .. controls (242.62,31.5) and (260.12,15.62) .. (266.33,11.67) ;
    \draw [shift={(216.33,51)}, rotate = 323.75] [fill={rgb, 255:red, 0; green, 0; blue, 0 }  ][line width=0.08]  [draw opacity=0] (8.93,-4.29) -- (0,0) -- (8.93,4.29) -- cycle    ;
    
    \draw (154,58.4) node [anchor=north west][inner sep=0.75pt]    {$=$};
    \draw (260,6.4) node [anchor=north west][inner sep=0.75pt]  [font=\scriptsize]  {$\times $};
    \draw (111.38,35.24) node [anchor=north west][inner sep=0.75pt]  [font=\footnotesize,rotate=-338.04]  {$\vdots $};
    \draw (119,9.4) node [anchor=north west][inner sep=0.75pt]  [font=\scriptsize]  {$k_{1}$};
    \draw (114.33,123.4) node [anchor=north west][inner sep=0.75pt]  [font=\scriptsize]  {$k_{n}$};
    \draw (117.37,92.27) node [anchor=north west][inner sep=0.75pt]  [font=\footnotesize,rotate=-32.77,xslant=0]  {$\vdots $};
    \draw (139,50.4) node [anchor=north west][inner sep=0.75pt]  [font=\scriptsize]  {$k_{p}$};
    \draw (134.33,82.4) node [anchor=north west][inner sep=0.75pt]  [font=\scriptsize]  {$k_{p+1}$};
    \draw (19,65.4) node [anchor=north west][inner sep=0.75pt]  [font=\footnotesize]  {$k$};
    \draw (182,62.4) node [anchor=north west][inner sep=0.75pt]  [font=\footnotesize]  {$k$};
    \draw (261.96,41.41) node [anchor=north west][inner sep=0.75pt]  [font=\footnotesize,rotate=-338.04]  {$\vdots $};
    \draw (276.33,17.4) node [anchor=north west][inner sep=0.75pt]  [font=\scriptsize]  {$k_{1}$};
    \draw (267.33,123.4) node [anchor=north west][inner sep=0.75pt]  [font=\scriptsize]  {$k_{n}$};
    \draw (270.37,92.27) node [anchor=north west][inner sep=0.75pt]  [font=\footnotesize,rotate=-32.77,xslant=0]  {$\vdots $};
    \draw (291,52.4) node [anchor=north west][inner sep=0.75pt]  [font=\scriptsize]  {$k_{p}$};
    \draw (288.33,88.4) node [anchor=north west][inner sep=0.75pt]  [font=\scriptsize]  {$k_{p+1}$};
    \draw (305,62.4) node [anchor=north west][inner sep=0.75pt]    {$+$};
    \draw (395.38,44.24) node [anchor=north west][inner sep=0.75pt]  [font=\footnotesize,rotate=-338.04]  {$\vdots $};
    \draw (403,18.4) node [anchor=north west][inner sep=0.75pt]  [font=\scriptsize]  {$k_{1}$};
    \draw (390.33,126.4) node [anchor=north west][inner sep=0.75pt]  [font=\scriptsize]  {$k_{n}$};
    \draw (393.37,95.27) node [anchor=north west][inner sep=0.75pt]  [font=\footnotesize,rotate=-32.77,xslant=0]  {$\vdots $};
    \draw (415,53.4) node [anchor=north west][inner sep=0.75pt]  [font=\scriptsize]  {$k_{p}$};
    \draw (408.33,82.4) node [anchor=north west][inner sep=0.75pt]  [font=\scriptsize]  {$k_{p+1}$};
    \draw (321,74.4) node [anchor=north west][inner sep=0.75pt]  [font=\footnotesize]  {$k$};
    \draw (496,62.4) node [anchor=north west][inner sep=0.75pt]    {$+$};
    \draw (586.38,44.24) node [anchor=north west][inner sep=0.75pt]  [font=\footnotesize,rotate=-338.04]  {$\vdots $};
    \draw (594,18.4) node [anchor=north west][inner sep=0.75pt]  [font=\scriptsize]  {$k_{1}$};
    \draw (589.33,132.4) node [anchor=north west][inner sep=0.75pt]  [font=\scriptsize]  {$k_{n}$};
    \draw (592.37,101.27) node [anchor=north west][inner sep=0.75pt]  [font=\footnotesize,rotate=-32.77,xslant=0]  {$\vdots $};
    \draw (622,69.4) node [anchor=north west][inner sep=0.75pt]  [font=\scriptsize]  {$k_{p}$};
    \draw (612.33,90.4) node [anchor=north west][inner sep=0.75pt]  [font=\scriptsize]  {$k_{p+1}$};
    \draw (512,74.4) node [anchor=north west][inner sep=0.75pt]  [font=\footnotesize]  {$k$};
    \draw (453,62.4) node [anchor=north west][inner sep=0.75pt]    {$+$};
    \draw (468,62.4) node [anchor=north west][inner sep=0.75pt]    {$\dotsc $};
    \draw (361,198.4) node [anchor=north west][inner sep=0.75pt]  [font=\scriptsize]  {$\frac{1}{2!}$};
    \draw (469.38,179.24) node [anchor=north west][inner sep=0.75pt]  [font=\footnotesize,rotate=-338.04]  {$\vdots $};
    \draw (477,153.4) node [anchor=north west][inner sep=0.75pt]  [font=\scriptsize]  {$k_{1}$};
    \draw (472.33,267.4) node [anchor=north west][inner sep=0.75pt]  [font=\scriptsize]  {$k_{n}$};
    \draw (475.37,236.27) node [anchor=north west][inner sep=0.75pt]  [font=\footnotesize,rotate=-32.77,xslant=0]  {$\vdots $};
    \draw (497,194.4) node [anchor=north west][inner sep=0.75pt]  [font=\scriptsize]  {$k_{p}$};
    \draw (493.33,226.4) node [anchor=north west][inner sep=0.75pt]  [font=\scriptsize]  {$k_{p+1}$};
    \draw (374,210.4) node [anchor=north west][inner sep=0.75pt]  [font=\footnotesize]  {$k$};
    \draw (345,201.4) node [anchor=north west][inner sep=0.75pt]    {$+$};
    \draw (160,200.4) node [anchor=north west][inner sep=0.75pt]    {$+$};
    \draw (288.38,179.24) node [anchor=north west][inner sep=0.75pt]  [font=\footnotesize,rotate=-338.04]  {$\vdots $};
    \draw (296,153.4) node [anchor=north west][inner sep=0.75pt]  [font=\scriptsize]  {$k_{1}$};
    \draw (291.33,267.4) node [anchor=north west][inner sep=0.75pt]  [font=\scriptsize]  {$k_{n}$};
    \draw (294.37,236.27) node [anchor=north west][inner sep=0.75pt]  [font=\footnotesize,rotate=-32.77,xslant=0]  {$\vdots $};
    \draw (316,194.4) node [anchor=north west][inner sep=0.75pt]  [font=\scriptsize]  {$k_{p}$};
    \draw (312.33,226.4) node [anchor=north west][inner sep=0.75pt]  [font=\scriptsize]  {$k_{p+1}$};
    \draw (193,210.4) node [anchor=north west][inner sep=0.75pt]  [font=\footnotesize]  {$k$};

    \end{tikzpicture} \ .
\end{equation}

We can write down the corresponding Schwinger-Dyson equations for the generating function of the connected correlators $W_{\rm SK}[J_{\Pb}, J_{\Fb}]$ defined as
\begin{equation}
    Z_{\rm SK}[J_{\Pb}, J_{\Fb}] \equiv e^{i W_{\rm SK}[J_{\Pb}, J_{\Fb}]} \ .
\end{equation}
The one-point Schwinger-Dyson equations for $W_{\rm SK}[J_{\Pb}, J_{\Fb}]$ are as follows:
\begin{equation}
    \begin{split}
        &\frac{1}{i}\frac{\delta}{\delta J_{\Pb}(p_1)}i W_{\rm SK}\\
        &= -i G(\bar{p}_1) \Bigg[i J_{\Fb} (\bar{p}_1)-i \int_{p_{2,3}}(2\pi)^{d+1} \delta^{d+1}(p_1+p_2+p_3)\\
        &\times \frac{1}{2!} \lambda_{\Fsmall \Fsmall \Psmall}(p_2,p_3,p_1) \Bigg(\frac{1}{i}\frac{\delta}{\delta J_{\Pb}(-p_3)} \frac{1}{i}\frac{\delta}{\delta J_{\Pb}(-p_2)}i W_{\rm SK} + \frac{1}{i}\frac{\delta}{\delta J_{\Pb}(-p_3)} i W_{\rm SK} \frac{1}{i}\frac{\delta}{\delta J_{\Pb}(-p_2)}i W_{\rm SK} \Bigg)\\
        &+  \lambda_{\Fsmall \Psmall \Psmall}(p_2,p_1,p_3) \Bigg( \frac{1}{i}\frac{\delta}{\delta J_{\Fb}(-p_3)} \frac{1}{i}\frac{\delta}{\delta J_{\Pb}(-p_2)} i W_{\rm SK} + \frac{1}{i}\frac{\delta}{\delta J_{\Fb}(-p_3)}i W_{\rm SK} \frac{1}{i}\frac{\delta}{\delta J_{\Pb}(-p_2)} i W_{\rm SK} \Bigg)\Bigg]\ ,
    \end{split}
\end{equation}
\begin{equation}
    \begin{split}
        &\frac{1}{i}\frac{\delta}{\delta J_{\Fb}(p_1)}i W_{\rm SK}\\
        &= -i G({p}_1) \Bigg[i J_{\Pb} (\bar{p}_1)-i \int_{p_{2,3}}(2\pi)^{d+1} \delta^{d+1}(p_1+p_2+p_3)\\
        &\times \lambda_{\Fsmall \Fsmall \Psmall}(p_1,p_2,p_3) \Bigg(\frac{1}{i}\frac{\delta}{\delta J_{\Pb}(-p_2)} \frac{1}{i}\frac{\delta}{\delta J_{\Fb}(-p_3)}i W_{\rm SK} + \frac{1}{i}\frac{\delta}{\delta J_{\Pb}(-p_2)}i W_{\rm SK} \frac{1}{i}\frac{\delta}{\delta J_{\Fb}(-p_3)}i W_{\rm SK} \Bigg)\\
        &+\frac{1}{2!} \lambda_{\Fsmall \Psmall \Psmall}(p_1,p_2,p_3) \Bigg( \frac{1}{i}\frac{\delta}{\delta J_{\Fb}(-p_2)} \frac{1}{i}\frac{\delta}{\delta J_{\Fb}(-p_3)}i W_{\rm SK} + \frac{1}{i}\frac{\delta}{\delta J_{\Fb}(-p_2)}i W_{\rm SK} \frac{1}{i}\frac{\delta}{\delta J_{\Fb}(-p_3)}i W_{\rm SK} \Bigg) \Bigg]\ ,
    \end{split}
\end{equation}
The equations look more and more complicated at higher-point but are, in fact, simple as the following diagrammatic equations demonstrate. We simply get the connected (and multi-blob) descendants of the Schwinger-Dyson equations for $Z_{\rm SK}[J_{\Pb}, J_{\Fb}]$.
\begin{equation}
    \tikzset{every picture/.style={line width=0.75pt}} 
 \ .
\end{equation}
Similarly, we can find the higher-point Schwinger-Dyson equations for the connected generating functional. Diagrammatically, it is very clear how these can be written down. For illustration, the two-point and three-point SDEs are given below.
\begin{equation}
    \tikzset{every picture/.style={line width=0.75pt}} 
    %
 \ .
\end{equation}

\section{SDEs for grSK and exterior quantum field theories}\label{app:grSKExtEFTSDE}

Here, we write down the Schwinger-Dyson equations to quantise the scalar theory on the grSK spacetime. We then discuss the classical limit of these equations, and check their consistency with what is already known. Furthermore, we will set up the perturbation theory by doing a perturbative expansion of the Schwinger-Dyson equations.

The object we are interested in computing is the boundary SK generating functional. In the case where the bulk field theory is classical, we used the GKPW condition to obtain this quantity from the bulk on-shell action. The bulk quantum field theory generalisation is to instead compute the bulk on-shell 1PI effective action, and impose the GKPW boundary conditions on the 1PI field. This is once again, an application of the GKPW condition \cite{Gubser:1998bc, Witten:1998qj}. The bulk on-shell 1PI effective action will then give us the boundary SK generating functional, and similarly the bulk connected generating functional will give us the connected generating functional of the boundary theory.

Our aim in this section is to compute the bulk extremum 1PI effective action. To this end, it is useful to first study the generating functional of the connected Green functions of the bulk field theory, denoted as $W[\mathfrak{J}]$. Here $\mathfrak{J}(Y)$ is the source at the bulk point $Y$. The bulk 1PI effective action is then just the Legendre transform of $W[\mathfrak{J}]$ with respect to the source $\mathfrak{J}(Y)$.

We define the generating functional of connected bulk Green functions to satisfy the Schwinger-Dyson equation for cubic interactions\footnote{The reader unfamiliar with this type of definition is referred to Appendix \ref{app:SDEsfromPI} for a derivation of the Schwinger-Dyson equations from the path-integral formalism. A field theory book that uses the Schwinger-Dyson equation starting point is \cite{Cvitanovic:1983eb}.}
\begin{equation}
    \begin{split}
     \dJh{Y} i W &= \phi_{(0)}(Y)+ \int_{Y_0} (-i)\bbG(Y|Y_0) \ i \mathfrak{J}(Y_0) + \int_{Y_0}(-i)\bbG(Y|Y_0)\\
     & \times  -i \lambda_{3\rm B}  \left\{ \frac{1}{2!} \left[\dJh{Y_0}\right]^2 i W+ \frac{1}{2!} \left[\dJh{Y_0} i W\right]^2  \right\}   \ .
    \end{split}
    \label{eq:SDEConnected}
\end{equation}
Note that this is simply the diagrammatic Schwinger-Dyson equation that we have already met in Eq.\eqref{eq:SchDysonDiag}. This is our starting point. Here $-i \bbG(Y|Y_0)$ is the propagator of the scalar field that we introduced in Eq.~\eqref{eq:PDEFormalBlkBlk} and Eq.~\eqref{eq:BCBlkBlk}. The symbol $\phi_{(0)}$ denotes the solution of the free bulk equations of motion that we introduced in Eq.~\eqref{eq:LeadingOrderSolutionPF}, and $\lambda_{3 \rm B}$ is the bulk three-point coupling. 

The Schwinger-Dyson equations alone do not uniquely specify the theory. They have to be supplemented with appropriate boundary conditions. We take these to be the usual GKPW boundary conditions appropriate to the grSK context:
\begin{equation}\label{eq:qGKPW}
    \lim_{\z \to 0}\frac{\delta W[\mathfrak{J}, J]}{\delta \mathfrak{J}(\z,k)}\Bigg|_{\mathfrak{J}=0} = J_{\sL}(k)  \ , \qquad \lim_{\z \to 1}\frac{\delta W[\mathfrak{J}, J]}{\delta \mathfrak{J}(\z,k)}\Bigg|_{\mathfrak{J}=0} = J_{\sR}(k) \ .
\end{equation}
Here, we have gone to the Fourier domain to write the boundary conditions. We will do this often. Note that we use the notation $ W[\mathfrak{J}, J]$ for the bulk generating functional since it also depends on the boundary sources $J$. There is an implicit dependence through $\phi_{(0)}$ as well as an explicit dependence through the above boundary condition.\footnote{In usual QFT, we generally take the boundary limit of the bulk field to vanish and thus we have $W[\mathfrak{J}]$.}

Let us take a moment to demystify these boundary conditions. Note that the usual GKPW boundary conditions must be imposed on the 1PI field \cite{Gubser:1998bc}. And this is exactly what we are doing here. Recall that the 1PI field is defined as the derivative of the generating functional of connected Green functions with respect to the source. 

The above Schwinger-Dyson equations and the corresponding GKPW boundary conditions completely specify the theory we want to study. 

\subsection*{Classical limit of the grSK SDEs}\label{sec:ConsCheckgrSKSDE}

First of all, we expect that the SDEs should reduce to the usual grSK prescription (see Sec.(\ref{sec:grSKeFTReview})) in the classical field theory limit. In this limit, we expect the SDEs along with the GKPW boundary conditions to give the classical equations of motion for the bulk field with the grSK GKPW boundary conditions.

Let us see how this is the case. In the classical field theory limit, the Schwinger-Dyson equation in Eq.~\eqref{eq:SDEConnected} takes the form
\begin{equation}\label{eq:W1ptSDEhto0}
    \begin{split}
     \frac{\delta \,  W_{\rm cl}}{\delta \mathfrak{J}(Y)}  &= \phi_{(0)}(Y)+ \int_{Y_0} \bbG(Y|Y_0)  \mathfrak{J}(Y_0) - \frac{\lambda_{3 \rm B}}{2} \int_{Y_0} \bbG(Y|Y_0)   \left[\frac{\delta \,  W_{\rm cl}}{\delta \mathfrak{J}(Y_0)}\right]^2  +\mathcal{O}(\hbar) \ ,
    \end{split}
\end{equation}  
where $W_{\rm cl}$ is the leading term in $W$ in the $\hbar \to 0$ classical field expansion. What we have done here is to expand the generating functional in $\hbar$, substituted in the SDE, and read off the leading answer. The leading term $W_{\rm cl}$ is an expansion in tree diagrams as usual \cite{Siegel:1999ew}.

The above equation is nothing but the classical equation of motion (EOM) once the $\mathcal{O}(\hbar)$ corrections are neglected. To see this, we first write the above SDE in terms of the classical field $\phi_{\rm cl}$ as
\begin{equation}\label{eq:phicl}
    \begin{split}
     \phi_{\rm cl}(Y)  &= \phi_{(0)}(Y)+ \int_{Y_0} \bbG(Y|Y_0)  \mathfrak{J}(Y_0) - \frac{\lambda_{3 \rm B}}{2} \int_{Y_0} \bbG(Y|Y_0)   \left[\phi_{\rm cl}(Y_0)\right]^2   \ , 
    \end{split}
\end{equation}
where we have  defined $\phi_{\rm cl}$
\begin{equation}
    \phi_{\rm cl}(Y) \equiv \frac{\delta \,  W_{\rm cl}}{\delta \mathfrak{J}(Y)} \ .
    \label{eq:SolutionClassical}
\end{equation}
Note that this is the 1PI field at leading order in the $\hbar \to 0$ limit. But this is just what is commonly called the classical field.
Now, we will apply the Klein-Gordon operator\footnote{We are restoring the mass term for time being.} on Eq.~(\ref{eq:phicl}) to get
\begin{equation}\label{eq:classEOM}
   \nabla_Y^2\phi_{\rm cl}(Y) =   -\mathfrak{J}(Y)+\frac{\lambda_{3 \rm B}}{2} \phi_{\rm cl}^2(Y)   \ ,
\end{equation}
where we have used the definitions of the free solution $\phi_{(0)}$ and the bulk-to-bulk Green function $\bbG(Y|Y_0)$ in Eq.~\eqref{eq:LeadingOrderSolutionPF}, and Eq.~\eqref{eq:PDEFormalBlkBlk}, respectively. Here we point out that Eq.~(\ref{eq:classEOM}) is the classical equation of motion of $\phi^3$ theory in the presence of a linearly coupled bulk source $\mathfrak{J}$. Turning off the bulk source, we reproduce exactly the classical equation of motion in Eq.~\eqref{eq:EOMClassical}.

\subsection*{Perturbative expansion of the grSK SDEs}

To find $W$ in perturbation theory, all we need to do is to perturbatively expand the SDE in Eq.~\eqref{eq:SDEConnected} and integrate it with respect to the bulk source. Focussing on the Schwinger-Dyson equation (SDE) in Eq.~\eqref{eq:SchDysonDiag} at leading order in the bulk coupling, we get
\begin{equation}
\tikzset{every picture/.style={line width=0.75pt}} 
\begin{tikzpicture}[x=0.75pt,y=0.75pt,yscale=-1,xscale=1]

\draw   (112.93,102.51) .. controls (112.93,94.25) and (119.65,87.55) .. (127.93,87.55) .. controls (136.22,87.55) and (142.93,94.25) .. (142.93,102.51) .. controls (142.93,110.77) and (136.22,117.47) .. (127.93,117.47) .. controls (119.65,117.47) and (112.93,110.77) .. (112.93,102.51) -- cycle ;
\draw    (113.21,99.45) -- (127.93,87.55) ;
\draw    (131.59,88.02) -- (112.97,103.14) ;
\draw    (113.45,106.48) -- (134.82,89.1) ;
\draw    (114.64,109.57) -- (137.44,90.88) ;
\draw    (116.41,112.03) -- (123.98,105.77) -- (139.45,92.98) ;
\draw    (118.58,114.1) -- (141.14,95.4) ;
\draw    (120.97,115.76) -- (142.22,98.02) ;
\draw    (123.95,116.83) -- (142.93,101.12) ;
\draw    (127.06,117.55) -- (142.69,104.81) ;
\draw    (132.07,116.83) -- (141.14,109.45) ;
\draw    (122.28,88.62) -- (115.72,93.98) ;
\draw    (80.93,102.55) -- (111.93,102.47) ;
\draw    (180.93,101.55) -- (224.93,101.47) ;
\draw    (258.93,101.55) -- (302.93,101.47) ;

\draw (149.6,96.95) node [anchor=north west][inner sep=0.75pt]    {$=$};
\draw (218.6,95.95) node [anchor=north west][inner sep=0.75pt]  [font=\scriptsize]  {$\otimes $};
\draw (72.6,103.95) node [anchor=north west][inner sep=0.75pt]  [font=\footnotesize]  {$Y$};
\draw (176.6,102.95) node [anchor=north west][inner sep=0.75pt]  [font=\footnotesize]  {$Y$};
\draw (316.6,91.95) node [anchor=north west][inner sep=0.75pt]    {$ \ + \   \mathcal{O}( \lambda_{3\rm B} )$};
\draw (254.6,102.95) node [anchor=north west][inner sep=0.75pt]  [font=\footnotesize]  {$Y$};
\draw (235,92.4) node [anchor=north west][inner sep=0.75pt]    {$+$};
\draw (297,95.4) node [anchor=north west][inner sep=0.75pt]  [font=\scriptsize]  {$\times $};

\end{tikzpicture}
\end{equation}
We now encounter a problem. It seems like to expand to higher orders, we need also the perturbative expansion of the second derivative of $W$ due to the third term in Eq.~\eqref{eq:SDEConnected}. Thus, we need a separate SDE for the second derivative. This can be easily obtained by differentiating the SDE for the first derivative. Similarly, to go to higher and higher orders, we will require further derivatives of the Schwinger-Dyson equations. The problem of setting up the perturbative expansion is equivalent to the problem of self-consistently solving the SDE in perturbation theory. We refer the reader to \cite{Cvitanovic:1983eb} for the details and provide the answer here. We tabulate all the diagrams that contribute to the generating functional, with vanishing bulk source, in table \ref{tab:PertExpgrSK} below.

\begin{table}[H]
    \centering
    \begin{tabular}{c|c|c|c|c}
        \ & 
\tikzset{every picture/.style={line width=0.75pt}} 

\begin{tikzpicture}[x=0.75pt,y=0.75pt,yscale=-1,xscale=1,baseline=(current bounding box.center)]
\centering

\draw    (82.33,123) -- (126.33,122.92) ;

\draw (119,118.4) node [anchor=north west][inner sep=0.75pt]  [font=\scriptsize]  {$\otimes $};
\draw (75,118.4) node [anchor=north west][inner sep=0.75pt]  [font=\scriptsize]  {$\otimes $};

\end{tikzpicture}
\\
 & \ & \ & \ \\
 \hline
\tikzset{every picture/.style={line width=0.75pt}} 

\begin{tikzpicture}[x=0.75pt,y=0.75pt,yscale=-1,xscale=1,baseline=(current bounding box.center)]
\centering

\draw    (313.33,159) -- (344.33,158.96) ;
\draw   (344.33,158.96) .. controls (344.33,150.7) and (351.05,144) .. (359.33,144) .. controls (367.62,144) and (374.33,150.7) .. (374.33,158.96) .. controls (374.33,167.22) and (367.62,173.92) .. (359.33,173.92) .. controls (351.05,173.92) and (344.33,167.22) .. (344.33,158.96) -- cycle ;

\draw (294,143.4) node [anchor=north west][inner sep=0.75pt]  [font=\footnotesize]  {$\frac{1}{2!}$};
\draw (306.33,154.4) node [anchor=north west][inner sep=0.75pt]  [font=\scriptsize]  {$\otimes $};

\end{tikzpicture}
 & \  & 

\tikzset{every picture/.style={line width=0.75pt}} 

\begin{tikzpicture}[x=0.75pt,y=0.75pt,yscale=-1,xscale=1,baseline=(current bounding box.center)]
\centering

\draw    (445.29,124) -- (474.29,124) ;
\draw    (474.29,124) -- (490.81,101.93) ;
\draw    (474.29,124) -- (492.57,144.9) ;

\draw (485,95.4) node [anchor=north west][inner sep=0.75pt]  [font=\scriptsize]  {$\otimes $};
\draw (487.57,140.3) node [anchor=north west][inner sep=0.75pt]  [font=\scriptsize]  {$\otimes $};
\draw (438,118.4) node [anchor=north west][inner sep=0.75pt]  [font=\scriptsize]  {$\otimes $};

\end{tikzpicture}
 & \ & \ \\
\hline
        \ & 

\tikzset{every picture/.style={line width=0.75pt}} 

\begin{tikzpicture}[x=0.75pt,y=0.75pt,yscale=-1,xscale=1,baseline=(current bounding box.center)]
\centering

\draw    (159,50.6) -- (181.37,50.6) ;
\draw   (181.37,50.6) .. controls (181.37,44.35) and (186.67,39.3) .. (193.21,39.3) .. controls (199.74,39.3) and (205.04,44.35) .. (205.04,50.6) .. controls (205.04,56.84) and (199.74,61.9) .. (193.21,61.9) .. controls (186.67,61.9) and (181.37,56.84) .. (181.37,50.6) -- cycle ;
\draw    (205.04,51.6) -- (227,51.6) ;
\draw    (165.33,77) -- (194.33,77) ;
\draw    (194.33,77) -- (221,77.6) ;
\draw    (194.33,77) -- (194,97.6) ;
\draw   (185.31,106.1) .. controls (185.31,101.41) and (189.2,97.6) .. (194,97.6) .. controls (198.8,97.6) and (202.69,101.41) .. (202.69,106.1) .. controls (202.69,110.79) and (198.8,114.6) .. (194,114.6) .. controls (189.2,114.6) and (185.31,110.79) .. (185.31,106.1) -- cycle ;

\draw (220,46.4) node [anchor=north west][inner sep=0.75pt]  [font=\scriptsize]  {$\otimes $};
\draw (158,72.4) node [anchor=north west][inner sep=0.75pt]  [font=\scriptsize]  {$\otimes $};
\draw (151,45.4) node [anchor=north west][inner sep=0.75pt]  [font=\scriptsize]  {$\otimes $};
\draw (215,72.4) node [anchor=north west][inner sep=0.75pt]  [font=\scriptsize]  {$\otimes $};
\draw (132,35.4) node [anchor=north west][inner sep=0.75pt]  [font=\footnotesize]  {$\frac{1}{2!}$};
\draw (134,79.4) node [anchor=north west][inner sep=0.75pt]  [font=\footnotesize]  {$\frac{1}{2!}$};

\end{tikzpicture}
 & \ & 

\tikzset{every picture/.style={line width=0.75pt}} 

\begin{tikzpicture}[x=0.75pt,y=0.75pt,yscale=-1,xscale=1,baseline=(current bounding box.center)]
\centering

\draw    (562.57,239.73) -- (594.29,240) ;
\draw    (594.29,240) -- (610.81,217.93) ;
\draw    (594.29,240) -- (612.57,260.9) ;
\draw    (562.57,239.73) -- (546.47,262.11) ;
\draw    (562.57,239.73) -- (543.9,219.18) ;

\draw (605,211.4) node [anchor=north west][inner sep=0.75pt]  [font=\scriptsize]  {$\otimes $};
\draw (607.57,256.3) node [anchor=north west][inner sep=0.75pt]  [font=\scriptsize]  {$\otimes $};
\draw (552.4,268.53) node [anchor=north west][inner sep=0.75pt]  [font=\scriptsize,rotate=-178.92]  {$\otimes $};
\draw (548.98,223.68) node [anchor=north west][inner sep=0.75pt]  [font=\scriptsize,rotate=-178.92]  {$\otimes $};
\end{tikzpicture}
 & \ \\
 \hline

\tikzset{every picture/.style={line width=0.75pt}} 

\begin{tikzpicture}[x=0.75pt,y=0.75pt,yscale=-1,xscale=1,baseline=(current bounding box.center)]

\draw    (269.33,51) -- (300.33,50.96) ;
\draw   (300.33,50.96) .. controls (300.33,42.7) and (307.05,36) .. (315.33,36) .. controls (323.62,36) and (330.33,42.7) .. (330.33,50.96) .. controls (330.33,59.22) and (323.62,65.92) .. (315.33,65.92) .. controls (307.05,65.92) and (300.33,59.22) .. (300.33,50.96) -- cycle ;
\draw    (315.33,36) -- (315.33,65.92) ;
\draw    (270.33,98) -- (301.33,97.96) ;
\draw   (301.33,97.7) .. controls (301.33,92.23) and (305.74,87.8) .. (311.17,87.8) .. controls (316.6,87.8) and (321,92.23) .. (321,97.7) .. controls (321,103.17) and (316.6,107.6) .. (311.17,107.6) .. controls (305.74,107.6) and (301.33,103.17) .. (301.33,97.7) -- cycle ;
\draw    (321.33,99) -- (335.33,98.96) ;
\draw   (335.33,98.7) .. controls (335.33,93.23) and (339.74,88.8) .. (345.17,88.8) .. controls (350.6,88.8) and (355,93.23) .. (355,98.7) .. controls (355,104.17) and (350.6,108.6) .. (345.17,108.6) .. controls (339.74,108.6) and (335.33,104.17) .. (335.33,98.7) -- cycle ;
\draw    (269.33,144) -- (298.33,144) ;
\draw    (298.33,144) -- (320,131.6) ;
\draw    (298.33,144) -- (322,158.6) ;
\draw   (322,158.6) .. controls (322,153.93) and (326.03,150.15) .. (331,150.15) .. controls (335.97,150.15) and (340,153.93) .. (340,158.6) .. controls (340,163.27) and (335.97,167.05) .. (331,167.05) .. controls (326.03,167.05) and (322,163.27) .. (322,158.6) -- cycle ;
\draw   (320,131.6) .. controls (320,126.93) and (324.03,123.15) .. (329,123.15) .. controls (333.97,123.15) and (338,126.93) .. (338,131.6) .. controls (338,136.27) and (333.97,140.05) .. (329,140.05) .. controls (324.03,140.05) and (320,136.27) .. (320,131.6) -- cycle ;

\draw (250,35.4) node [anchor=north west][inner sep=0.75pt]  [font=\footnotesize]  {$\frac{1}{2!}$};
\draw (262.33,46.4) node [anchor=north west][inner sep=0.75pt]  [font=\scriptsize]  {$\otimes $};
\draw (236,79.4) node [anchor=north west][inner sep=0.75pt]  [font=\footnotesize]  {$\left(\frac{1}{2!}\right)^{2}$};
\draw (263.33,93.4) node [anchor=north west][inner sep=0.75pt]  [font=\scriptsize]  {$\otimes $};
\draw (236,123.4) node [anchor=north west][inner sep=0.75pt]  [font=\footnotesize]  {$\left(\frac{1}{2!}\right)^{3}$};
\draw (262.33,139.4) node [anchor=north west][inner sep=0.75pt]  [font=\scriptsize]  {$\otimes $};

\end{tikzpicture}
 & \ & 
\tikzset{every picture/.style={line width=0.75pt}} 

\begin{tikzpicture}[x=0.75pt,y=0.75pt,yscale=-0.75,xscale=0.75,baseline=(current bounding box.center)]
\centering

\draw    (190,158.6) -- (241.29,158) ;
\draw    (241.29,158) -- (257.81,135.93) ;
\draw    (241.29,158) -- (259.57,178.9) ;
\draw    (226.79,158) -- (227,135.6) ;
\draw   (226.45,134.97) .. controls (221.04,135.03) and (216.6,130.67) .. (216.54,125.24) .. controls (216.48,119.81) and (220.82,115.36) .. (226.23,115.3) .. controls (231.65,115.24) and (236.08,119.6) .. (236.14,125.03) .. controls (236.2,130.46) and (231.86,134.91) .. (226.45,134.97) -- cycle ;
\draw    (190.29,87) -- (219.29,87) ;
\draw    (234.29,79) -- (250.81,56.93) ;
\draw    (234.29,95) -- (252.57,115.9) ;
\draw   (229.2,96.73) .. controls (223.79,96.79) and (219.35,92.43) .. (219.29,87) .. controls (219.23,81.57) and (223.57,77.12) .. (228.99,77.06) .. controls (234.4,77) and (238.83,81.36) .. (238.89,86.79) .. controls (238.95,92.22) and (234.61,96.67) .. (229.2,96.73) -- cycle ;
\draw    (190,205.6) -- (212.37,205.6) ;
\draw   (212.37,205.6) .. controls (212.37,199.35) and (217.67,194.3) .. (224.21,194.3) .. controls (230.74,194.3) and (236.04,199.35) .. (236.04,205.6) .. controls (236.04,211.84) and (230.74,216.9) .. (224.21,216.9) .. controls (217.67,216.9) and (212.37,211.84) .. (212.37,205.6) -- cycle ;
\draw    (236.04,206.6) -- (258,206.6) ;
\draw    (257.29,207) -- (273.81,184.93) ;
\draw    (257.29,207) -- (275.57,227.9) ;

\draw (252,129.4) node [anchor=north west][inner sep=0.75pt]  [font=\footnotesize]  {$\otimes $};
\draw (254.57,174.3) node [anchor=north west][inner sep=0.75pt]  [font=\footnotesize]  {$\otimes $};
\draw (183,153.4) node [anchor=north west][inner sep=0.75pt]  [font=\footnotesize]  {$\otimes $};
\draw (168,143.4) node [anchor=north west][inner sep=0.75pt]  [font=\footnotesize]  {$\frac{1}{2!}$};
\draw (245,50.4) node [anchor=north west][inner sep=0.75pt]  [font=\footnotesize]  {$\otimes $};
\draw (247.57,111.3) node [anchor=north west][inner sep=0.75pt]  [font=\footnotesize]  {$\otimes $};
\draw (183,81.4) node [anchor=north west][inner sep=0.75pt]  [font=\footnotesize]  {$\otimes $};
\draw (166,193.4) node [anchor=north west][inner sep=0.75pt]  [font=\footnotesize]  {$\frac{1}{2!}$};
\draw (182,200.4) node [anchor=north west][inner sep=0.75pt]  [font=\scriptsize]  {$\otimes $};
\draw (268,178.4) node [anchor=north west][inner sep=0.75pt]  [font=\scriptsize]  {$\otimes $};
\draw (270.57,223.3) node [anchor=north west][inner sep=0.75pt]  [font=\scriptsize]  {$\otimes $};

\end{tikzpicture}
 & \ & 
\tikzset{every picture/.style={line width=0.75pt}} 
\begin{tikzpicture}[x=0.75pt,y=0.75pt,yscale=-1,xscale=1,baseline=(current bounding box.center)]
\centering

\draw    (222.6,227.84) -- (254.32,228.11) ;
\draw    (254.32,228.11) -- (270.84,206.04) ;
\draw    (254.32,228.11) -- (272.6,249.01) ;
\draw    (222.6,227.84) -- (206.5,250.22) ;
\draw    (222.6,227.84) -- (203.93,207.29) ;
\draw    (238.46,227.97) -- (238,205.6) ;

\draw (265.03,199.51) node [anchor=north west][inner sep=0.75pt]  [font=\scriptsize]  {$\otimes $};
\draw (267.6,244.41) node [anchor=north west][inner sep=0.75pt]  [font=\scriptsize]  {$\otimes $};
\draw (212.43,256.64) node [anchor=north west][inner sep=0.75pt]  [font=\scriptsize,rotate=-178.92]  {$\otimes $};
\draw (209.01,211.79) node [anchor=north west][inner sep=0.75pt]  [font=\scriptsize,rotate=-178.92]  {$\otimes $};
\draw (231.03,199.51) node [anchor=north west][inner sep=0.75pt]  [font=\scriptsize]  {$\otimes $};

\end{tikzpicture}
 \\
    \end{tabular}
    \caption{Diagrams contributing to $W[\mathfrak{J}=0, J_{\sR}, J_{\sL}]$ to $O(\lambda_{3\rm B}^3)$. The diagrams are arranged according to the number of boundary sources along the rows, and according to the number of coupling constant along the columns.}
    \label{tab:PertExpgrSK}
\end{table}

\subsection{Exterior QFT}\label{sec:ExtQFT}

Now that we understand how to formulate quantum field theory on the grSK spacetime, we turn to repeat the same analysis for the exterior field theory. This will yield a quantum field theory defined entirely outside the horizon. Importantly, we will independently propose the Schwinger-Dyson equations for this exterior theory --- separate from those proposed in the grSK geometry. By the end of this section, we will thus have two distinct theories --- one defined on the grSK geometry and the other in the black hole exterior --- both of which compute the boundary Schwinger-Keldysh generating functional.

In the following section, we will demonstrate that these two formulations are, in fact, equivalent: they yield identical boundary generating functionals in perturbation theory. We emphasize once more that this equivalence is the central result of this paper. Let us begin by proposing the SDEs for the \emph{exterior QFT} --- a QFT living outside of the black hole.

As always, the quantity we eventually aim to compute is the Schwinger-Keldysh (SK) generating functional for connected boundary Green functions. As in the previous section, we begin by introducing a bulk generating functional, $W$, which depends on sources defined in the bulk. However, the key difference here is that, instead of a single source distributed over the entire grSK geometry, we now consider two distinct sources --- both confined to one copy of the black hole exterior. The reason for this doubling of sources is rooted in the structure of the SK formalism itself: it inherently involves a duplication of fields and sources to correctly account for real-time dynamics. The same logic applies here to the construction of an exterior QFT within this formalism.

We will choose to call these two bulk sources $\mathfrak{J}_{\Pb}(r,k)$ and $\mathfrak{J}_{\Fb}(r,k)$. Here, we are simply using the past-future basis (see Eq.~\eqref{eq:DefBasisPF}) for the bulk sources as well. As the reader might recall from Sec.~(\ref{sec:grSKeFTReview}), this is the most convenient basis for the description of the classical exterior field theory. We continue to use this basis for the quantum theory as well. Note that we explicitly write the dependences of the sources on the boundary momenta as well as the radial coordinate for clarity. In particular, we use the Schwarzschild radial coordinate $r$ instead of the mock tortoise coordinate $\zeta$ to suggest that the sources are on a single copy of the spacetime rather than on the entire grSK geometry.

Since the bulk generating functional now has two distinct sources, there will naturally be two appropriate copies of Schwinger-Dyson equations; one with a derivative of $W$ with respect to the past source, and another with a derivative with respect to the future source. Instead of writing down tedious expressions, we will resort to the kind of diagrammatics we have found to be extremely useful in the previous section.

The two Schwinger-Dyson equations take the form
\begin{equation}
    \tikzset{every picture/.style={line width=0.75pt}} 
    \begin{tikzpicture}[x=0.75pt,y=0.75pt,yscale=-1,xscale=1]
    \centering
    
    \draw    (19,77.6) -- (72.33,77.94) ;
    \draw [shift={(75.33,77.96)}, rotate = 180.37] [fill={rgb, 255:red, 0; green, 0; blue, 0 }  ][line width=0.08]  [draw opacity=0] (8.93,-4.29) -- (0,0) -- (8.93,4.29) -- cycle    ;
    \draw    (177.33,76) -- (199.3,76.36) ;
    \draw [shift={(202.3,76.41)}, rotate = 180.94] [fill={rgb, 255:red, 0; green, 0; blue, 0 }  ][line width=0.08]  [draw opacity=0] (8.93,-4.29) -- (0,0) -- (8.93,4.29) -- cycle    ;
    \draw    (283.33,77) -- (309.33,77) ;
    \draw [shift={(312.33,77)}, rotate = 180] [fill={rgb, 255:red, 0; green, 0; blue, 0 }  ][line width=0.08]  [draw opacity=0] (8.93,-4.29) -- (0,0) -- (8.93,4.29) -- cycle    ;
    \draw   (530.33,55.96) .. controls (530.33,47.7) and (537.05,41) .. (545.33,41) .. controls (553.62,41) and (560.33,47.7) .. (560.33,55.96) .. controls (560.33,64.22) and (553.62,70.92) .. (545.33,70.92) .. controls (537.05,70.92) and (530.33,64.22) .. (530.33,55.96) -- cycle ;
    \draw    (530.61,52.9) -- (545.33,41) ;
    \draw    (548.99,41.48) -- (530.37,56.6) ;
    \draw    (530.85,59.93) -- (552.22,42.55) ;
    \draw    (532.04,63.02) -- (554.84,44.33) ;
    \draw    (533.81,65.48) -- (541.38,59.23) -- (556.85,46.44) ;
    \draw    (535.98,67.55) -- (558.54,48.86) ;
    \draw    (538.37,69.21) -- (559.62,51.48) ;
    \draw    (541.35,70.29) -- (560.33,54.57) ;
    \draw    (544.46,71) -- (560.09,58.26) ;
    \draw    (549.47,70.29) -- (558.54,62.9) ;
    \draw    (539.68,42.07) -- (533.12,47.43) ;
    \draw   (532.33,92.96) .. controls (532.33,84.7) and (539.05,78) .. (547.33,78) .. controls (555.62,78) and (562.33,84.7) .. (562.33,92.96) .. controls (562.33,101.22) and (555.62,107.92) .. (547.33,107.92) .. controls (539.05,107.92) and (532.33,101.22) .. (532.33,92.96) -- cycle ;
    \draw    (532.61,89.9) -- (547.33,78) ;
    \draw    (550.99,78.48) -- (532.37,93.6) ;
    \draw    (532.85,96.93) -- (554.22,79.55) ;
    \draw    (534.04,100.02) -- (556.84,81.33) ;
    \draw    (535.81,102.48) -- (543.38,96.23) -- (558.85,83.44) ;
    \draw    (537.98,104.55) -- (560.54,85.86) ;
    \draw    (540.37,106.21) -- (561.62,88.48) ;
    \draw    (543.35,107.29) -- (562.33,91.57) ;
    \draw    (546.46,108) -- (562.09,95.26) ;
    \draw    (551.47,107.29) -- (560.54,99.9) ;
    \draw    (541.68,79.07) -- (535.12,84.43) ;
    \draw    (312.33,77) -- (324.33,77) ;
    \draw [shift={(312.33,77)}, rotate = 180] [color={rgb, 255:red, 0; green, 0; blue, 0 }  ][line width=0.75]    (0,5.59) -- (0,-5.59)   ;
    \draw    (324.33,77) .. controls (325.29,71.24) and (333.32,56.82) .. (353.53,64.37) ;
    \draw [shift={(356.12,65.43)}, rotate = 203.96] [fill={rgb, 255:red, 0; green, 0; blue, 0 }  ][line width=0.08]  [draw opacity=0] (8.93,-4.29) -- (0,0) -- (8.93,4.29) -- cycle    ;
    \draw    (324.33,77) .. controls (325.94,83.37) and (336.24,98.49) .. (356.72,86.91) ;
    \draw [shift={(358.98,85.55)}, rotate = 147.41] [fill={rgb, 255:red, 0; green, 0; blue, 0 }  ][line width=0.08]  [draw opacity=0] (8.93,-4.29) -- (0,0) -- (8.93,4.29) -- cycle    ;
    \draw    (450.33,77) -- (476.33,77) ;
    \draw [shift={(479.33,77)}, rotate = 180] [fill={rgb, 255:red, 0; green, 0; blue, 0 }  ][line width=0.08]  [draw opacity=0] (8.93,-4.29) -- (0,0) -- (8.93,4.29) -- cycle    ;
    \draw    (479.33,77) -- (491.33,77) ;
    \draw [shift={(479.33,77)}, rotate = 180] [color={rgb, 255:red, 0; green, 0; blue, 0 }  ][line width=0.75]    (0,5.59) -- (0,-5.59)   ;
    \draw    (491.33,77) .. controls (502.53,64.14) and (505.75,58.09) .. (527.25,57.16) ;
    \draw [shift={(530.03,57.07)}, rotate = 178.74] [fill={rgb, 255:red, 0; green, 0; blue, 0 }  ][line width=0.08]  [draw opacity=0] (8.93,-4.29) -- (0,0) -- (8.93,4.29) -- cycle    ;
    \draw    (491.33,77) .. controls (492.31,71.15) and (495.82,99.14) .. (530.64,98.71) ;
    \draw [shift={(533.38,98.62)}, rotate = 176.97] [fill={rgb, 255:red, 0; green, 0; blue, 0 }  ][line width=0.08]  [draw opacity=0] (8.93,-4.29) -- (0,0) -- (8.93,4.29) -- cycle    ;
    \draw    (202.3,76.41) -- (232.33,76.67) ;
    \draw [shift={(202.3,76.41)}, rotate = 180.49] [color={rgb, 255:red, 0; green, 0; blue, 0 }  ][line width=0.75]    (0,5.59) -- (0,-5.59)   ;
    \draw   (75.33,77.96) .. controls (75.33,69.7) and (82.05,63) .. (90.33,63) .. controls (98.62,63) and (105.33,69.7) .. (105.33,77.96) .. controls (105.33,86.22) and (98.62,92.92) .. (90.33,92.92) .. controls (82.05,92.92) and (75.33,86.22) .. (75.33,77.96) -- cycle ;
    \draw    (75.61,74.9) -- (90.33,63) ;
    \draw    (93.99,63.48) -- (75.37,78.6) ;
    \draw    (75.85,81.93) -- (97.22,64.55) ;
    \draw    (77.04,85.02) -- (99.84,66.33) ;
    \draw    (78.81,87.48) -- (86.38,81.23) -- (101.85,68.44) ;
    \draw    (80.98,89.55) -- (103.54,70.86) ;
    \draw    (83.37,91.21) -- (104.62,73.48) ;
    \draw    (86.35,92.29) -- (105.33,76.57) ;
    \draw    (89.46,93) -- (105.09,80.26) ;
    \draw    (94.47,92.29) -- (103.54,84.9) ;
    \draw    (84.68,64.07) -- (78.12,69.43) ;
    \draw    (176.33,160) -- (198.3,160.36) ;
    \draw [shift={(201.3,160.41)}, rotate = 180.94] [fill={rgb, 255:red, 0; green, 0; blue, 0 }  ][line width=0.08]  [draw opacity=0] (8.93,-4.29) -- (0,0) -- (8.93,4.29) -- cycle    ;
    \draw    (201.3,160.41) -- (231.33,160.67) ;
    \draw [shift={(201.3,160.41)}, rotate = 180.49] [color={rgb, 255:red, 0; green, 0; blue, 0 }  ][line width=0.75]    (0,5.59) -- (0,-5.59)   ;
    \draw   (353.33,73.96) .. controls (353.33,65.7) and (360.05,59) .. (368.33,59) .. controls (376.62,59) and (383.33,65.7) .. (383.33,73.96) .. controls (383.33,82.22) and (376.62,88.92) .. (368.33,88.92) .. controls (360.05,88.92) and (353.33,82.22) .. (353.33,73.96) -- cycle ;
    \draw    (353.61,70.9) -- (368.33,59) ;
    \draw    (371.99,59.48) -- (353.37,74.6) ;
    \draw    (353.85,77.93) -- (375.22,60.55) ;
    \draw    (355.04,81.02) -- (377.84,62.33) ;
    \draw    (356.81,83.48) -- (364.38,77.23) -- (379.85,64.44) ;
    \draw    (358.98,85.55) -- (381.54,66.86) ;
    \draw    (361.37,87.21) -- (382.62,69.48) ;
    \draw    (364.35,88.29) -- (383.33,72.57) ;
    \draw    (367.46,89) -- (383.09,76.26) ;
    \draw    (372.47,88.29) -- (381.54,80.9) ;
    \draw    (362.68,60.07) -- (356.12,65.43) ;
    \draw    (284.33,163) -- (310.33,163) ;
    \draw [shift={(313.33,163)}, rotate = 180] [fill={rgb, 255:red, 0; green, 0; blue, 0 }  ][line width=0.08]  [draw opacity=0] (8.93,-4.29) -- (0,0) -- (8.93,4.29) -- cycle    ;
    \draw    (313.33,163) -- (325.33,163) ;
    \draw [shift={(313.33,163)}, rotate = 180] [color={rgb, 255:red, 0; green, 0; blue, 0 }  ][line width=0.75]    (0,5.59) -- (0,-5.59)   ;
    \draw    (325.33,163) .. controls (326.29,157.24) and (334.32,142.82) .. (354.53,150.37) ;
    \draw [shift={(357.12,151.43)}, rotate = 203.96] [fill={rgb, 255:red, 0; green, 0; blue, 0 }  ][line width=0.08]  [draw opacity=0] (8.93,-4.29) -- (0,0) -- (8.93,4.29) -- cycle    ;
    \draw    (325.33,163) .. controls (327,169.6) and (345,189.6) .. (359.98,171.55) ;
    \draw [shift={(359.98,171.55)}, rotate = 129.69] [color={rgb, 255:red, 0; green, 0; blue, 0 }  ][line width=0.75]    (0,5.59) -- (0,-5.59)   ;
    \draw   (354.33,159.96) .. controls (354.33,151.7) and (361.05,145) .. (369.33,145) .. controls (377.62,145) and (384.33,151.7) .. (384.33,159.96) .. controls (384.33,168.22) and (377.62,174.92) .. (369.33,174.92) .. controls (361.05,174.92) and (354.33,168.22) .. (354.33,159.96) -- cycle ;
    \draw    (354.61,156.9) -- (369.33,145) ;
    \draw    (372.99,145.48) -- (354.37,160.6) ;
    \draw    (354.85,163.93) -- (376.22,146.55) ;
    \draw    (356.04,167.02) -- (378.84,148.33) ;
    \draw    (357.81,169.48) -- (365.38,163.23) -- (380.85,150.44) ;
    \draw    (359.98,171.55) -- (382.54,152.86) ;
    \draw    (362.37,173.21) -- (383.62,155.48) ;
    \draw    (365.35,174.29) -- (384.33,158.57) ;
    \draw    (368.46,175) -- (384.09,162.26) ;
    \draw    (373.47,174.29) -- (382.54,166.9) ;
    \draw    (363.68,146.07) -- (357.12,151.43) ;
    \draw   (533.33,145.96) .. controls (533.33,137.7) and (540.05,131) .. (548.33,131) .. controls (556.62,131) and (563.33,137.7) .. (563.33,145.96) .. controls (563.33,154.22) and (556.62,160.92) .. (548.33,160.92) .. controls (540.05,160.92) and (533.33,154.22) .. (533.33,145.96) -- cycle ;
    \draw    (533.61,142.9) -- (548.33,131) ;
    \draw    (551.99,131.48) -- (533.37,146.6) ;
    \draw    (533.85,149.93) -- (555.22,132.55) ;
    \draw    (535.04,153.02) -- (557.84,134.33) ;
    \draw    (536.81,155.48) -- (544.38,149.23) -- (559.85,136.44) ;
    \draw    (538.98,157.55) -- (561.54,138.86) ;
    \draw    (541.37,159.21) -- (562.62,141.48) ;
    \draw    (544.35,160.29) -- (563.33,144.57) ;
    \draw    (547.46,161) -- (563.09,148.26) ;
    \draw    (552.47,160.29) -- (561.54,152.9) ;
    \draw    (542.68,132.07) -- (536.12,137.43) ;
    \draw   (535.33,182.96) .. controls (535.33,174.7) and (542.05,168) .. (550.33,168) .. controls (558.62,168) and (565.33,174.7) .. (565.33,182.96) .. controls (565.33,191.22) and (558.62,197.92) .. (550.33,197.92) .. controls (542.05,197.92) and (535.33,191.22) .. (535.33,182.96) -- cycle ;
    \draw    (535.61,179.9) -- (550.33,168) ;
    \draw    (553.99,168.48) -- (535.37,183.6) ;
    \draw    (535.85,186.93) -- (557.22,169.55) ;
    \draw    (537.04,190.02) -- (559.84,171.33) ;
    \draw    (538.81,192.48) -- (546.38,186.23) -- (561.85,173.44) ;
    \draw    (540.98,194.55) -- (563.54,175.86) ;
    \draw    (543.37,196.21) -- (564.62,178.48) ;
    \draw    (546.35,197.29) -- (565.33,181.57) ;
    \draw    (549.46,198) -- (565.09,185.26) ;
    \draw    (554.47,197.29) -- (563.54,189.9) ;
    \draw    (544.68,169.07) -- (538.12,174.43) ;
    \draw    (453.33,167) -- (479.33,167) ;
    \draw [shift={(482.33,167)}, rotate = 180] [fill={rgb, 255:red, 0; green, 0; blue, 0 }  ][line width=0.08]  [draw opacity=0] (8.93,-4.29) -- (0,0) -- (8.93,4.29) -- cycle    ;
    \draw    (482.33,167) -- (494.33,167) ;
    \draw [shift={(482.33,167)}, rotate = 180] [color={rgb, 255:red, 0; green, 0; blue, 0 }  ][line width=0.75]    (0,5.59) -- (0,-5.59)   ;
    \draw    (494.33,167) .. controls (505.53,154.14) and (508.75,148.09) .. (530.25,147.16) ;
    \draw [shift={(533.03,147.07)}, rotate = 178.74] [fill={rgb, 255:red, 0; green, 0; blue, 0 }  ][line width=0.08]  [draw opacity=0] (8.93,-4.29) -- (0,0) -- (8.93,4.29) -- cycle    ;
    \draw    (494.33,167) .. controls (495.33,161) and (499,190.6) .. (536.38,188.62) ;
    \draw [shift={(536.38,188.62)}, rotate = 176.97] [color={rgb, 255:red, 0; green, 0; blue, 0 }  ][line width=0.75]    (0,5.59) -- (0,-5.59)   ;
    
    \draw (144,68.4) node [anchor=north west][inner sep=0.75pt]    {$=$};
    \draw (227,71.4) node [anchor=north west][inner sep=0.75pt]  [font=\scriptsize]  {$\times $};
    \draw (252,67.4) node [anchor=north west][inner sep=0.75pt]    {$+$};
    \draw (266,61.4) node [anchor=north west][inner sep=0.75pt]  [font=\footnotesize]  {$\frac{1}{2!}$};
    \draw (413,66.4) node [anchor=north west][inner sep=0.75pt]    {$+$};
    \draw (431,60.4) node [anchor=north west][inner sep=0.75pt]  [font=\footnotesize]  {$\frac{1}{2!}$};
    \draw (253,151.4) node [anchor=north west][inner sep=0.75pt]    {$+$};
    \draw (412,155.4) node [anchor=north west][inner sep=0.75pt]    {$+$};
    \draw (226,155.4) node [anchor=north west][inner sep=0.75pt]  [font=\scriptsize]  {$\otimes $};
    \draw (150,149.4) node [anchor=north west][inner sep=0.75pt]    {$+$};

    \end{tikzpicture} \ ,
\end{equation}
\begin{equation}
    \tikzset{every picture/.style={line width=0.75pt}} 
    \begin{tikzpicture}[x=0.75pt,y=0.75pt,yscale=-1,xscale=1]
    \centering
    
    \draw    (35,87.6) -- (91.33,87.96) ;
    \draw [shift={(91.33,87.96)}, rotate = 180.37] [color={rgb, 255:red, 0; green, 0; blue, 0 }  ][line width=0.75]    (0,5.59) -- (0,-5.59)   ;
    \draw    (193.33,86) -- (218.3,86.41) ;
    \draw [shift={(218.3,86.41)}, rotate = 180.94] [color={rgb, 255:red, 0; green, 0; blue, 0 }  ][line width=0.75]    (0,5.59) -- (0,-5.59)   ;
    \draw    (299.33,87) -- (319.33,87) ;
    \draw [shift={(319.33,87)}, rotate = 180] [color={rgb, 255:red, 0; green, 0; blue, 0 }  ][line width=0.75]    (0,5.59) -- (0,-5.59)   ;
    \draw   (546.33,65.96) .. controls (546.33,57.7) and (553.05,51) .. (561.33,51) .. controls (569.62,51) and (576.33,57.7) .. (576.33,65.96) .. controls (576.33,74.22) and (569.62,80.92) .. (561.33,80.92) .. controls (553.05,80.92) and (546.33,74.22) .. (546.33,65.96) -- cycle ;
    \draw    (546.61,62.9) -- (561.33,51) ;
    \draw    (564.99,51.48) -- (546.37,66.6) ;
    \draw    (546.85,69.93) -- (568.22,52.55) ;
    \draw    (548.04,73.02) -- (570.84,54.33) ;
    \draw    (549.81,75.48) -- (557.38,69.23) -- (572.85,56.44) ;
    \draw    (551.98,77.55) -- (574.54,58.86) ;
    \draw    (554.37,79.21) -- (575.62,61.48) ;
    \draw    (557.35,80.29) -- (576.33,64.57) ;
    \draw    (560.46,81) -- (576.09,68.26) ;
    \draw    (565.47,80.29) -- (574.54,72.9) ;
    \draw    (555.68,52.07) -- (549.12,57.43) ;
    \draw   (548.33,102.96) .. controls (548.33,94.7) and (555.05,88) .. (563.33,88) .. controls (571.62,88) and (578.33,94.7) .. (578.33,102.96) .. controls (578.33,111.22) and (571.62,117.92) .. (563.33,117.92) .. controls (555.05,117.92) and (548.33,111.22) .. (548.33,102.96) -- cycle ;
    \draw    (548.61,99.9) -- (563.33,88) ;
    \draw    (566.99,88.48) -- (548.37,103.6) ;
    \draw    (548.85,106.93) -- (570.22,89.55) ;
    \draw    (550.04,110.02) -- (572.84,91.33) ;
    \draw    (551.81,112.48) -- (559.38,106.23) -- (574.85,93.44) ;
    \draw    (553.98,114.55) -- (576.54,95.86) ;
    \draw    (556.37,116.21) -- (577.62,98.48) ;
    \draw    (559.35,117.29) -- (578.33,101.57) ;
    \draw    (562.46,118) -- (578.09,105.26) ;
    \draw    (567.47,117.29) -- (576.54,109.9) ;
    \draw    (557.68,89.07) -- (551.12,94.43) ;
    \draw    (322.33,87) -- (340.33,87) ;
    \draw [shift={(319.33,87)}, rotate = 0] [fill={rgb, 255:red, 0; green, 0; blue, 0 }  ][line width=0.08]  [draw opacity=0] (8.93,-4.29) -- (0,0) -- (8.93,4.29) -- cycle    ;
    \draw    (340.33,87) .. controls (341.33,81) and (350,65.6) .. (372.12,75.43) ;
    \draw [shift={(372.12,75.43)}, rotate = 203.96] [color={rgb, 255:red, 0; green, 0; blue, 0 }  ][line width=0.75]    (0,5.59) -- (0,-5.59)   ;
    \draw    (340.33,87) .. controls (342,93.6) and (353,109.6) .. (374.98,95.55) ;
    \draw [shift={(374.98,95.55)}, rotate = 147.41] [color={rgb, 255:red, 0; green, 0; blue, 0 }  ][line width=0.75]    (0,5.59) -- (0,-5.59)   ;
    \draw    (466.33,87) -- (487.33,87) ;
    \draw [shift={(487.33,87)}, rotate = 180] [color={rgb, 255:red, 0; green, 0; blue, 0 }  ][line width=0.75]    (0,5.59) -- (0,-5.59)   ;
    \draw    (490.33,87) -- (502,87) -- (507.33,87) ;
    \draw [shift={(487.33,87)}, rotate = 0] [fill={rgb, 255:red, 0; green, 0; blue, 0 }  ][line width=0.08]  [draw opacity=0] (8.93,-4.29) -- (0,0) -- (8.93,4.29) -- cycle    ;
    \draw    (507.33,87) .. controls (519,73.6) and (522,67.6) .. (546.03,67.07) ;
    \draw [shift={(546.03,67.07)}, rotate = 178.74] [color={rgb, 255:red, 0; green, 0; blue, 0 }  ][line width=0.75]    (0,5.59) -- (0,-5.59)   ;
    \draw    (506.81,87.31) .. controls (507.81,81.31) and (511.47,110.91) .. (548.85,108.93) ;
    \draw [shift={(548.85,108.93)}, rotate = 176.97] [color={rgb, 255:red, 0; green, 0; blue, 0 }  ][line width=0.75]    (0,5.59) -- (0,-5.59)   ;
    \draw    (221.3,86.44) -- (248.33,86.67) ;
    \draw [shift={(218.3,86.41)}, rotate = 0.49] [fill={rgb, 255:red, 0; green, 0; blue, 0 }  ][line width=0.08]  [draw opacity=0] (8.93,-4.29) -- (0,0) -- (8.93,4.29) -- cycle    ;
    \draw   (91.33,87.96) .. controls (91.33,79.7) and (98.05,73) .. (106.33,73) .. controls (114.62,73) and (121.33,79.7) .. (121.33,87.96) .. controls (121.33,96.22) and (114.62,102.92) .. (106.33,102.92) .. controls (98.05,102.92) and (91.33,96.22) .. (91.33,87.96) -- cycle ;
    \draw    (91.61,84.9) -- (106.33,73) ;
    \draw    (109.99,73.48) -- (91.37,88.6) ;
    \draw    (91.85,91.93) -- (113.22,74.55) ;
    \draw    (93.04,95.02) -- (115.84,76.33) ;
    \draw    (94.81,97.48) -- (102.38,91.23) -- (117.85,78.44) ;
    \draw    (96.98,99.55) -- (119.54,80.86) ;
    \draw    (99.37,101.21) -- (120.62,83.48) ;
    \draw    (102.35,102.29) -- (121.33,86.57) ;
    \draw    (105.46,103) -- (121.09,90.26) ;
    \draw    (110.47,102.29) -- (119.54,94.9) ;
    \draw    (100.68,74.07) -- (94.12,79.43) ;
    \draw   (369.33,83.96) .. controls (369.33,75.7) and (376.05,69) .. (384.33,69) .. controls (392.62,69) and (399.33,75.7) .. (399.33,83.96) .. controls (399.33,92.22) and (392.62,98.92) .. (384.33,98.92) .. controls (376.05,98.92) and (369.33,92.22) .. (369.33,83.96) -- cycle ;
    \draw    (369.61,80.9) -- (384.33,69) ;
    \draw    (387.99,69.48) -- (369.37,84.6) ;
    \draw    (369.85,87.93) -- (391.22,70.55) ;
    \draw    (371.04,91.02) -- (393.84,72.33) ;
    \draw    (372.81,93.48) -- (380.38,87.23) -- (395.85,74.44) ;
    \draw    (374.98,95.55) -- (397.54,76.86) ;
    \draw    (377.37,97.21) -- (398.62,79.48) ;
    \draw    (380.35,98.29) -- (399.33,82.57) ;
    \draw    (383.46,99) -- (399.09,86.26) ;
    \draw    (388.47,98.29) -- (397.54,90.9) ;
    \draw    (378.68,70.07) -- (372.12,75.43) ;
    \draw    (300.33,173) -- (321.33,173) ;
    \draw [shift={(321.33,173)}, rotate = 180] [color={rgb, 255:red, 0; green, 0; blue, 0 }  ][line width=0.75]    (0,5.59) -- (0,-5.59)   ;
    \draw    (324.33,173) -- (341.33,173) ;
    \draw [shift={(321.33,173)}, rotate = 0] [fill={rgb, 255:red, 0; green, 0; blue, 0 }  ][line width=0.08]  [draw opacity=0] (8.93,-4.29) -- (0,0) -- (8.93,4.29) -- cycle    ;
    \draw    (341.33,173) .. controls (342.29,167.24) and (350.32,152.82) .. (370.53,160.37) ;
    \draw [shift={(373.12,161.43)}, rotate = 203.96] [fill={rgb, 255:red, 0; green, 0; blue, 0 }  ][line width=0.08]  [draw opacity=0] (8.93,-4.29) -- (0,0) -- (8.93,4.29) -- cycle    ;
    \draw    (341.33,173) .. controls (343,179.6) and (361,199.6) .. (375.98,181.55) ;
    \draw [shift={(375.98,181.55)}, rotate = 129.69] [color={rgb, 255:red, 0; green, 0; blue, 0 }  ][line width=0.75]    (0,5.59) -- (0,-5.59)   ;
    \draw   (370.33,169.96) .. controls (370.33,161.7) and (377.05,155) .. (385.33,155) .. controls (393.62,155) and (400.33,161.7) .. (400.33,169.96) .. controls (400.33,178.22) and (393.62,184.92) .. (385.33,184.92) .. controls (377.05,184.92) and (370.33,178.22) .. (370.33,169.96) -- cycle ;
    \draw    (370.61,166.9) -- (385.33,155) ;
    \draw    (388.99,155.48) -- (370.37,170.6) ;
    \draw    (370.85,173.93) -- (392.22,156.55) ;
    \draw    (372.04,177.02) -- (394.84,158.33) ;
    \draw    (373.81,179.48) -- (381.38,173.23) -- (396.85,160.44) ;
    \draw    (375.98,181.55) -- (398.54,162.86) ;
    \draw    (378.37,183.21) -- (399.62,165.48) ;
    \draw    (381.35,184.29) -- (400.33,168.57) ;
    \draw    (384.46,185) -- (400.09,172.26) ;
    \draw    (389.47,184.29) -- (398.54,176.9) ;
    \draw    (379.68,156.07) -- (373.12,161.43) ;
    \draw   (549.33,155.96) .. controls (549.33,147.7) and (556.05,141) .. (564.33,141) .. controls (572.62,141) and (579.33,147.7) .. (579.33,155.96) .. controls (579.33,164.22) and (572.62,170.92) .. (564.33,170.92) .. controls (556.05,170.92) and (549.33,164.22) .. (549.33,155.96) -- cycle ;
    \draw    (549.61,152.9) -- (564.33,141) ;
    \draw    (567.99,141.48) -- (549.37,156.6) ;
    \draw    (549.85,159.93) -- (571.22,142.55) ;
    \draw    (551.04,163.02) -- (573.84,144.33) ;
    \draw    (552.81,165.48) -- (560.38,159.23) -- (575.85,146.44) ;
    \draw    (554.98,167.55) -- (577.54,148.86) ;
    \draw    (557.37,169.21) -- (578.62,151.48) ;
    \draw    (560.35,170.29) -- (579.33,154.57) ;
    \draw    (563.46,171) -- (579.09,158.26) ;
    \draw    (568.47,170.29) -- (577.54,162.9) ;
    \draw    (558.68,142.07) -- (552.12,147.43) ;
    \draw    (469.33,177) -- (488.33,177) ;
    \draw [shift={(488.33,177)}, rotate = 180] [color={rgb, 255:red, 0; green, 0; blue, 0 }  ][line width=0.75]    (0,5.59) -- (0,-5.59)   ;
    \draw    (491.33,177) -- (510.33,177) ;
    \draw [shift={(488.33,177)}, rotate = 0] [fill={rgb, 255:red, 0; green, 0; blue, 0 }  ][line width=0.08]  [draw opacity=0] (8.93,-4.29) -- (0,0) -- (8.93,4.29) -- cycle    ;
    \draw    (510.33,177) .. controls (521.53,164.14) and (524.75,158.09) .. (546.25,157.16) ;
    \draw [shift={(549.03,157.07)}, rotate = 178.74] [fill={rgb, 255:red, 0; green, 0; blue, 0 }  ][line width=0.08]  [draw opacity=0] (8.93,-4.29) -- (0,0) -- (8.93,4.29) -- cycle    ;
    \draw    (510.33,177) .. controls (511.33,171) and (515,200.6) .. (552.38,198.62) ;
    \draw [shift={(552.38,198.62)}, rotate = 176.97] [color={rgb, 255:red, 0; green, 0; blue, 0 }  ][line width=0.75]    (0,5.59) -- (0,-5.59)   ;
    \draw    (194.33,169) -- (219.3,169.41) ;
    \draw [shift={(219.3,169.41)}, rotate = 180.94] [color={rgb, 255:red, 0; green, 0; blue, 0 }  ][line width=0.75]    (0,5.59) -- (0,-5.59)   ;
    \draw    (222.3,169.44) -- (249.33,169.67) ;
    \draw [shift={(219.3,169.41)}, rotate = 0.49] [fill={rgb, 255:red, 0; green, 0; blue, 0 }  ][line width=0.08]  [draw opacity=0] (8.93,-4.29) -- (0,0) -- (8.93,4.29) -- cycle    ;
    \draw   (553.33,197.96) .. controls (553.33,189.7) and (560.05,183) .. (568.33,183) .. controls (576.62,183) and (583.33,189.7) .. (583.33,197.96) .. controls (583.33,206.22) and (576.62,212.92) .. (568.33,212.92) .. controls (560.05,212.92) and (553.33,206.22) .. (553.33,197.96) -- cycle ;
    \draw    (553.61,194.9) -- (568.33,183) ;
    \draw    (571.99,183.48) -- (553.37,198.6) ;
    \draw    (553.85,201.93) -- (575.22,184.55) ;
    \draw    (555.04,205.02) -- (577.84,186.33) ;
    \draw    (556.81,207.48) -- (564.38,201.23) -- (579.85,188.44) ;
    \draw    (558.98,209.55) -- (581.54,190.86) ;
    \draw    (561.37,211.21) -- (582.62,193.48) ;
    \draw    (564.35,212.29) -- (583.33,196.57) ;
    \draw    (567.46,213) -- (583.09,200.26) ;
    \draw    (572.47,212.29) -- (581.54,204.9) ;
    \draw    (562.68,184.07) -- (556.12,189.43) ;
    
    \draw (160,78.4) node [anchor=north west][inner sep=0.75pt]    {$=$};
    \draw (243,81.4) node [anchor=north west][inner sep=0.75pt]  [font=\scriptsize]  {$\times $};
    \draw (268,77.4) node [anchor=north west][inner sep=0.75pt]    {$+$};
    \draw (282,71.4) node [anchor=north west][inner sep=0.75pt]  [font=\footnotesize]  {$\frac{1}{2!}$};
    \draw (429,76.4) node [anchor=north west][inner sep=0.75pt]    {$+$};
    \draw (447,70.4) node [anchor=north west][inner sep=0.75pt]  [font=\footnotesize]  {$\frac{1}{2!}$};
    \draw (269,161.4) node [anchor=north west][inner sep=0.75pt]    {$+$};
    \draw (428,165.4) node [anchor=north west][inner sep=0.75pt]    {$+$};
    \draw (166,159.4) node [anchor=north west][inner sep=0.75pt]    {$+$};
    \draw (244,164.4) node [anchor=north west][inner sep=0.75pt]  [font=\scriptsize]  {$\otimes $};

    \end{tikzpicture} \ .
\end{equation}
Here, once again, as in the previous section, we use the blob to denote the bulk generating functional:
\begin{equation}
   i W[\mathfrak{J}, J] \equiv
\tikzset{every picture/.style={line width=0.75pt}} 
\begin{tikzpicture}[baseline={([yshift=-0.5ex]current bounding box.center)},x=0.75pt,y=0.75pt,yscale=-1,xscale=1]

\draw   (305.33,129.96) .. controls (305.33,121.7) and (312.05,115) .. (320.33,115) .. controls (328.62,115) and (335.33,121.7) .. (335.33,129.96) .. controls (335.33,138.22) and (328.62,144.92) .. (320.33,144.92) .. controls (312.05,144.92) and (305.33,138.22) .. (305.33,129.96) -- cycle ;
\draw    (305.61,126.9) -- (320.33,115) ;
\draw    (323.99,115.48) -- (305.37,130.6) ;
\draw    (305.85,133.93) -- (327.22,116.55) ;
\draw    (307.04,137.02) -- (329.84,118.33) ;
\draw    (308.81,139.48) -- (316.38,133.23) -- (331.85,120.44) ;
\draw    (310.98,141.55) -- (333.54,122.86) ;
\draw    (313.37,143.21) -- (334.62,125.48) ;
\draw    (316.35,144.29) -- (335.33,128.57) ;
\draw    (319.46,145) -- (335.09,132.26) ;
\draw    (324.47,144.29) -- (333.54,136.9) ;
\draw    (314.68,116.07) -- (308.12,121.43) ;
\end{tikzpicture}
 \ .
\end{equation}
Functional derivatives of the generating functional come with a line attached to this blob. Since there are two sources now, there must be two different kinds of lines. We denote them as follows:
\begin{equation}
    \frac{1}{i} \frac{\delta}{\delta \mathfrak{J}_{\Pb}(r,k)} {i W} \equiv
\tikzset{every picture/.style={line width=0.75pt}} 
\begin{tikzpicture}[baseline={([yshift=-0.5ex]current bounding box.center)},x=0.75pt,y=0.75pt,yscale=-1,xscale=1]

\draw    (367,170.6) -- (423.33,170.96) ;
\draw [shift={(423.33,170.96)}, rotate = 180.37] [color={rgb, 255:red, 0; green, 0; blue, 0 }  ][line width=0.75]    (0,5.59) -- (0,-5.59)   ;
\draw   (423.33,170.96) .. controls (423.33,162.7) and (430.05,156) .. (438.33,156) .. controls (446.62,156) and (453.33,162.7) .. (453.33,170.96) .. controls (453.33,179.22) and (446.62,185.92) .. (438.33,185.92) .. controls (430.05,185.92) and (423.33,179.22) .. (423.33,170.96) -- cycle ;
\draw    (423.61,167.9) -- (438.33,156) ;
\draw    (441.99,156.48) -- (423.37,171.6) ;
\draw    (423.85,174.93) -- (445.22,157.55) ;
\draw    (425.04,178.02) -- (447.84,159.33) ;
\draw    (426.81,180.48) -- (434.38,174.23) -- (449.85,161.44) ;
\draw    (428.98,182.55) -- (451.54,163.86) ;
\draw    (431.37,184.21) -- (452.62,166.48) ;
\draw    (434.35,185.29) -- (453.33,169.57) ;
\draw    (437.46,186) -- (453.09,173.26) ;
\draw    (442.47,185.29) -- (451.54,177.9) ;
\draw    (432.68,157.07) -- (426.12,162.43) ;
\end{tikzpicture} \ ,
\qquad
\frac{1}{i} \frac{\delta}{\delta \mathfrak{J}_{\Fb}(r,k)} {i W} \equiv
\tikzset{every picture/.style={line width=0.75pt}} 
\begin{tikzpicture}[baseline={([yshift=-0.5ex]current bounding box.center)},x=0.75pt,y=0.75pt,yscale=-1,xscale=1]

\draw    (335,133.6) -- (388.33,133.94) ;
\draw [shift={(391.33,133.96)}, rotate = 180.37] [fill={rgb, 255:red, 0; green, 0; blue, 0 }  ][line width=0.08]  [draw opacity=0] (8.93,-4.29) -- (0,0) -- (8.93,4.29) -- cycle    ;
\draw   (391.33,133.96) .. controls (391.33,125.7) and (398.05,119) .. (406.33,119) .. controls (414.62,119) and (421.33,125.7) .. (421.33,133.96) .. controls (421.33,142.22) and (414.62,148.92) .. (406.33,148.92) .. controls (398.05,148.92) and (391.33,142.22) .. (391.33,133.96) -- cycle ;
\draw    (391.61,130.9) -- (406.33,119) ;
\draw    (409.99,119.48) -- (391.37,134.6) ;
\draw    (391.85,137.93) -- (413.22,120.55) ;
\draw    (393.04,141.02) -- (415.84,122.33) ;
\draw    (394.81,143.48) -- (402.38,137.23) -- (417.85,124.44) ;
\draw    (396.98,145.55) -- (419.54,126.86) ;
\draw    (399.37,147.21) -- (420.62,129.48) ;
\draw    (402.35,148.29) -- (421.33,132.57) ;
\draw    (405.46,149) -- (421.09,136.26) ;
\draw    (410.47,148.29) -- (419.54,140.9) ;
\draw    (400.68,120.07) -- (394.12,125.43) ;
\end{tikzpicture} \ .
\end{equation}
For higher functional derivatives with respect to any of the sources, we just attach the appropriate line to the blob. We are left to specify the bulk sources, propagators and vertices. Let's start with the boundary-to-bulk propagators. We have already discussed these during the review of the classical exterior field theory in Sec.(\ref{sec:grSKeFTReview}) (see Eq.~\eqref{eq:BndryBlkPropEFT}). The same holds for the vertices (Tab.(\ref{tab:Vertices})) as well as the bulk-to-bulk propagator (Eq.~\eqref{eq:BlkBlkPropEFT}).

We have one final ingredient to introduce: the bulk source. The symbol $\times$ denotes a bulk source. Whenever a lines ends on it, it gives a bulk source, i.e., 
\begin{equation}
    \tikzset{every picture/.style={line width=0.75pt}}
    \begin{tikzpicture}[baseline={([yshift=-0.5ex]current bounding box.center)},x=0.75pt,y=0.75pt,yscale=-1,xscale=1]
    \draw    (474.33,136) -- (526.3,136.41) ;
    \draw [shift={(474.33,136)}, rotate = 180.45] [color={rgb, 255:red, 0; green, 0; blue, 0 }  ][line width=0.75]    (0,5.59) -- (0,-5.59)   ;
    \draw (521,130.4) node [anchor=north west][inner sep=0.75pt]    {$\times $};
    \end{tikzpicture} = i \mathfrak{J}_{\Pb}(r,k)\ , \qquad
    \tikzset{every picture/.style={line width=0.75pt}}
    \begin{tikzpicture}[baseline={([yshift=-0.5ex]current bounding box.center)},x=0.75pt,y=0.75pt,yscale=-1,xscale=1]
    \draw    (477.33,136.02) -- (526.3,136.41) ;
    \draw [shift={(474.33,136)}, rotate = 0.45] [fill={rgb, 255:red, 0; green, 0; blue, 0 }  ][line width=0.08]  [draw opacity=0] (8.93,-4.29) -- (0,0) -- (8.93,4.29) -- cycle    ;
    \draw (521,130.4) node [anchor=north west][inner sep=0.75pt]    {$\times $};
    \end{tikzpicture}= i \mathfrak{J}_{\Fb}(r,k) \ .
\end{equation}
We have already noted that we have two distinct bulk sources. We distinguish between these by the type of lines that emerge from the $\times$. If the emanating line ends in a semi-capacitor, the source is a past source. On the other hand, if the emanating line ends in a diode, then it is a future source.

We will now perturbatively expand the above Schwinger-Dyson equations in the bulk coupling $\lambda_{3\rm B}$. Furthermore, as we did in the previous section, we will integrate with respect to the bulk sources and write down the diagrams that contribute to the boundary generating functional. To set up the perturbative expansion, just as in the case of the grSK QFT, one needs to write down the higher-point SDEs by differentiating the ont-point SDEs with respect to the bulk sources. We do not do this explicitly here, but refer the interested reader to Appendix \ref{app:SDEsfromPI} for the diagrammatic equations.

After the dust settles, once again we will see that a structure as in table \ref{tab:PertExpgrSK} emerges. The uppermost diagonal is all of tree-level diagrams, the next diagonal one-loop, and so on. The major difference is that we have many more diagrams here since there are two distinct vertices rather than one. We do not draw these diagrams explicitly here, but note that all the topologies in each row and column of table \ref{tab:PertExpgrSK} will have descendants with arrows on the propagators and the corresponding two distinct vertices.

We can now look at the classical limit of the above SDEs, but we will postpone this discussion for now, since in the next section, we will show that the Exterior QFT SDEs produce exactly the same results as the grSK SDEs. This will in turn prove that the classical limit works out correctly.

\section{Monodromy integrals over the grSK contour}\label{app:MonInt}

In showing that the QFT defined by the SDEs on the grSK contour is equivalent to the QFT on the exterior, we required results for the monodromy integrals for propagators on the grSK contour. In this appendix, we will provide a general conjecture to perform such monodromy integrals for an arbitrary loop diagram on the grSK spacetime. We will also explicitly check this conjecture for simple diagrams with one and two loops.

We are interested in multiple integrals over the grSK contour given in Fig.~(\ref{fig:grSKsaddle}). The general form of the multiple discontinuity integral we would like to compute is
\begin{equation}
\oint_{\zeta_1} \ldots \oint_{\zeta_{n_\text{v}}} \prod_{i = 1}^{n_\text{v}} e^{\beta \Cnst_i (1- \zeta_i)} \prod_{\ell =1}^{n_{\text{e}}} \bbG(\zeta_{\ell_{\rm f}}|\zeta_{\ell_{\rm i}}, p_{\ell}) \mathscr{F}(\zeta_1, \ldots, \zeta_{n_{\text{v}}}) \ .
\label{eq:MonIntExpression}
\end{equation}
Here $\oint_{\zeta}$ denotes an integral over the grSK contour, $n_{\rm v}$ is the number of vertices, i.e., the number of bulk vertices that are being integrated, and $n_{\rm e}$ is the number of edges, i.e., the number of bulk-to-bulk propagators that are being integrated. The symbol $\mathscr{F}(\zeta_1,\ldots, \zeta_{n_{\rm v}})$ denotes a function that is analytic in the grSK keyhole in the $r_1, \ldots, r_{n_{\rm v}}$ planes. The symbols $\zeta_{\ell_{\rm i}}$ and $\zeta_{\ell_{\rm f}}$ denote the initial and the final vertices respectively of the $\ell$th edge, where the momentum $p_{\ell}$ flows from $\zeta_{\ell_{\rm i}}$ to $\zeta_{\ell_{\rm f}}$.

Such monodromy integrals were computed for tree Witten diagrams on the grSK contour in \cite{Loganayagam:2024mnj}. The authors conjectured that these complicated monodromy integrals over multiple grSK contours can be reduced to integrals over the exterior of the black hole. They also checked that the conjecture works for the simplest tree level diagrams.  This is a very useful result since it avoids the repeated computation of the monodromies of products of theta functions and $e^{-\beta \Cnst \zeta}$ over the grSK contour. Here, we would like to generalise this result to loop diagrams as well. We find that the conjecture of \cite{Loganayagam:2024mnj} works for loop diagrams as well. Thus we reproduce their results here for the convenience of the reader.

It was shown in \cite{Loganayagam:2024mnj} that it is convenient to think of the conjecture in terms of directed graphs. The expression in Eq.~\eqref{eq:MonIntExpression} can be viewed as an integral over the vertices in a directed graph, where the directions of the edges are specified by the momentum flow. For example, the integral 
\begin{equation}
    \oint_{\zeta_1} \oint_{\zeta_2}e^{\beta \Cnst_1 (1-\zeta_1)}e^{\beta (1-\Cnst_2) \zeta_2}\ \bbG(\zeta_2|\zeta_1, p_1)\bbG(\zeta_2|\zeta_1, p_2)\bbG(\zeta_2|\zeta_1, p_3) \mathscr{F}(\zeta_1, \zeta_2)
\end{equation}
would be associated with the directed graph
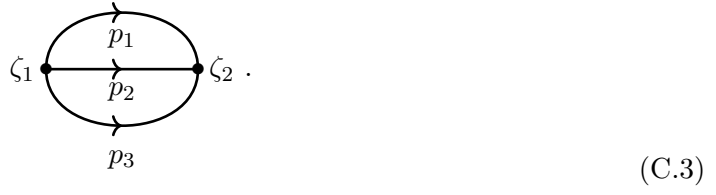
\begin{equation}
    \begin{tikzpicture}[line width=1pt]
        \begin{scope}[decoration={
        markings,
        mark=at position 0.5 with {\arrow{>}}}
        ] 
        \centering
          \coordinate (A) at (-1,0);  
          \coordinate (B) at (1,0);   
        
          \draw[postaction={decorate}] (A) -- (B);       
          \draw[postaction={decorate}] (A) .. controls (-1,1) and (1,1) .. (B);      
          \draw[postaction={decorate}] (A) .. controls (-1,-1) and (1,-1) .. (B);
          \node at (A){$\bullet$};
          \node at (B){$\bullet$};
          \node at (0,0.4) {$p_1$};
          \node at (0,-0.3) {$p_2$};
          \node at (0,-1.2) {$p_3$};
          \node[left] at (A) {$\zeta_1$};
          \node[right] at (B) {$\zeta_2  \ .$};
        \end{scope}
    \end{tikzpicture}
\end{equation}

According to the conjecture in \cite{Loganayagam:2024mnj}, we have to now consider the coloured descendants of this graph with two colours (red and blue). By this we mean that we have to consider all graphs that are obtained from a given graph by colouring its edges either red or blue. For instance, for the graph in the example above, there are eight coloured descendants given in the folllowing table.
\begin{table}[H]
\centering
\begin{tabular}{cccc}
  \begin{tikzpicture}[line width=1pt]
        \begin{scope}[decoration={
        markings,
        mark=at position 0.5 with {\arrow{>}}}
        ] 
        \centering
          \coordinate (A) at (-1,0);  
          \coordinate (B) at (1,0);   
        
          \draw[blue,postaction={decorate}] (A) -- (B);       
          \draw[blue,postaction={decorate}] (A) .. controls (-1,1) and (1,1) .. (B);      
          \draw[blue,postaction={decorate}] (A) .. controls (-1,-1) and (1,-1) .. (B);
          \node at (A){$\bullet$};
          \node at (B){$\bullet$};
          \node at (0,0.4) {$p_1$};
          \node at (0,-0.3) {$p_2$};
          \node at (0,-1.2) {$p_3$};
          \node[left] at (A) {$\zeta_1$};
          \node[right] at (B) {$\zeta_2  \ .$};
        \end{scope}
    \end{tikzpicture}  & \begin{tikzpicture}[line width=1pt]
        \begin{scope}[decoration={
        markings,
        mark=at position 0.5 with {\arrow{>}}}
        ] 
        \centering
          \coordinate (A) at (-1,0);  
          \coordinate (B) at (1,0);   
        
          \draw[blue,postaction={decorate}] (A) -- (B);       
          \draw[red,postaction={decorate}] (A) .. controls (-1,1) and (1,1) .. (B);      
          \draw[blue,postaction={decorate}] (A) .. controls (-1,-1) and (1,-1) .. (B);
          \node at (A){$\bullet$};
          \node at (B){$\bullet$};
          \node at (0,0.4) {$p_1$};
          \node at (0,-0.3) {$p_2$};
          \node at (0,-1.2) {$p_3$};
          \node[left] at (A) {$\zeta_1$};
          \node[right] at (B) {$\zeta_2  \ .$};
        \end{scope}
    \end{tikzpicture}   & \begin{tikzpicture}[line width=1pt]
        \begin{scope}[decoration={
        markings,
        mark=at position 0.5 with {\arrow{>}}}
        ] 
        \centering
          \coordinate (A) at (-1,0);  
          \coordinate (B) at (1,0);   
        
          \draw[red,postaction={decorate}] (A) -- (B);       
          \draw[blue,postaction={decorate}] (A) .. controls (-1,1) and (1,1) .. (B);      
          \draw[blue,postaction={decorate}] (A) .. controls (-1,-1) and (1,-1) .. (B);
          \node at (A){$\bullet$};
          \node at (B){$\bullet$};
          \node at (0,0.4) {$p_1$};
          \node at (0,-0.3) {$p_2$};
          \node at (0,-1.2) {$p_3$};
          \node[left] at (A) {$\zeta_1$};
          \node[right] at (B) {$\zeta_2  \ .$};
        \end{scope}
    \end{tikzpicture}   & \begin{tikzpicture}[line width=1pt]
        \begin{scope}[decoration={
        markings,
        mark=at position 0.5 with {\arrow{>}}}
        ] 
        \centering
          \coordinate (A) at (-1,0);  
          \coordinate (B) at (1,0);   
        
          \draw[red,postaction={decorate}] (A) -- (B);       
          \draw[red,postaction={decorate}] (A) .. controls (-1,1) and (1,1) .. (B);      
          \draw[blue,postaction={decorate}] (A) .. controls (-1,-1) and (1,-1) .. (B);
          \node at (A){$\bullet$};
          \node at (B){$\bullet$};
          \node at (0,0.4) {$p_1$};
          \node at (0,-0.3) {$p_2$};
          \node at (0,-1.2) {$p_3$};
          \node[left] at (A) {$\zeta_1$};
          \node[right] at (B) {$\zeta_2  \ .$};
        \end{scope}
    \end{tikzpicture}   \\
  \begin{tikzpicture}[line width=1pt]
        \begin{scope}[decoration={
        markings,
        mark=at position 0.5 with {\arrow{>}}}
        ] 
        \centering
          \coordinate (A) at (-1,0);  
          \coordinate (B) at (1,0);   
        
          \draw[blue,postaction={decorate}] (A) -- (B);       
          \draw[blue,postaction={decorate}] (A) .. controls (-1,1) and (1,1) .. (B);      
          \draw[red,postaction={decorate}] (A) .. controls (-1,-1) and (1,-1) .. (B);
          \node at (A){$\bullet$};
          \node at (B){$\bullet$};
          \node at (0,0.4) {$p_1$};
          \node at (0,-0.3) {$p_2$};
          \node at (0,-1.2) {$p_3$};
          \node[left] at (A) {$\zeta_1$};
          \node[right] at (B) {$\zeta_2  \ .$};
        \end{scope}
    \end{tikzpicture}   & \begin{tikzpicture}[line width=1pt]
        \begin{scope}[decoration={
        markings,
        mark=at position 0.5 with {\arrow{>}}}
        ] 
        \centering
          \coordinate (A) at (-1,0);  
          \coordinate (B) at (1,0);   
        
          \draw[blue,postaction={decorate}] (A) -- (B);       
          \draw[red,postaction={decorate}] (A) .. controls (-1,1) and (1,1) .. (B);      
          \draw[red,postaction={decorate}] (A) .. controls (-1,-1) and (1,-1) .. (B);
          \node at (A){$\bullet$};
          \node at (B){$\bullet$};
          \node at (0,0.4) {$p_1$};
          \node at (0,-0.3) {$p_2$};
          \node at (0,-1.2) {$p_3$};
          \node[left] at (A) {$\zeta_1$};
          \node[right] at (B) {$\zeta_2  \ .$};
        \end{scope}
    \end{tikzpicture}   & \begin{tikzpicture}[line width=1pt]
        \begin{scope}[decoration={
        markings,
        mark=at position 0.5 with {\arrow{>}}}
        ] 
        \centering
          \coordinate (A) at (-1,0);  
          \coordinate (B) at (1,0);   
        
          \draw[red,postaction={decorate}] (A) -- (B);       
          \draw[blue,postaction={decorate}] (A) .. controls (-1,1) and (1,1) .. (B);      
          \draw[red,postaction={decorate}] (A) .. controls (-1,-1) and (1,-1) .. (B);
          \node at (A){$\bullet$};
          \node at (B){$\bullet$};
          \node at (0,0.4) {$p_1$};
          \node at (0,-0.3) {$p_2$};
          \node at (0,-1.2) {$p_3$};
          \node[left] at (A) {$\zeta_1$};
          \node[right] at (B) {$\zeta_2  \ .$};
        \end{scope}
    \end{tikzpicture}   & \begin{tikzpicture}[line width=1pt]
        \begin{scope}[decoration={
        markings,
        mark=at position 0.5 with {\arrow{>}}}
        ] 
        \centering
          \coordinate (A) at (-1,0);  
          \coordinate (B) at (1,0);   
        
          \draw[red,postaction={decorate}] (A) -- (B);       
          \draw[red,postaction={decorate}] (A) .. controls (-1,1) and (1,1) .. (B);      
          \draw[red,postaction={decorate}] (A) .. controls (-1,-1) and (1,-1) .. (B);
          \node at (A){$\bullet$};
          \node at (B){$\bullet$};
          \node at (0,0.4) {$p_1$};
          \node at (0,-0.3) {$p_2$};
          \node at (0,-1.2) {$p_3$};
          \node[left] at (A) {$\zeta_1$};
          \node[right] at (B) {$\zeta_2  \ .$};
        \end{scope}
    \end{tikzpicture}   \\
\end{tabular}
\caption{Coloured descendants of the sunset loop graph.}
\end{table}

The blue coloured edges denote the retarded bulk-to-bulk propagators, while the red edges denote the advanced bulk-to-bulk propagators. The multiple disconinuity integral conjecture states that the grSK integral over the sum of exterior integrals over the coloured graphs. More precisely,
\begin{equation}
    \begin{split}
        &\oint_{\zeta_1} \ldots \oint_{\zeta_{n_\text{v}}} \prod_{i = 1}^{n_\text{v}} e^{\beta \Cnst_i (1- \zeta_i)} \prod_{\ell =1}^{n_{\text{e}}} \bbG(\zeta_{\ell_{f}}|\zeta_{\ell_{i}}, p_{\ell}) \mathscr{F}(\zeta_1, \ldots, \zeta_{n_{\text{v}}})\\
        &\qquad= \sum_{\text{graphs}} \int_{\zeta_1} \ldots \int_{\zeta_{n_{\text{v}}}}\prod_{i = 1}^{n_\text{v}} e^{\beta \Cnst_i (1- \zeta_i)} \left[1 - \exp \left(-\beta \Cnst_i +\beta \sum_{j\in \text{out. blue}}p_j-\sum_{j\in \text{in. red}}p_j\right)\right]\\
        &\qquad \quad \times \prod_{\ell = 1}^{n_{\text{e, blue}}} (-n_{p_\ell})\bbG_{\text{ret}}(\zeta_{\ell_{f}}|\zeta_{\ell_{i}}, p_{\ell}) \prod_{m = 1}^{n_{\text{e, red}}} (1+n_{p_m})\bbG_{\text{adv}}(\zeta_{m_{f}}|\zeta_{m_{i}}, p_{m}) \mathscr{F}(\zeta_1, \ldots, \zeta_{n_{\text{v}}}) \ .
    \end{split}
    \label{eq:MultipleDiscontGenFormula}
\end{equation}
Here, $\sum_{\rm graphs}$ denotes the sum over the coloured descendant graphs, and $\rm (out.\ blue)$ denotes all the blue edges that start at the given vertex and have outgoing momentum arrows, and $\rm (in.\ red)$ denotes all the red edges that start at the given vertex and have incoming momentum arrows.

In the present work, we have checked that this conjecture also works for loop diagrams. We did this by explicitly performing the contour integrals for the simplest loop diagrams. In particular, we have checked that this conjecture is valid for one loop graphs with the loops having two, three, and three vertices, as well as for the two-loop and three-loop melon graphs.

\subsection*{Double discontinuity for one loop with two propagators}

We will now take a moment to explain how the conjecture is checked for the one-loop graph with two propagators. In this case, the integral we are interested in of the form
\begin{equation}
    \oint_{\zeta_1} \oint_{\zeta_2} e^{\beta \Cnst_1 \zeta_1} e^{\beta \Cnst_2 \zeta_2}\ \bbG(\zeta_2|\zeta_1, p_1)\bbG(\zeta_2|\zeta_1, p_2) \mathscr{F}(\zeta_1, \zeta_2) \ .
\end{equation}
Since the function $\mathscr{F}(\zeta, \zeta_2)$ is analytic in the grSK keyhole, it has no poles at the horizon, in particular. Neither do the bulk-to-bulk propagators. Thus the contour integral reduces to an integral over the exterior with an appropriate monodromy, given by
\begin{equation}
    \begin{split}
        &\oint_{\zeta_1} \oint_{\zeta_2}e^{\beta \Cnst_1 (1-\zeta_1)}e^{\beta \Cnst_2 (1-\zeta_)}\ \bbG(\zeta_2|\zeta_1, p_1)\bbG(\zeta_2|\zeta_1, p_2) \mathscr{F}(\zeta_1, \zeta_2)\\
        &\hspace{2cm}= \int_{\rm Ext_1} \int_{\rm Ext_2}e^{\beta \Cnst_1 (1-\zeta_1)}e^{\beta \Cnst_2 (1-\zeta_)}\ \bbG_{\rm DD}(\zeta_2|\zeta_1, p_1, p_2) \mathscr{F}(\zeta_1, \zeta_2)\ ,
    \end{split}
\end{equation}
where $\bbG_{\rm DD}(\zeta_2|\zeta_1, p_1)$ is the function that encapsulates the double discontinuity, and is given by
\begin{equation}
\begin{split}
&\bbG_{\rm DD}(\z_2,\z_1|,p_1,p_2) \equiv  \bbG(\z_2|\z_1,p_1) \bbG(\z_2|\z_1,p_2)-e^{-\beta \kappa_1}\bbG(\z_2|\z_1+1,p_1) \bbG(\z_2|\z_1+1,p_2)\\
&\hspace{3.7cm}-e^{-\beta \kappa_2}\bbG(\z_2+1|\z_1,p_1) \bbG(\z_2+1|\z_1,p_2)\\
&\hspace{3.7cm}+e^{-\beta(\kappa_1 +\kappa_2)}\bbG(\z_2+1|\z_1+1,p_1) \bbG(\z_2+1|\z_1+1,p_2) \ ,
\end{split}
\end{equation}
which can be found to be equal to
\begin{equation}
    \begin{split}
        \bbG_{\rm DD}(\zeta_2|\zeta_1,p_1,p_2) &= \frac{1}{1+n_{\Cnst_1-p_1-p_2}}\frac{1}{1+n_{\Cnst_2}} [-n_{p_1}\bbGR(\zeta_2|\zeta_1, p_1)]\ [-n_{p_2}\bbGR(\zeta_2|\zeta_1, p_2)]\\
        &+\frac{1}{1+n_{\Cnst_1-p_1}}\frac{1}{1+n_{\Cnst_2+p_2}} [-n_{p_1}\bbGR(\zeta_2|\zeta_1, p_1)]\ [(1+n_{p_2})\bbGA(\zeta_2|\zeta_1, p_2)]\\
        &+\frac{1}{1+n_{\Cnst_1-p_2}}\frac{1}{1+n_{\Cnst_2+p_1}} [(1+n_{p_1})\bbGA(\zeta_2|\zeta_1, p_1)]\ [-n_{p_2}\bbGR(\zeta_2|\zeta_1, p_2)]\\
        &+\frac{1}{1+n_{\Cnst_1}}\frac{1}{1+n_{\Cnst_2+p_1+p_2}} [(1+n_{p_1})\bbGA(\zeta_2|\zeta_1, p_1)]\ [(1+n_{p_2})\bbGA(\zeta_2|\zeta_1, p_2)] \ .
    \end{split}
    \label{eq:DoubleDisc}
\end{equation}

\addcontentsline{toc}{section}{References}
\bibliographystyle{JHEP}

\bibliography{LoopReferences}

\end{document}